\documentclass{aa}  

\usepackage{graphicx}
\usepackage{array}
\usepackage{txfonts}
\usepackage{caption}
\usepackage{lipsum}
\usepackage{xcolor}
\usepackage{url}
\usepackage{natbib}
\usepackage{breqn}
\usepackage{diagbox}
\usepackage{cuted} 
\usepackage{amsmath}
\usepackage[mathscr]{euscript}
\usepackage{multirow}

\newcolumntype{C}{>{\centering\arraybackslash}p{0.2\textwidth}}
\usepackage{enumitem}
\usepackage{booktabs}
\bibpunct{(}{)}{;}{a}{}{,} 
\begin{document} 

\title{A deep-learning approach to the 3D reconstruction of  dust density and temperature  in star-forming regions}

\titlerunning{Reconstruction of 3D dust distributions}
\authorrunning{Victor F. Ksoll \& Stefan Reissl}
\author{Victor~F.~Ksoll \inst{\ref{inst1}} \and 
        Stefan~Reissl \inst{\ref{inst1}} \and 
        Ralf~S.~Klessen \inst{\ref{inst1}, \ref{inst2}} \and 
        Ian~W.~Stephens \inst{\ref{inst3},\ref{inst4}} \and 
        Rowan~J.~Smith \inst{\ref{inst9},\ref{inst10}} \and
        Juan D. Soler \inst{\ref{inst5}} \and
        Alessio Traficante \inst{\ref{inst5}} \and
        Philipp Girichidis \inst{\ref{inst1}} \and
        Leonardo Testi \inst{\ref{inst6}, \ref{inst7}}\and
        Patrick Hennebelle \inst{\ref{inst8}} \and
        Sergio Molinari \inst{\ref{inst5}}
        }
\institute{
\centering Universität Heidelberg, Zentrum für Astronomie, Institut für Theoretische Astrophysik, Albert-Ueberle-Str. 2,\\ D-69120 Heidelberg, Germany \label{inst1} \and
\centering Universität Heidelberg, Interdisziplinäres Zentrum für Wissenschaftliches Rechnen, Im Neuenheimer Feld 205,\\ D-69120 Heidelberg, Germany\label{inst2} \and
\centering Department of Earth, Environment, and Physics, Worcester State University, Worcester, MA 01602, USA\label{inst3} \and
\centering Center for Astrophysics, Harvard \& Smithsonian, 60 Garden Street, Cambridge, MA 02138, USA\label{inst4} \and
\centering Istituto di Astrofisica e Planetologia Spaziali (IAPS). INAF. Via Fosso del Cavaliere 100, 00133 Roma, Italy \label{inst5} \and
\centering Alma Mater Studiorum Università di Bologna, Dipartimento di Fisica e Astronomia (DIFA), Via Gobetti 93/2, I-40129, Bologna, Italy \label{inst6} \and
\centering INAF-Osservatorio Astrofisico di Arcetri, Largo E. Fermi 5, I-50125, Firenze, Italy \label{inst7} \and
\centering Université Paris Cité, Université Paris-Saclay, CEA, CNRS, AIM, F-91191, Gif-sur-Yvette, France \label{inst8} \and 
\centering School of Physics and Astronomy, University of St Andrews, North Haugh, St Andrews, KY16 9SS UK\label{inst9} \and
\centering Jodrell Bank Centre for Astrophysics, Department of Physics and Astronomy, University of Manchester, Oxford Road, Manchester M13 9PL, UK\label{inst10}
}
                                        
\abstract
   {}
    {We introduce a new deep-learning approach for the reconstruction of 3D dust density and temperature distributions from multi-wavelength dust emission observations on the scale of individual star-forming cloud cores ($< 0.2$\,pc).}
    {We constructed a training data set by processing cloud cores from the Cloud Factory simulations with the {\sc POLARIS}  radiative transfer code to produce synthetic dust emission observations at 23 wavelengths between $12$ and $1300\,\mu\mathrm{m}$. We simplified the task by reconstructing the cloud structure along individual lines of sight (LoSs) and trained a conditional invertible neural network (cINN) for this purpose. The cINN belongs to the group of normalising flow methods and it is able to predict full posterior distributions for the target dust properties. We tested different cINN setups, ranging from a scenario that includes all 23 wavelengths down to a more realistically limited case with observations at only seven wavelengths. We evaluated the predictive performance of these models on synthetic test data.}
    {We report an excellent reconstruction performance for the 23-wavelength cINN model, achieving median absolute relative errors of about $1.8$\% in $\log(n/\mathrm{m}^{-3})$ and $1\%$ in $\log(T_{\mathrm{dust}}/\mathrm{K})$, respectively. We identify trends towards an overestimation at the low end of the density range and towards an underestimation at the high end of both the density and temperature values, which may be related to a bias in the training data. After limiting our coverage to a combination of only seven wavelengths, we still find a satisfactory performance with average absolute relative errors of about $2.8$\% and $1.7$\% in $\log(n/\mathrm{m}^{-3})$ and $\log(T_{\mathrm{dust}}/\mathrm{K})$.} 
    {This proof-of-concept study shows that the cINN-based approach for 3D reconstruction of dust density and temperature is very promising and it is even compatible with a more realistically constrained wavelength coverage.} 
  \maketitle
%
\section{Introduction}
A fundamental limitation of astronomical observations is that they only give access to the two-dimensional (2D) projection of cosmic structures onto the plane of the sky. Consequently, it is a central theme of modern astronomical and astrophysical research to resolve this degeneracy and try to reconstruct the underlying three-dimensional (3D) structures. When measuring lines, for example, we can take spectra at equally spaced positions within the area of interest and build a 3D cube of position-position-velocity (i.e. line-of-sight velocity) information. This PPV data is often used as an approximation for the intrinsic 3D position-position-position (PPP) structure of the emitting region \cite[for further discussions, see e.g.][]{BallesterosParedes2002, Beaumont2013}. However, what to do when we are relying on continuum radiation, such as the thermal emission from interstellar dust grains \citep[e.g.][]{Molinari2010, planck2013-XI}, is less clear. To address this challenge, we present a new deep learning approach for the reconstruction of 3D morphological information and employ multi-band observations in the wavelength regime from  $12$ to $1300\,\mu\mathrm{m}$ to estimate the 3D spatial distribution of dust density and temperature of star-forming cloud cores.

Thermal emission from interstellar dust grains is the dominant source of radiation across the sky at mid- and far-infrared wavelengths \citep[see][and references therein]{hill2018}. This is the result of dust grains distributed throughout the interstellar medium (ISM), which are heated primarily by starlight and cool through thermal radiation \citep{Tielens2010,Draine2011,Klessen2016}. The dust grains are heated to temperatures between roughly 20 and 200\,K depending on the spectrum and intensity of the interstellar radiation field and the size and optical properties of the grains \citep[see][for a recent review]{galliano2018}.

Dust emission remains optically thin at relatively high column densities \citep[see, e.g.][]{planck2011-7.0}, hence, it provides a crucial observable to study regions of the universe that are not accessible at visible wavelengths. Observations of the dust thermal emission have been used to characterise the evolution of the universe through the study of the cosmic infrared background \citep[CIB, see][]{hauserANDdwek2001}. Radiation at far infrared wavelengths registered by the Herschel satellite has allowed for the reconstruction of star formation activity across the Milky Way disk \cite[][]{molinari2016,elia2021} and within nearby star-forming regions \cite[][]{andre2010}. Polarised dust thermal emission observed by the Planck satellite has provided the opportunity to infer the first whole-sky map of the projected Galactic magnetic field \cite[see, e.g.][]{planck2015-XXXV}.

Interstellar dust also plays many critical roles in galactic evolution. It is a catalyser for the formation of molecular hydrogen (H$_{2}$), while sequestering select elements in solid grains (see, e.g. \citealt{gouldANDsalperter1963} or \citealt{jones2019}). The interaction between dust grains and ultraviolet (UV) starlight releases electrons that can be the dominant source of heating for interstellar gas \citep{wolfire2003,gloverANDmaclow2011}.
Dust grains also transfer the  radiation pressure from starlight to the gas and couple it to the interstellar magnetic field through collisions (see, e.g. \citealt{Draine2003} for a review; or \citealt{Reissl2018} and \citealt{Reissl2023} for a microphysical model).  Thus, reconstructing the distribution of the dust is crucial for understanding the physical conditions of the ISM.

Existing attempts to reconstruct the 3D distribution of dust on galactic scales often take into account distance information from stars in combination with individual extinction measurements (or the combination of Gaia with auxiliary data; see e.g. \citealt{Lallement2018, Lallement2019, Lallement2022}, \citealt{Leike2020, Leike2022}, or \citealt{Zhang2023}). A similar approach has also been adopted for assessing the 3D structure of individual molecular clouds \cite[e.g.][]{RezaeiKh2017, RezaeiKh2020, Zucker2021, RezaeiKh2022} or for building a realistic model of the matter distribution in the solar neighbourhood \citep[e.g.][]{Zucker2022Nature, Zucker2023PPVII}. On the smaller scales of individual star-forming clumps, there have also been forward-modeling approaches introduced for the 3D dust distribution based on the combination with line data \citep[see, e.g.][]{Liseau2015} or from multi-frequency dust emission data by fitting overlapping Gaussian ellipsoids  \citep{Steinacker2005}. Previous attempts to use machine learning to invert the radiative transfer problem have also been reported \citep[e.g.][]{Garcia-Cuesta2009}.

In this study, we introduce a novel deep learning approach for the 3D reconstruction task, which employs a conditional invertible neural network \citep[][]{Ardizzone2019a, Ardizzone2019b}. The latter belongs to the group of methods based on normalising flows \citep[e.g.][]{Kobyzev2021} and has the advantage of giving access to the full posterior distribution function. For this reason, the cINN architecture is particularly well suited for solving degenerate inverse problems and has been successfully applied to a range of subjects in astronomy. These include: the characterisation of stellar properties from photometry \citep{Ksoll2020} or spectra \citep{Kang2023b}, prediction of exoplanet properties \citep{Haldemann2023}, analysis of emission lines in HII regions \citep{Kang2022, Kang2023a}, cosmic ray origin studies \citep{Bister2022}, and the reconstruction of galaxy assembly histories from numerical simulations \citep{Eisert2023}. Due to the lack of an observational ground truth sample, we have trained the cINN with data taken from numerical models of the turbulent multi-phase ISM, based on the Cloud Factory suite of simulations introduced by \cite{Smith2020_CF1}, which we post-processed using detailed radiative transfer calculations \citep[employing POLARIS, see][]{Reissl2016, Reissl2019} to bring them closer to the observational domain. 

This paper is structured as follows. In Section~\ref{sec:TrainingData}, we outline the construction of the training data for our method from synthetic dust cloud simulations, including our setup for radiative transfer. Section~\ref{sec:Method} provides a summary of the invertible neural network approach, specifications of the inverse problem, implementation details, and our analysis methods. In Section~\ref{sec:Results}, we present the evaluation of our trained models on synthetic test data and discuss the predictive performance of our approach. Lastly, Section~\ref{sec:Summary} summarises our main results.

\section{Training data}\label{sec:TrainingData}

The main goal of this study is to reconstruct 3D dust distributions from the observed dust emission for sites of star formation on the scale of individual cloud cores. 
As we want to tackle this task with a supervised deep-learning approach, we therefore require a training data set consisting of 3D dust distributions with their properties and corresponding dust emission observations. 
As such a data set does not yet exist for real observations, we have turned to simulations to build a suitable database for training. 

\subsection{Simulation data}
As a basis for our training data set we chose the AREPO-based \citep[an adaptive Voronoi mesh hydrosolver, see][]{Springel2010}, galactic-scale ISM hydrodynamics simulation suite Cloud Factory, introduced by \citet{Smith2020_CF1} and \citet{Smith2021_CF2}. The Cloud Factory self-consistently follows the formation of dense gas and molecular hydrogen  with an average molecular weight of $\mu_{\mathrm{g}}=2.4$ in a Milky Way-like galactic gas disc at radii $4\,\mathrm{kpc} < r < 12\,\mathrm{kpc,}$ including the effects of galactic scale forces, gas chemistry and cooling, and supernova feedback. The time-dependent chemical evolution is modelled as in \citet{Smith14a} using the hydrogen chemistry of \citet{Glover07a} and the simplified CO treatment of \citet{Nelson97}. It includes gas self-shielding from a UV field equivalent to that seen in the solar neighbourhood and cosmic-ray ionisation at the local rate as well. 

The Cloud Factory employs a series of nested zooms with a base mass resolution of $1000\,\mathrm{M}_\odot$ smoothly increased to $10\,\mathrm{M}_\odot$ within a co-rotating box of size 3 kpc. In the co-rotating box individual cloud complexes are then selected and their resolution further increased to $0.25\,\mathrm{M}_\odot$, which is equivalent to a spatial resolution better than 0.1pc in gas with number densities higher than $10^9$\,m$^{-3}$. By including the galactic scale forces that form the clouds, the Cloud Factory suite reproduces the turbulent gas motions on multiple scales as observed in the ISM. However, the current version of the suite does not include magnetic fields or other forms of stellar feedback, such as stellar winds, jets, or photoionisation.

For our analysis, we used Complex C and D, as shown in Figure 5 of \citet{Smith2020_CF1}, and only included gas at the highest resolution ($0.25\,\mathrm{M}_\odot$ or four cells per local jeans length, whichever is higher). These molecular cloud complexes were formed in regions that had previously experienced supernova feedback and, therefore, they already contained a well developed turbulent energy cascade. They are filamentary in structure, and extend for more than 100 pc along their longest axis. The density probability density function of the entire cloud complex peaks at number densities of around $10^8$\,m$^{-3}$, but extends beyond number densities of $10^{10}$\,m$^{-3}$, which marks the point when sink particles may form (see \citealt{Tress20a}). Sink particles represent regions of star formation and at this resolution, they correspond to single star systems that may actually be multiples. While sinks may form above densities of $10^{10}$\,m$^{-3}$, they will only be created if the gas passes energy checks to ensure it is bound and converging, so in practice the simulations include densities up to $10^{10}-10^{12}$\,m$^{-3}$. Similarly, neighbouring gas cells can have material accreted by the sinks, but only when the gas becomes bound to the sink.

A high spatial resolution allows us to select the compact cloud cores on the scale of individual star-forming clumps that we want to analyse in this study. As a starting point, we extracted compact pre-stellar cloud cores (i.e. dense, cloud-like structures that have not formed a sink and are smaller than $0.2$~pc) for this purpose from the Cloud Factory. However, we did not just want to model these early phases of star formation, but also more evolved star forming regions, particularly those affected by nearby stars. As the star formation prescription in the Cloud Factory forms sink particles on cluster scales (rather than individual stars), we chose to modify the raw simulation data in a post processing step by manually adding a star to model the types of evolved star-forming regions that we want to consider in this study. We guided these modifications on the example of a well-observed real-world counterpart of this type of star-forming cores, such as the nearby \citep[$d=120$~pc, ][]{Loinard2008}, compact ($0.1$ pc), star-forming core $\rho$ Oph A \citep{Loren1990} in the Ophiuchus star-forming region. 

We began our training set construction by cutting out cubes centered on high-density, core-like gas aggregations from the Cloud Factory in the complexes C and D, so that each cube contains a mass between $5$ to $40\,\mathrm{M}_\odot$ and a substructure with a gas number density of at least $10^{10}$ to $10^{11}\,\mathrm{m}^{-3}$ (i.e. the lower density end of low-mass star-forming regions). In practise, the candidate positions for these cubes are determined by applying a density threshold to column density maps generated for three different (perpendicular) viewing angles of the complexes. We note that beyond these criteria, the cubes were randomly selected and do not represent any specific real-world counterpart star-forming core. For simplicity, we matched the Cloud Factory data to a regular grid. Initially, these cubes are selected on a $64\times 64 \times 64$ pixel resolution, corresponding to a physical cube size of $0.4$ pc. In total, we prepared a sample set of $11\ 036$ individual cubes from The Cloud Factory in this first step. The initial large cube size serves primarily to avoid edge artefacts that can occur in the radiative transfer simulation when synthesising the dust temperature in post-processing for the later synthetic dust emission observations. For the final training data, we actually cropped out the inner $32 \times 32 \times 32$ pixel cubes, corresponding to a total $0.2$ pc edge length. This resolution and size were chosen to keep the problem simple for this proof-of-concept and to reduce the overall training data size to facilitate the data handling during the training phase of our approach.

\subsection{Synthetic images}
\label{sec:TrainingData_SyntheticImages}
To produce synthetic dust emission observations from the selected cloud model cubes, we employed the Monte Carlo (MC) radiative transfer (RT) code {\sc POLARIS}\footnote{{\sc POLARIS} website: \url{https://portia.astrophysik.uni-kiel.de/polaris/}} \citep{Reissl2016, Reissl2019}. {\sc POLARIS} calculates a dust temperature based on a given 3D density distribution and a specific dust composition, assuming an instantaneous temperature correction and a thermal equilibrium between the dust and its surroundings (for details we refer to \citealt{Lucy1999} and \citealt{Bjorkman2001}). For the subsequent RT dust heating and emission simulations, we assumed a dust mass to gas mass ratio of $\delta_{\mathrm{gd}} = 1\%$ and a material composition of $37.5\%$ graphite and $62.5\%$ (astro)silicate for the grains typical for the ISM. The applied grain sizes are ${ a\in [5\ \mathrm{nm}, 250\ \mathrm{nm}] }$ and the number of grains, $N,$ follows a power-law $N(a) \propto a^{-3.5}$ \citep[see e.g.][for further details]{Mathis1977,LiDraine2001}. We emphasise that the selected dust parameters in combination with the average molecular weight of the gas, $\mu_{\mathrm{g}}$, define an exact conversion factor from gas to dust density. Because of this, we use the gas density as a measure of the dust density in the following, without explicitly performing the conversion to remain consistent with the Cloud Factory.

We began by preparing the models for two distinct RT setups. In the first one, we considered the dust clouds to be only subject to the diffuse interstellar radiation field (ISRF). Here, we used the parameterisation of \cite{Mathis1983} for the spectral energy distribution with an intensity of $G_0=3, $ which is typical for star-forming cores \citep{Liseau2015}. In the second scenario, we also added a single star inside the cube in addition to the background ISRF. We used the parameters of a typical B4-type star ($R=4.33\ R_\odot$, $T_{\mathrm{eff}}=16\ 000\ \mathrm{K}$) in our MC dust heating simulation. We note that the dust MC heating by {\sc POLARIS} in this step does not modify the ionisation state of the gas or redistribute the gas by means of radiatiave feedback. In each individual cube, we simply placed the B4-analogue star inside the inner $32 \times 32 \times 32$ pixels, selecting a point of low gas density. This procedure roughly emulates the fact that the feedback of such a star would likely clear out its immediate surroundings.

We generated synthetic, monochromatic dust emission observations with a $32 \times 32$ pixel resolution (matching the resolution of the underlying dust distribution) at 23 wavelengths between $12$ and $1300\,\mu\mathrm{m}$, matching the central wavelengths of bands available at various observational facilities (see Table~\ref{tab:FilterBeamSizes} in the Appendix for a full list).
We note that we only generate synthetic observations for one viewing angle for each cube. In principle, it would be possible to include multiple viewing angles to increase the size of the training dataset, but given that our simulation suite provides a sufficiently large dataset for training from different physical regions, this is not necessary here. Nevertheless, after training, we also conducted a performance test of our algorithm on an individual region observed from different viewing angles (Appendix \ref{app:RotationTest}), which confirms that the reconstructed 3D structure is nearly identical.

The choice for monochromatic emission observations is again made for simplicity, as modelling the full instrument responses of the considered bands is quite complex and beyond the scope of this proof of concept. Nevertheless, we wanted to select wavelengths that are actually accessible with current observational facilities; thus, we employed the corresponding central wavelengths. Henceforth, we refer to the different wavelengths by the names of the respective instrument bands in the following. We note that we did not consider wavelengths shorter than $12\,\mu\mathrm{m}$ because the influence of scattered light becomes non-negligible in this regime, adding extra complexity. At the current stage of our development, we have not considered instrumental effects related to the point spread functions (PSFs) of the various telescopes or observational noise. Thus, we treat our synthetic observations as fully resolved at all wavelengths and uncertainty free. Properly modelling these effects is  not trivial either, particularly with respect to interferometric observations with ALMA, where simulations with a dedicated processing tool such as {\sc CASA}\footnote{{\sc CASA} website: \url{https://casa.nrao.edu/}} \citep[][]{CASATeam2022} would be necessary. Thus, we reserve a proper treatment of these effects for a follow-up work. Still, we want to note that accounting for uncertainties is well within the capabilities of the invertible neural network architecture used in this work, as demonstrated by \cite{Kang2023a}.

We initially generated our synthetic dust emission observations assuming a distance of $3.703 \times 10^{18}$\,m ($120$\,pc). To build a more generally applicable approach, we then rescaled the synthetic fluxes following:
\begin{equation}
    \hat{f} = f \times \frac{d^2}{d_{\mathrm{ref}}^2},
\end{equation}
where $d$ denotes the actual distance and $d_{\mathrm{ref}}$ is the reference distance (which the flux is scaled to) to determine a distance independent absolute flux measure. The choice of $d_{\mathrm{ref}}$ is arbitrary and since we operate on the logarithm of the fluxes in the following (see Section~\ref{sec:Methods_DataPreprocessing}) only represents  a linear offset to all fluxes, which will not notably affect the training outcome of our neural network approach. For simplicity, we set $d_{\mathrm{ref}} = 1$\,m, so that the offset is zero in logarithmic space.

Having a complement of observations at $23$ different wavelengths for a single real cloud core is typically unrealistic. Given the complexity of the 3D reconstruction task, we started out with this unrealistically large wavelength coverage to emulate a perfect information scenario and determined the best predictive performance our approach could achieve for this proof of concept. In addition, we also investigated a second, more realistically limited scenario, where we considered synthetic observations at the  central wavelengths of the following bands: WISE $22\ \mu$m, SOFIA $89\ \mu$m and $154\ \mu$m, Herschel PACS $100\ \mu$m and $160\ \mu$m, Herschel SPIRE $350\ \mu$m, and LABOCA $870\mu$m (see Table~\ref{tab:FilterBeamSizes}). This particular combination of bands is inspired by real observational data, which is, for instance, available for the $\rho$ Oph A star forming cloud \citep[][]{Liseau2015,Santos2019}. We emphasise here that this particular wavelength selection does not necessarily preserve the most information for the given inverse problem and that there may be a much more optimal subset of seven wavelengths among our total of 23 to maximise the reconstructive performance of the approach outlined below. Determining this combination is beyond the scope of this proof of concept and we reserve this to a dedicated follow-up study.

\section{Reconstruction approach}
\label{sec:Method}

\begin{figure*}
    \centering
    \includegraphics[width=\textwidth]{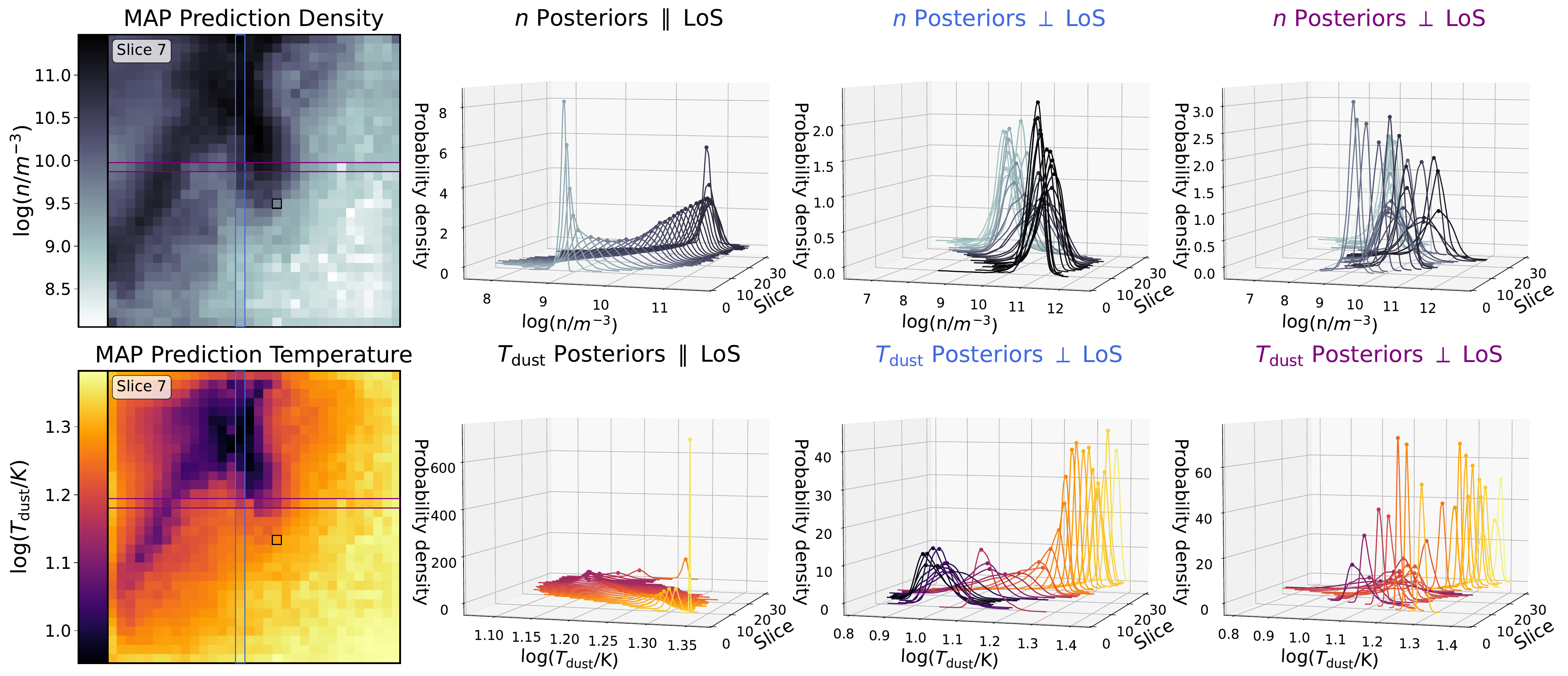}
    \caption{Comparison of the predicted posterior distributions along vs. perpendicular to the LoS (into the plane). The left column shows an example slice with the MAP estimates for dust density (top) and dust temperature (bottom). The other three columns show the posterior distributions of dust density (top) and temperature (bottom) for the lines indicated in the left panels in black, blue and purple, respectively. Here, the black square denotes a LoS going into the plane of the image, whereas the blue and purple lines are perpendicular to the LoS along the x and y axis, respectively.} 
    \label{fig:SlicePosteriorExample}
\end{figure*}

\begin{figure*}
    \centering
    \includegraphics[width = 0.75\textwidth]{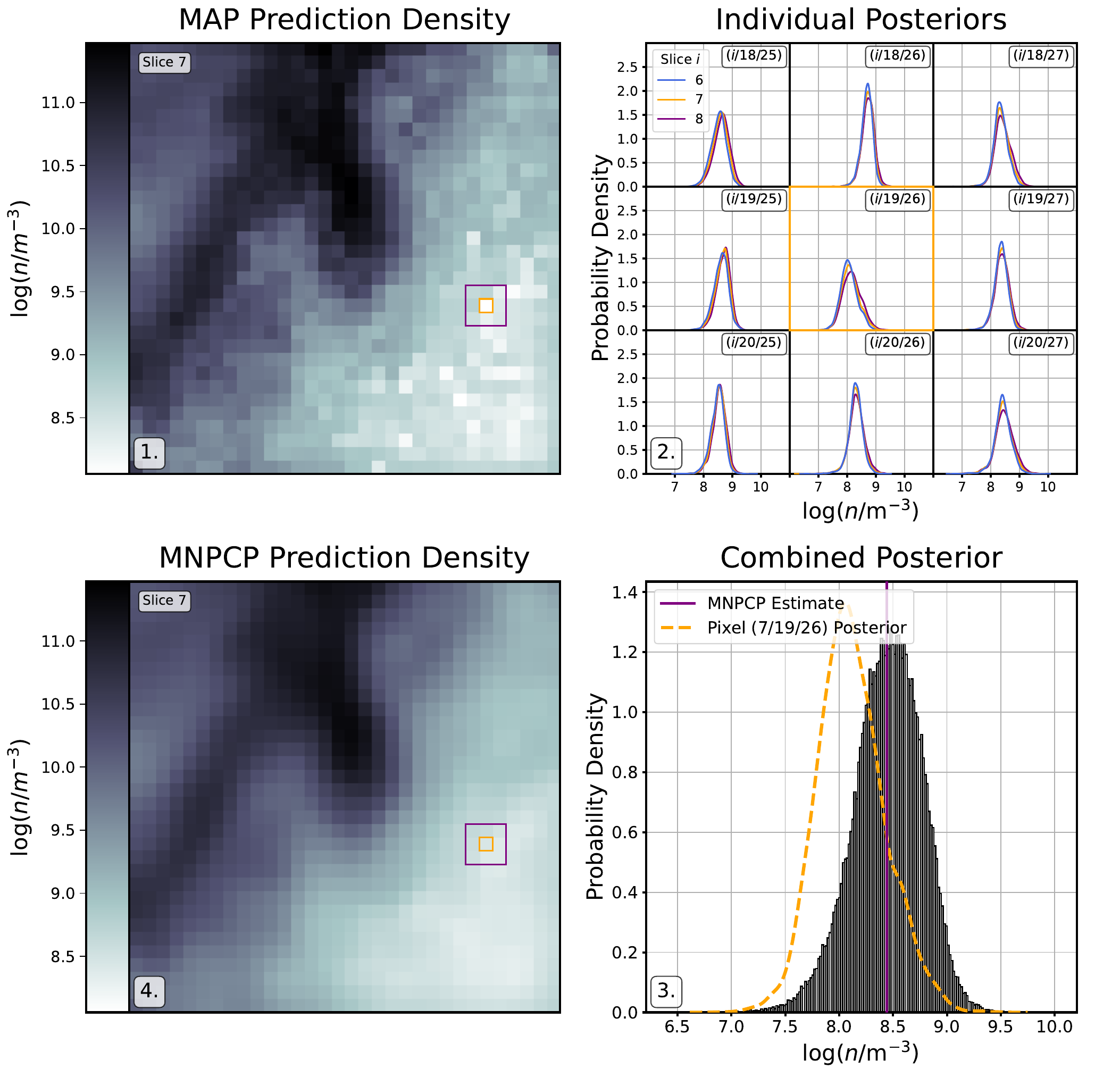}
    \caption{Schematic outline of the median neighbour pixel combined posterior approach for point estimations based on the example of dust density. The procedure follows the panels from top left to bottom left in clockwise order. The top-left panel shows the MAP density estimates for a single slice of a dust cube (perpendicular to the LoS), highlighting the discontinuities that can occur in the MAP estimate.  The query pixel, for which the MNPCP estimate will be computed, and its neighbourhood are indicated in orange and purple, respectively. The top-right panel shows the predicted posterior distributions for the dust density for the query pixel with index 7/11/8 (center subpanel indicated by the orange outline) and all 26 neighbouring pixels. The colours of the curves indicate the slice, $i,$ that they come from, whereas the indices in the top right corner denote the pixel index perpendicular to the LoS. The bottom-right panel shows a histogram of the all posterior samples accumulated from the query pixel and its neighbours. Here, the orange dotted curve indicates the original (marginalised 1D) posterior distribution of the query pixel and the purple line marks the MNPCP estimate, which is the city-block distance weighted median of all posterior samples (the distance in pixels when allowing only right angle moves, no diagonals). Finally, the bottom-left panel presents the MNPCP dust density estimates of the slice shown in the top left with the orange and purple boxes indicating again the query pixel and its neighbourhood.}
    \label{fig:mnpcp_example}
\end{figure*}

To solve the inverse problem of recovering the 3D dust temperature and dust density distribution from the observed dust emission maps, we employed a supervised deep learning approach called an invertible neural network (INN). In the following, we provide a short summary of this methodology and outline our specific setup for the 3D dust reconstruction task.

\subsection{The conditional invertible neural network}
The INN \citep{Ardizzone2019a} belongs to the greater family of normalising flows \citep[NFs,][]{Tabak2010, Tabak2013, Dinh2015, Rezende2015, Kobyzev2021}. More specifically, these are deep learning approaches that model complex distributions through sequences of invertible transformations of simpler known probability distributions \citep[see also][for a review]{Kobyzev2021}. Among the NF methods, the INN stands as a neural network (NN) architecture that is particularly well suited for solving degenerate inverse problems. Introducing a set of latent variables $\mathbf{z}$ to encode the information loss in the forward mapping $\mathbf{x} \rightarrow \mathbf{y}$ from the physical parameters $\mathbf{x}$ to a set of observables $\mathbf{y}$, which renders the inverse problem $\mathbf{y} \rightarrow \mathbf{x}$ degenerate, the INN can estimate full posterior distributions $p(\mathbf{x}|\mathbf{y})$ for the target parameters. This allows this method to both highlight and in some cases even break degeneracies in solving the inverse problem.

In this study we employ an INN architecture called conditional invertible neural network \citep[cINN,][]{Ardizzone2019b}. During training, this method learns a mapping of the physical parameters $\mathbf{x}$ to the latent variables $\mathbf{z}$, conditioned on the observables $\mathbf{y}$, that is the forward mapping denoted as:
\begin{equation}
    \mathbf{z} = f(\mathbf{x}; \mathbf{c} = \mathbf{y}).
\end{equation}
In doing so, the cINN encodes all variance of the physical parameters that is not explained by the corresponding observables in the latent variables, while the training process explicitly maintains a prescribed prior distribution $P(\mathrm{z})$ for the latent variables.
At prediction time, the cINN can then query this encoded variance by drawing samples from the known prior distribution $P(\mathbf{z})$ of the latent variables and once it has been conditioned on a new query observation $\mathbf{y}'$, it can make use of its fully invertible architecture to generate corresponding samples of the posterior distribution $p(\mathbf{x}|\mathbf{y}')$ following:
\begin{equation}
    p(\mathbf{x}|\mathbf{y}') \sim g(\mathbf{z}; \mathbf{c} = \mathbf{y}') \,\,\,\mathrm{with}\,\,\, \mathbf{z} \propto P(\mathbf{z}),
\end{equation}
where $g(\cdot;\mathbf{c}) = f^{-1}(\cdot;\mathbf{c})$ denotes the inverse of the forward mapping, $f,$ for fixed condition, $\mathbf{c}$. For simplicity, $P(\mathbf{z})$ is usually prescribed to be a multivariate normal distribution with zero mean and unit covariance. The dimension of the latent space $\dim(\mathbf{z})$ is per construction equal to the dimension of the target parameter space $\dim(\mathbf{x})$. On the other hand, as the observations are treated as a condition their dimension can become arbitrarily large. In fact, the architecture of the cINN allows for the introduction of a feature extraction network, trained in tandem with the cINN itself, to transform the input observations into a more useful (learned) representation \citep{Ardizzone2019b}. 

The invertibility of the cINN is achieved by employing so called conditional affine coupling blocks \citep{Dinh2017, Ardizzone2019b}. After splitting their input vector $\mathbf{u}$ into two halves $\mathbf{u}_1$ and $\mathbf{u}_2$, these coupling blocks perform two complementary affine transformations:
\begin{equation}
    \label{eq:coupling_forward}
    \begin{split}
        \mathbf{v}_1 &= \mathbf{u}_1 \odot \exp\left(s_2(\mathbf{u}_2, \mathbf{c})\right) \oplus t_2(\mathbf{u}_2, \mathbf{c}), \\
        \mathbf{v}_2 &= \mathbf{u}_2 \odot \exp\left(s_1(\mathbf{v}_1, \mathbf{c})\right) \oplus t_1(\mathbf{v}_1, \mathbf{c}), 
    \end{split}
\end{equation}
to compute the halves, $\mathbf{v}_1$ and $\mathbf{v}_2$, of the output vector, $\mathbf{v}$, where $\odot$ and $\oplus$ denote elementwise multiplication and addition, respectively. Here, $s_i$ and $t_i$ represent arbitrarily complex transformations of the concatenation of $\mathbf{u}_i$/$\mathbf{v}_i$ and the conditioning input $\mathbf{c}$. To run the network in reverse,  Equation~\ref{eq:coupling_forward} is then trivially inverted given the output vector $\mathbf{v} = (\mathbf{v_1}, \mathbf{v_2})$ following:

\begin{equation}
    \label{eq:coupling_backward}
    \begin{split}
        \mathbf{u}_2 &= \left(\mathbf{v}_2 \ominus t_1(\mathbf{v}_1, \mathbf{c})\right) \odot \exp\left(-s_1(\mathbf{v}_1, \mathbf{c})\right), \\
        \mathbf{u}_1 &= \left(\mathbf{v}_1 \ominus t_2(\mathbf{u}_2, \mathbf{c})\right) \odot \exp\left(-s_2(\mathbf{u}_2, \mathbf{c})\right),
    \end{split}
\end{equation}
where $\ominus$ denotes elementwise subtraction. As the transformations $s_i$ and $t_i$ are always evaluated in the same direction in both the forward, Eq.~\eqref{eq:coupling_forward} and backward pass, Eq.~\eqref{eq:coupling_backward}, of the coupling block, it is not necessary to choose them to be  invertible themselves. In fact, $s_i$ and $t_i$ do not even need to be prescribed, but can be learned instead during the training of the cINN by representing them with small sub-networks; for instance, a fully connected neural network \citep{Ardizzone2019a, Ardizzone2019b}. The specific setup of the cINN and the coupling block architecture used in this work is described in Section~\ref{sec:Method_ImplementationDetails}.

\subsection{Single LoS reformulation}
\label{Sec:Method_SLoS}
The full inverse problem of the 3D reconstruction task consists of predicting the $2 \times N \times N \times N$ hypercube $\mathbf{X}$ of dust densities and temperatures from the $K \times N \times N$ cube, $\mathbf{Y,}$ containing the corresponding observed dust emission in the  $K$ different wavelength. Evidently, even for the small resolution of $N=32$ that we selected, this is a very high-dimensional problem. 
To simplify our approach and mitigate the difficulties of high dimensionality, we therefore decided to reformulate the inverse problem. 
In particular, we reduced the 3D reconstruction problem to a matter of individual LoSs under the main assumption that the emission measured in any given pixel is independent of its neighbouring pixels. This independence assumption is especially valid for the perfectly resolved observation scenario that we consider here, but it should also hold (at least to first order) for real observations unless the PSF of the instrument is significantly larger than the pixel size, where smearing could become an issue. With that, we aim to use the vector of the $K$ measured emission fluxes $(e_{\lambda_{1}}, \ldots, e_{\lambda_{K}})$ of a given pixel to recover the corresponding  dust density $(n_{1}, \ldots, n_{N})$ and temperature $(T_{\mathrm{dust, 1}}, \ldots, T_{\mathrm{dust, N}})$ vectors. To avoid having to train two networks, we further combined the line-of-sight (LoS) dust densities and temperatures into a single vector, formulating the inverse problem as:
\begin{equation}
    \begin{split}
        \mathbb{R}^K &\rightarrow \mathbb{R}^{2N} \\
        (e_{\lambda_{1}}, \ldots, e_{\lambda_{K}}) &\rightarrow (n_{1}, \ldots, n_{N}, T_{\mathrm{dust, 1}}, \ldots, T_{\mathrm{dust, N}}),
    \end{split}
\end{equation}
for a given LoS. 

Within the cINN framework, the LoS emission vectors, $\mathbf{y} = (e_{\lambda_{1}}, \ldots, e_{\lambda_{K}}),$ correspond to the conditioning input and the combined vector of dust densities and temperatures, while $\mathbf{x} = (n_{1}, \ldots, n_{N}, T_{\mathrm{dust, 1}}, \ldots, T_{\mathrm{dust, N}})$ denotes the target parameters. Consequently, the latent space introduced by the cINN has a dimension equal to that of $\mathbf{x}$, that is: $2N$.

At prediction time, a given query $K \times N \times N$ cube of emission maps is first decomposed into the $N^2$ line of sight emission vectors of length $K$. For each of these LoSs $i$ (with $i \in \{1, \ldots, N^2\}$), the corresponding emission vector, $\mathbf{y}_i$, is then processed by the trained cINN, generating $S$ samples of the full $2N$-dimensional joint posterior distribution $p(\mathbf{x}_i|\mathbf{y}_i)$ by sampling the latent space according to the known Gaussian prior distribution $P(z)$.
Afterwards, we reassemble these $N^2$ LoS prediction results of size $2N \times S$ into the $2 \times N \times N \times N \times S$ hypercube, $\mathbf{X}_{\mathrm{samp}}$, of the density and temperature posterior samples. 

It is worth noting that with this LoS decomposition approach, our method is not limited to the $32 \times 32$ pixel resolution in the plane of sky, so that larger dust emission maps can be processed as long as they match the physical resolution of $6.25 \times 10^{-3}$ pc per pixel of the training data. With regard to depth, however, the presented approach is always limited to the 32 pixel depth corresponding to a physical size of $0.2$ pc that the cINN is trained on. A possible avenue to create a more depth flexible extension of the method presented here could be to train the cINN on data with varying per pixel resolution and providing the physical resolution as an additional conditioning input. Because this would require a substantially larger training data set and notably increases the complexity of the inverse problem (perhaps even beyond the point of feasibility), we reserve this experiment to our follow-up studies and focus on the fixed depth scenario here. In any case, with the presented approach it is always possible to tailor the training data towards the characteristic sizes of the objects that are to be analysed and train a correspondingly specialised model, as we have done here for the example of very compact, star-forming cores.

\subsubsection{Final training data set}
Following the prescription of the reformulated inverse problem, we decomposed the $2\times 11036$ training cubes into their respective LoSs, netting a total of $2 \times 11036 \times 32 \times 32 = 22,601,728$ vectors. Figure~\ref{fig:TS_priors_targets} in the appendix shows the corresponding effective prior distributions prescribed by the training data for dust density and temperature across all pixels of these lines of sights, as well as a correlation diagram indicating the coverage in the density-temperature space. In particular, we have covered a total density range from $3.3 \times 10^6$ to $2.2 \times 10^{13}\,\mathrm{m}^{-3}$, although most of the data is concentrated between $10^8$ and $10^{12}\,\mathrm{m}^{-3}$. The effective prior distribution for dust temperature ranges from $6.3$ to $240$ K, but is fairly skewed towards the $13$ to $24$ K interval, so that there are comparatively a lot fewer training pixels above a temperature of $32$~K. This is a direct consequence of the fact that such high dust temperatures only occur in the relatively few pixels in the vicinity of a star. Given that half of our training cubes do not contain a star, the per-pixel dust temperature prior distribution is  naturally biased towards this intermediate dust temperature regime because there are simply much more pixels that are either only subject to the ISRF to begin with or far enough away from the star to avoid being heated to very high temperatures. A corresponding diagram of the prior distributions of the measured fluxes at the 23 considered wavelengths across all pixels is provided in Figure~\ref{fig:TS_priors_observables} in the appendix. We emphasise that as a data-driven approach, the cINN is mostly limited to the parameter space covered by the training data. Although the cINN does exhibit some capability for extrapolation beyond the limits of the learned parameter space, there is in general no guarantee for a (physically) sound prediction outcome for inputs and targets that fall outside the described ranges. 

We further split this data set randomly into a training (80\% of the data) and test set (20\% of the data). The latter serves as held-out data that is not seen during training of the cINN to later evaluate the convergence and performance of the model. While the split is in general randomly chosen, we make sure that the held-out test set contains a subset of 100 complete cubes. For this subset, we selected the same 50 cubes twice: once  subject solely to the ISRF and once in the ISRF + star configuration. The aim is to evaluate how much the radiation setup affects the prediction outcome for the dust density and temperature. While we verified the model convergence on the greater test data set, the reported performance and all diagrams presented in Section~\ref{sec:Results} are based on this subset of 100 coherent cubes. Although this set of $100 \times 32 \times 32 = 102,400$ LoSs only represents $0.5\%$ of the total data, it has been selected as a representative subset of the test data set in order to keep the memory requirements at a manageable level. For instance, storing~the predicted posterior samples for these 100 cubes as an uncompressed csv table following the setup outlined further below already requires $\sim 360$\,GB of memory.  

\subsection{Implementation details}
\label{sec:Method_ImplementationDetails}
We employed the Python deep learning module {\sc pytorch} \citep{Paszke2017_pytorch} and the dedicated Framework for Easily Invertible Architectures \citep[FrEIA\footnote{Available at https://github.com/vislearn/FrEIA},][]{Ardizzone2019a, Ardizzone2019b} package to implement the cINN approach. For the affine coupling blocks, we employed the Generative Flow \citep[GLOW;][]{Kingma2018} configuration, in which the transformations $s_1$, $t_1$ and $s_2$, $t_2$ were jointly estimated by one sub-network each, which reduces the number of sub-networks in each coupling block from four to two. As sub-networks we utilised simple, fully connected networks with three layers of size $1024$ and the rectified linear unit (ReLU) activation function. As in \citet{Ardizzone2019b}, we also introduced a clamping procedure in the affine transformations in Eqs.~\eqref{eq:coupling_forward} and Eqs.~\eqref{eq:coupling_backward} to the argument, $s,$ of the exponential functions of the form: 
\begin{equation}
    s_{\mathrm{clamp}} = \frac{2\alpha}{\pi} \arctan\left(\frac{s}{\alpha}\right),
\end{equation}
with $\alpha = 1.9$. This procedure avoids instabilities arising from exploding magnitudes of $\exp(s)$. Furthermore, we alternated the affine coupling layers with random permutation layers, which randomly (but in a fixed and thus invertible manner) permute the output vector between each coupling layer to better intermix the information between the two streams $\mathrm{u}_1$ and $\mathrm{u}_2$. Our final network architecture (as determined via hyperparameter optimisation) is made up of nine coupling blocks in total. We also employed a simple feature extraction network, consisting of a three-layer (with $512$, $512,$ and $256$ nodes, respectively) fully connected network with ReLU activation functions, trained jointly with the cINN, to process the input observations.

\subsubsection{Additional data preprocessing}
\label{sec:Methods_DataPreprocessing}
Prior to training, we converted both the dust density and temperature to logarithmic space. This serves to prevent issues during training that can occur when the target parameters have a large dynamic range. This is particularly notable in the case of the dust density, which covers almost seven orders of magnitude. In addition, this implicitly ensures that the predicted dust densities and temperatures are always strictly positive. Afterwards, we performed two linear scaling operations on the training data. Each element $x_i$ of the target parameters, $\mathbf{x,}$ was rescaled by subtraction of its mean (over the entire training set) and then by division by its standard deviation, so that the resulting distribution of the rescaled $\hat{x}_i$ has zero mean and unit standard deviation. For the observables, we applied a matrix whitening procedure \citep{Hyvarinen2000} to the $M \times K$ matrix of training observations, $\mathbf{Y}$, where $M$ is the number of training examples and $K$ is the dimension of a single observation, $\mathbf{y}$, such that the rescaled observable matrix, $\hat{\mathbf{Y,}}$ has a unit covariance matrix. Given they are linear transformations, these scaling operations are easily inverted to convert the cINN output back to the true target parameter space. The coefficients of these scaling operations were determined on the training data and at the prediction time applied in the same fashion to the new query input. 

\subsubsection{Training setup and sampling strategy}
We trained our cINN approach via minimisation of the maximum likelihood loss, $\mathcal{L}$, as described in \cite{Ardizzone2019b}, namely:
\begin{equation}
    \mathcal{L} = \mathbb{E}_i \left(\frac{||f(\mathbf{x}_i; \mathbf{c}_i, \Theta)||_2^2}{2} - J_i \right),
\end{equation}
where $J_i = \det(\partial f/\partial \mathbf{x}|_{\mathbf{x} = \mathbf{x}_i})$ denotes the determinant of the Jacobian matrix evaluated at training instance $\mathbf{x}_i$ and $\Theta$ represents the network weights.
During training, the network weights, $\Theta,$ that minimise the loss function, $\mathcal{L,}$ are determined using a standard stochastic gradient descent approach. This means that after making an initial random guess for the weights, they are iteratively updated in the direction of the gradient $\nabla_\Theta \mathcal{L}$ based on randomly drawn subsets (batches) of the training data until a convergence is reached. In particular, we employ the adaptive learning rate, momentum-based Adam \citep[adaptive moment,][]{Kingma2018} optimiser for this purpose (with $\beta_1 = \beta_2 = 0.8$).
Here, we start with an initial learning rate (for Adam this is a scaling factor for the adaptive  step size in the weight updates along the loss gradient) of $l_{\mathrm{init}} = 9.642 \times 10^{-5}$ and then we reduced it by a factor of $\gamma = 0.831$ every $11$ epochs. In total, our models were trained for $250$ epochs, using a batch size of $512$ and processing $4096$ batches per epoch. We also employed an L2 weight regularisation with $\lambda = 6.093 \times 10^{-5}$. This setup was determined via hyperparameter optimisation, using the Hyperband algorithm \citep{Li2018_Hyperband}, a procedure that combines a random grid search approach with adaptive resource allocation and an early stopping criterion. Hyperband provides an efficient framework to test a large number of (randomly generated) hyperparameter configurations that finds a balance between running the training in full only for configurations that appear promising early (i.e.~converge fast), while also allowing for some slower converging models that might reach a better final result. For more details on the logistics of Hyperband we refer to \cite{Li2018_Hyperband}. Training a single network with the final setup described above takes about 19 hours using GPU acceleration on a NVIDIA RTX 2080Ti graphics card. 

At the prediction time, we then generated $S = 4096$ posterior samples for each new query LoS. This number of samples is chosen as a compromise between storage requirements and sample density, although experiments with even larger sample numbers have actually not shown a notable difference in the predicted posterior distributions, so this did not seem necessary within the framework of our analysis. A trained cINN can generate this amount of samples for $1024$ LoSs (that is a single cube) in about $28$ seconds (on a NVIDIA RTX 2080Ti), making the inference of the posterior distributions of the dust properties very efficient. 

\subsection{Making point estimates}
To better compare the cINN predictions to the ground truth hypercubes, $\mathbf{X,}$ in our synthetic test set, we computed a point estimate $\hat{\mathbf{X}}$ from the hypercube of posterior distribution samples, $\mathbf{X}_{\mathrm{samp}}$, returned by the cINN. The most straightforward approach for this is to derive the maximum a posteriori (MAP) prediction values for the dust density and temperature in every pixel of the 3D cube, which consists of determining the most likely value of the target parameters from the corresponding posterior distribution. In the following, we describe  how we tested two methods for computing the MAP estimate given the predicted posterior distribution from the cINN.

In the first approach, we treated the posterior distributions for density and temperature of each pixel individually, marginalising over all other pixels along the LoS, for which the cINN generated samples of the joint posterior distribution. From the corresponding set of posterior samples for each pixel, we identified the MAP estimate for density and temperature by employing a kernel density estimate (KDE) to first explicitly derive the probability density curve of the posterior and then find the peak of this curve. In practise, we used a Gaussian kernel function, determining the kernel bandwidth automatically with Silverman's rule of thumb \citep{Silverman1986} and evaluating it on an evenly spaced grid of 1024 points (between the minimum and maximum value of the posterior samples) to determine the MAP point estimates. 

The cINN does not actually generate samples from the posterior distributions of dust density and temperature of each individual pixel but, rather, from the full joint posterior for density and temperature for all pixels along a given LoS. Therefore, to be completely correct, the MAP has to be determined as the most probable combination of values in the full 64-dimensional space that the cINN constructs the posterior samples in. To find the maximum of the probability density in this very high-dimensional space and then compare it to the marginalised MAP estimate, we employed the MeanShift algorithm \citep{Fukunaga1975_MeanShift, Comaniciu2002_MeanShift}. It is a gradient ascent approach whereby, given a set of $N$ samples, the modes of the underlying density distribution can be found. MeanShift is an iterative procedure, in which the center of a kernel window is continuously moved into the direction of the maximum increase in density until convergence is reached. Given a kernel function $K(\mathbf{x})$ (e.g.~a Gaussian kernel) and an initial position for the center $\mathbf{x}$ of the kernel window, the algorithm computes the so-called mean shift: 
\begin{equation}
    \mathbf{m}(\mathbf{x}) = \frac{\sum_{i=1}^N K\left(\mathbf{x} - \mathbf{x_i}\right) \mathbf{x}_i}{\sum_{i=1}^N K\left(\mathbf{x} - \mathbf{x_i}\right)} - \mathbf{x},
\end{equation}
which is the difference between the kernel weighted mean and the center of the kernel window. As demonstrated by \cite{Comaniciu2002_MeanShift}, this vector is proportional to the estimate of the density gradient estimate obtained with the same kernel; thus, it always points in the direction of maximum increase in density. Iteratively translating the kernel window in direction of the mean shift will therefore find a (local) maximum for the underlying density distribution \citep{Comaniciu2002_MeanShift}. To find all modes of the distribution (and ideally the global maximum) this approach is then repeated for other initial kernel positions, scoring the identified peaks by their corresponding (kernel) density estimate. In a post-processing step, any spurious mode detections (such as plateaus in the distribution) or very close-by modes can then be further pruned \citep[see for example][for further details]{Comaniciu2002_MeanShift}. 

In practise, we employed the scikit-learn \citep{scikit-learn} Python implementation of the MeanShift algorithm, which uses a flat kernel:
\begin{equation}
    K(\mathbf{x}) = 
            \begin{cases}
            1\,\mathrm{if}\, ||\mathbf{x}|| \leq \lambda\\
            0\,\mathrm{if}\, ||\mathbf{x}|| > \lambda,
        \end{cases}
\end{equation}
where $\lambda$ denotes the bandwidth. 
To speed up computation, this implementation provides a binned seeding strategy, where the initial guesses for the kernel starting position are selected on a discretised grid instead of testing all of the individual sample points. 
The coarseness of this grid is determined by the bandwidth selected for the kernel. The automatic bandwidth selection that comes with this implementation (based on a nearest neighbour distance estimation) has, however, proven not to be robust enough for our very high-dimensional parameter space and would often select bandwidths that are too small for the kernel windows to find any data points inside of them (when used in combination with the binned-seeding approach). 
Since the computation time becomes prohibitively large without the binned seeding, we adopted a simple bandwidth selection procedure where we iteratively doubled an initial bandwidth guess of 32 until a bandwidth is found, with which the MeanShift algorithm converges. 
In practise, this simple approach leads to the selection of a bandwidth of 64 or 128 in most cases.

\subsection{Spatial consistency}
\label{sec:Method_SpatialConsistency}

As we outline above, the cINN approach predicts the posterior distributions for density and temperature for a single LoS jointly. 
Consequently, the prediction preserves the consistency of the predicted posteriors along the LoS. Perpendicular to the LoS, however, we have (by construction) no such spatial consistency guarantee. Figure~\ref{fig:SlicePosteriorExample} provides an example of this behaviour, highlighting the gradual shift of the posteriors along the LoS, whereas perpendicular to it, they are not necessarily consistent. 
As a consequence, the MAP or MeanShift point estimates can often exhibit sharp discontinuities in the predicted densities and temperatures. This can be seen, for example, in the MAP estimates in the left panels of Figure~\ref{fig:SlicePosteriorExample}. As these discontinuities and sharp jumps are rather unphysical, we experimented with two approaches in order to mitigate the spatial consistency issue.

\subsubsection{MNPCP point estimator}
Our first approach consists of introducing a third, alternative point estimator that enforces a degree of spatial consistency perpendicular to the LoSs, which we refer to as the median neighbour pixel combined posterior (MNPCP) in the following. Figure~\ref{fig:mnpcp_example} outlines the steps of the MNPCP approach. Looping over all pixels in the 3D cube of generated posterior samples, $\mathbf{X}_{\mathrm{samp}}$, we first collected the samples for the current pixel and its 26 neighbouring pixels. We then determined the $n$ and $T_{\mathrm{dust}}$ point estimates for the current pixel as the weighted median of this combined set of posterior samples. Here, each sample has been weighted according to the distance of the pixel to the query pixel using the city-block distance metric (Manhattan distance). For edge cases, we accumulate only samples from the existing neighbour pixels, meaning that no form of padding was applied. Taking, for example, a corner pixel, this means that samples from only the seven neighbour pixels are accumulated, as compared to the 26 neighbours available for an interior pixel.

\subsubsection{Neighbour LoS reformulation}
Aside from introducing an alternative point estimator to combat the spatial consistency issue, we also investigated whether a different reformulation of the inverse problem may improve the situation, in comparison to our primary formulation (introduced in Section~\ref{Sec:Method_SLoS}). We refer to cINNs trained on the primary reformulation as a single LoS cINN (SLoS-cINN) in the following, whereas models for the alternative formulation outlined below shall be denoted as a neighbour LoS cINN (NLoS-cINN). To directly compare the SLoS and NLoS approaches, we tested both of them with all three introduced point estimators, namely: MAP, MeanShift, and MNPCP.

The NLoS reformulation aims at improving the spatial consistency perpendicular to the LoS by adding information of the neighbouring LoSs to the observables. Instead of taking only the vector of fluxes corresponding to the pixel of a given LoS, we go on to also consider the observed dust emission in the eight neighbouring pixels, so that the inverse problem becomes:
\begin{equation}
    \begin{split}
        \mathbb{R}^{9K} &\rightarrow \mathbb{R}^{2N} \\
        (\mathbf{y}_1, \ldots, \mathbf{y}_9) &\rightarrow (n_{1}, \ldots, n_{N}, T_{\mathrm{dust, 1}}, \ldots, T_{\mathrm{dust, N}}),
    \end{split}
\end{equation}
where $\mathbf{y}_i = (e_{i, \lambda_1}, \ldots, e_{i, \lambda_K})$ denote the nine emission flux vectors corresponding to the pixel ($\mathbf{y}_5$) and its  neighbourhood. This reformulation has one immediate drawback, however, in that we lose some of the available training data. As we now require every LoS to have eight neighbours, we can no longer consider the edge cases in our training cubes, reducing the total amount of LoSs available for training data to $2 \times 11036 \times 30 \times 30 = 19,864,800$. Another disadvantage is a notably increased memory requirement when storing the training data as a simple csv-table, since we increased the size of the observables vector by a factor of 9. In our case, the table size increases from $55$ to $138$\,GB, even though the latter set contains $2,736,928$ fewer LoSs. Nevertheless, this is the most straightforward approach to providing the cINN with information on the vicinity of a given query pixel. 

\subsection{Performance evaluation}
To quantify the overall performance on the held-out test set of 100 coherent cubes, we computed two metrics for the three different point estimation approaches as an average over all $N_{\mathrm{test}} = 100 \times 32 \times 32 \times 32 = 3,276,800$ test pixels. The first one is the normalised root mean squared error (NRMSE), defined as:
\begin{equation}
    \mathrm{NRMSE} = \frac{1}{\Delta x_{\mathrm{TS}}} \sqrt{\frac{1}{N_{\mathrm{test}}} \sum_{i=1}^{N_{\mathrm{test}}} \left(x_{\mathrm{i,pred}} - x_{\mathrm{i,true}}\right)^2},
\end{equation}
where $x_{\mathrm{i,true}}$ and $x_{\mathrm{i,pred}}$ refer to the ground truth and point estimate prediction of target parameter $x$ for pixel $i$, and $\Delta x_{\mathrm{TS}} = \max(x_{\mathrm{TS}}) - \min(x_{\mathrm{TS}})$ denotes the range of target parameter, $x,$ in the training data ($6.82$ and $1.58$ for $\log(n/\mathrm{m}^{-3})$ and $\log(T_{\mathrm{dust}}/\mathrm{K})$, respectively). The second metric that we computed is the median $|\bar{e}_{\mathrm{rel}}|$ (and 25\% and 75\% quantiles) of the absolute relative error $|e_{\mathrm{i,rel}}|$, defined as:
\begin{equation}
    |e_{\mathrm{i,rel}}| = \left|\frac{x_{\mathrm{i,pred}} - x_{\mathrm{i,true}}}{x_{\mathrm{i,true}}}\right|
,\end{equation}
for pixel $i$.

\section{Results}\label{sec:Results}

\begin{table*}
    \centering
    \caption{Summary of the predictive performance for our three different cINN setups and the three different point estimation methods. Listed are respectively the NRMSE and the median absolute relative error $|\bar{e}_{\mathrm{rel}}|$ (along with the 25\% and 75\% quantiles) for the dust density and temperature evaluated across all pixels of the set of test cubes indicated in the first column.}
    \begin{tabular}{llllccc}
         \toprule
         & & & & \multicolumn{3}{c}{cINN} \\
         \cmidrule(rl){5-7} 
         Cube selection & Point estimator & Measure & Parameter & Single LoS & Neighbour LoS & Single LoS (7 wavelengths) \\
         \midrule
         \multirow{12}{*}{All Cubes} & \multirow{4}{*}{MAP} & \multirow{2}{*}{NRMSE} & $\log(n/\mathrm{m}^{-3})$ & $0.0707$ & $0.0608$ & $0.0981$\\
         &&& $\log(T_{\mathrm{dust}}/\mathrm{K})$ & $0.0300$ & $0.0260$ & $0.0458$\\
         \cmidrule(rl){3-7} 
         && \multirow{2}{*}{$|\bar{e}_{\mathrm{rel}}|\,(\%)$} & $\log(n/\mathrm{m}^{-3})$ & $1.85_{-1.07}^{+1.99}$ & $1.52_{-0.89}^{+1.62}$ & $2.61_{-1.50}^{+3.13}$ \\
         &&& $\log(T_{\mathrm{dust}}/\mathrm{K})$ & $0.87_{-0.56}^{+1.39}$ & $0.75_{-0.48}^{+1.15}$ & $1.47_{-0.94}^{+2.49}$ \\
         \cmidrule(rl){2-7} 
         & \multirow{4}{*}{MeanShift} & \multirow{2}{*}{NRMSE} & $\log(n/\mathrm{m}^{-3})$ & $0.0640$ & $0.0555$ & $0.0855$\\
         &&& $\log(T_{\mathrm{dust}}/\mathrm{K})$ & $0.0251$ & $0.0221$ & $0.0372$\\
         \cmidrule(rl){3-7} 
         && \multirow{2}{*}{$|\bar{e}_{\mathrm{rel}}|\,(\%)$} & $\log(n/\mathrm{m}^{-3})$ & $2.01_{-1.15}^{+1.87}$ & $1.63_{-0.93}^{+1.57}$ & $3.10_{-1.68}^{+2.52}$\\
         &&& $\log(T_{\mathrm{dust}}/\mathrm{K})$ & $1.01_{-0.64}^{+1.39}$ & $0.84_{-0.53}^{+1.19}$ & $1.97_{-1.14}^{+2.04}$\\
         \cmidrule(rl){2-7} 
         & \multirow{4}{*}{MNPCP} & \multirow{2}{*}{NRMSE} & $\log(n/\mathrm{m}^{-3})$ & $0.0620$ & $0.0536$ & $0.0838$\\
         &&& $\log(T_{\mathrm{dust}}/\mathrm{K})$ & $0.0245$ & $0.0213$ & $0.0359$\\
         \cmidrule(rl){3-7} 
         && \multirow{2}{*}{$|\bar{e}_{\mathrm{rel}}|\,(\%)$} & $\log(n/\mathrm{m}^{-3})$ & $1.84_{-1.05}^{+1.78}$ & $1.54_{-0.88}^{+1.48}$ & $2.81_{-1.54}^{+2.47}$\\
         &&& $\log(T_{\mathrm{dust}}/\mathrm{K})$ & $0.96_{-0.58}^{+1.26}$ & $0.82_{-0.49}^{+1.07}$ & $1.72_{-0.99}^{+1.89}$\\
         
         \midrule
         \multirow{12}{*}{ISRF-only} & \multirow{4}{*}{MAP} & \multirow{2}{*}{NRMSE} & $\log(n/\mathrm{m}^{-3})$ & $0.0510$ & $0.0441$ & $0.0917$\\
         &&& $\log(T_{\mathrm{dust}}/\mathrm{K})$ & $0.0633$ & $0.0550$ & $0.1135$\\
         \cmidrule(rl){3-7} 
         && \multirow{2}{*}{$|\bar{e}_{\mathrm{rel}}|\,(\%)$} & $\log(n/\mathrm{m}^{-3})$ & $1.41_{-0.83}^{+1.45}$ & $1.18_{-0.69}^{+1.20}$ & $2.25_{-1.29}^{2.82}$\\
         &&& $\log(T_{\mathrm{dust}}/\mathrm{K})$ & $0.61_{-0.39}^{+1.03}$ & $0.53_{-0.34}^{+0.89}$ & $1.33_{-0.87}^{+2.54}$\\
         \cmidrule(rl){2-7}
         & \multirow{4}{*}{MeanShift} & \multirow{2}{*}{NRMSE} & $\log(n/\mathrm{m}^{-3})$ & $0.0480$ & $0.0409$ & $0.0804$ \\
         &&& $\log(T_{\mathrm{dust}}/\mathrm{K})$ & $0.0522$ & $0.0455$ & $0.0906$ \\
         \cmidrule(rl){3-7}
         && \multirow{2}{*}{$|\bar{e}_{\mathrm{rel}}|\,(\%)$} & $\log(n/\mathrm{m}^{-3})$ & $1.53_{-0.88}^{+1.42}$ & $1.24_{-0.72}^{+1.18}$ & $2.86_{-1.56}^{+2.40}$\\
         &&& $\log(T_{\mathrm{dust}}/\mathrm{K})$ & $0.70_{-0.45}^{+1.12}$ & $0.58_{-0.37}^{+0.95}$ & $1.90_{-1.12}^{+2.08}$ \\
         \cmidrule(rl){2-7}
         & \multirow{4}{*}{MNPCP} & \multirow{2}{*}{NRMSE} & $\log(n/\mathrm{m}^{-3})$ & $0.0456$ & $0.0395$ & $0.0772$\\
         &&& $\log(T_{\mathrm{dust}}/\mathrm{K})$ & $0.0506$ & $0.0439$ & $0.0859$\\
         \cmidrule(rl){3-7}
         && \multirow{2}{*}{$|\bar{e}_{\mathrm{rel}}|\,(\%)$} & $\log(n/\mathrm{m}^{-3})$ & $1.40_{-0.80}^{+1.33}$ & $1.19_{-0.69}^{+1.12}$ & $2.51_{-1.38}^{+2.26}$\\
         &&& $\log(T_{\mathrm{dust}}/\mathrm{K})$ & $0.70_{-0.42}^{+1.02}$ & $0.60_{-0.36}^{+0.88}$ & $1.64_{-0.97}^{+1.93}$\\
         
         \midrule
         \multirow{12}{*}{ISRF + Star} & \multirow{4}{*}{MAP} & \multirow{2}{*}{NRMSE} & $\log(n/\mathrm{m}^{-3})$ & $0.0859$ & $0.0739$ & $0.1041$\\
         &&& $\log(T_{\mathrm{dust}}/\mathrm{K})$ & $0.0387$ & $0.0336$ & $0.0538$\\
         \cmidrule(rl){3-7}
         && \multirow{2}{*}{$|\bar{e}_{\mathrm{rel}}|\,(\%)$} & $\log(n/\mathrm{m}^{-3})$ & $2.45_{-1.38}^{+2.55}$ & $1.99_{-1.12}^{+2.05}$ & $3.01_{-1.71}^{+3.34}$\\
         &&& $\log(T_{\mathrm{dust}}/\mathrm{K})$ & $1.20_{-0.74}^{+1.67}$ & $1.02_{-0.62}^{+1.32}$ & $1.61_{-0.98}^{+2.43}$\\
         \cmidrule(rl){2-7}
         & \multirow{4}{*}{MeanShift} & \multirow{2}{*}{NRMSE} & $\log(n/\mathrm{m}^{-3})$ & $0.0767$ & $0.0670$ & $0.0903$\\
         &&& $\log(T_{\mathrm{dust}}/\mathrm{K})$ & $0.0327$ & $0.0289$ & $0.0441$\\
         \cmidrule(rl){3-7}
         && \multirow{2}{*}{$|\bar{e}_{\mathrm{rel}}|\,(\%)$} & $\log(n/\mathrm{m}^{-3})$ & $2.66_{-1.44}^{+2.19}$ & $2.15_{-1.18}^{+1.89}$ & $3.35_{-1.78}^{+2.61}$\\
         &&& $\log(T_{\mathrm{dust}}/\mathrm{K})$ & $1.38_{-0.80}^{+1.50}$ & $1.14_{-0.67}^{+1.29}$ & $2.05_{-1.15}^{+2.01}$\\
         \cmidrule(rl){2-7}
         & \multirow{4}{*}{MNPCP} & \multirow{2}{*}{NRMSE} & $\log(n/\mathrm{m}^{-3})$ & $0.0748$ & $0.0648$ & $0.0899$\\
         &&& $\log(T_{\mathrm{dust}}/\mathrm{K})$ & $0.0320$ & $0.0279$ & $0.0431$\\
         \cmidrule(rl){3-7}
         && \multirow{2}{*}{$|\bar{e}_{\mathrm{rel}}|\,(\%)$} & $\log(n/\mathrm{m}^{-3})$ & $2.45_{-1.34}^{+2.13}$ & $1.99_{-1.10}^{+1.80}$ & $3.13_{-1.69}^{+2.64}$\\
         &&& $\log(T_{\mathrm{dust}}/\mathrm{K})$ & $1.27_{-0.73}^{+1.39}$ & $1.08_{-0.61}^{+1.16}$ & $1.79_{-1.01}^{+1.85}$\\
         \bottomrule
    \end{tabular}
    \label{tab:PerformanceSummary}
\end{table*}

\begin{figure*}
    \centering
    \includegraphics[width=0.69\textwidth]{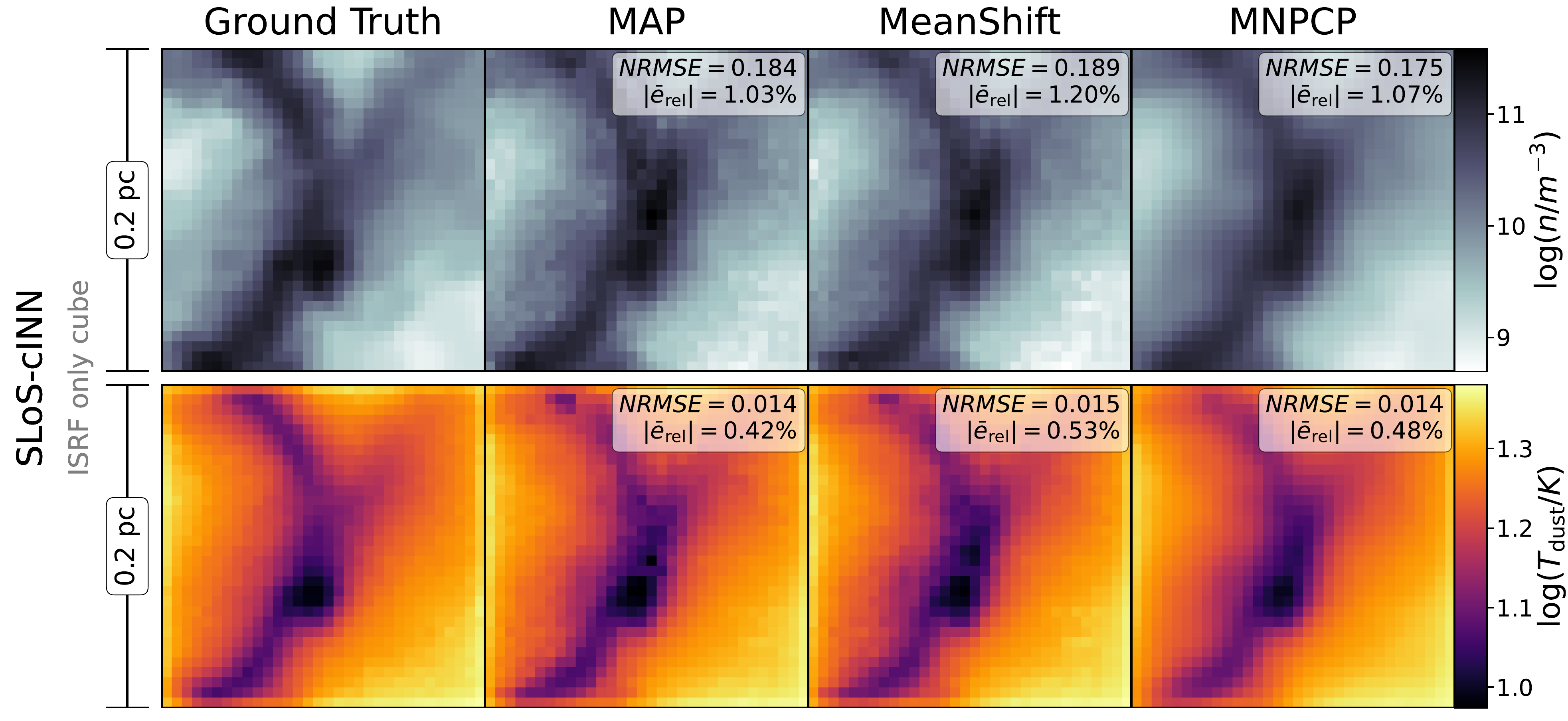}
    \includegraphics[width=0.69\textwidth]{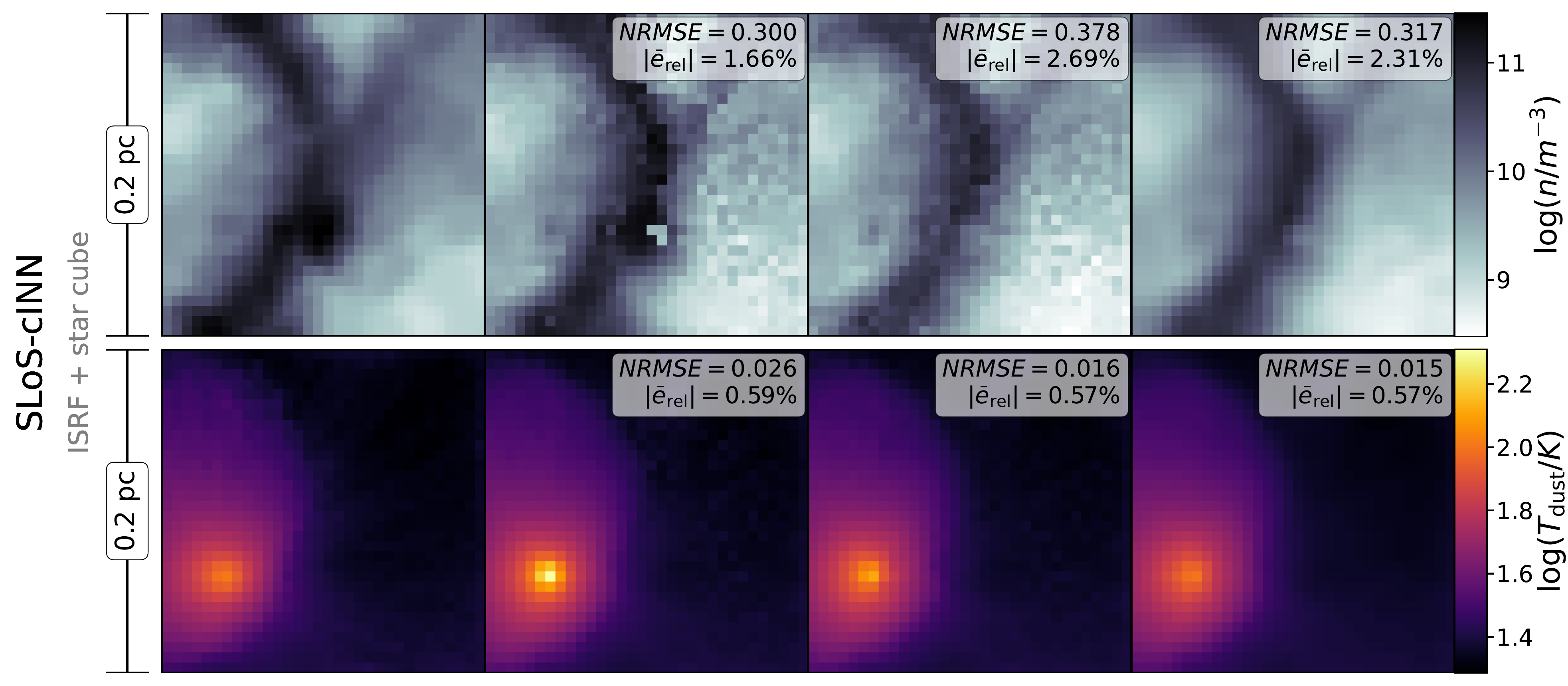}
    \includegraphics[width=0.69\textwidth]{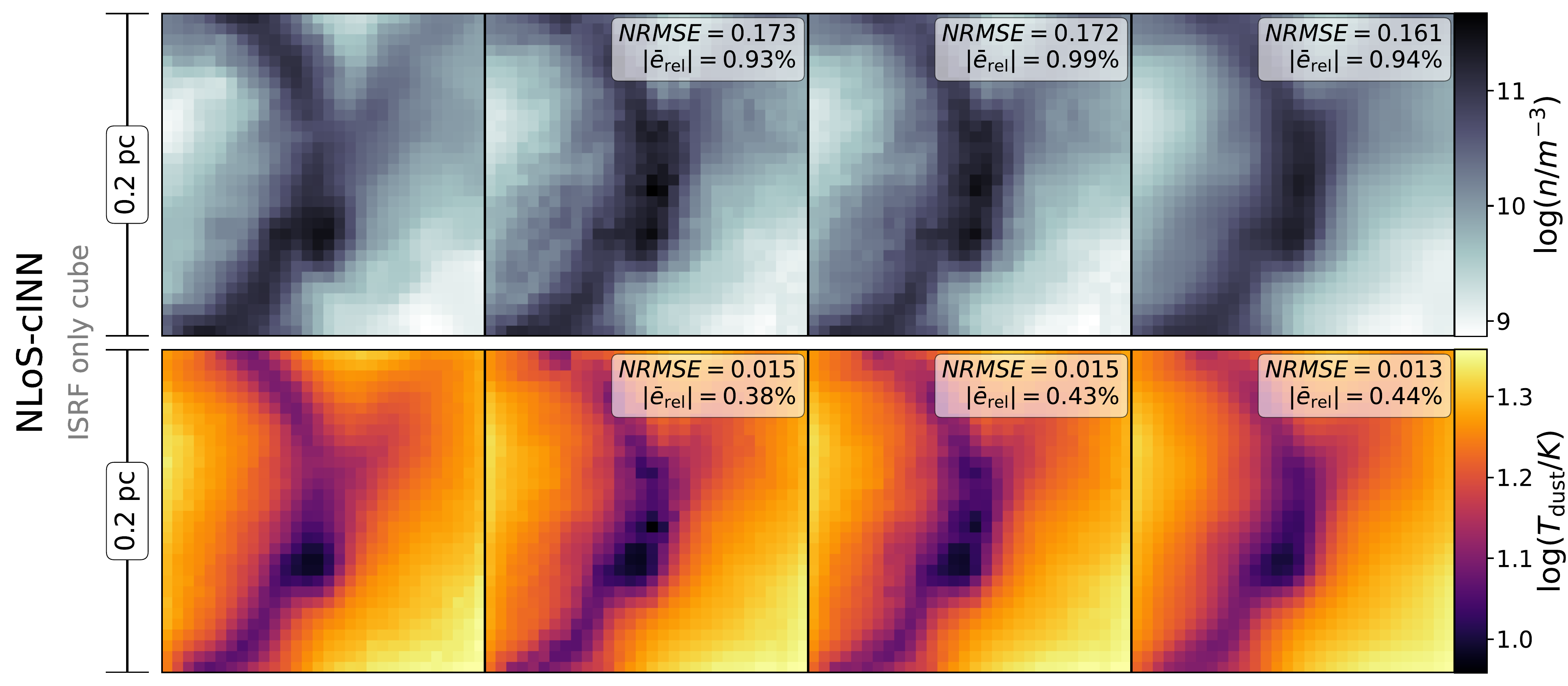}
    \includegraphics[width=0.69\textwidth]{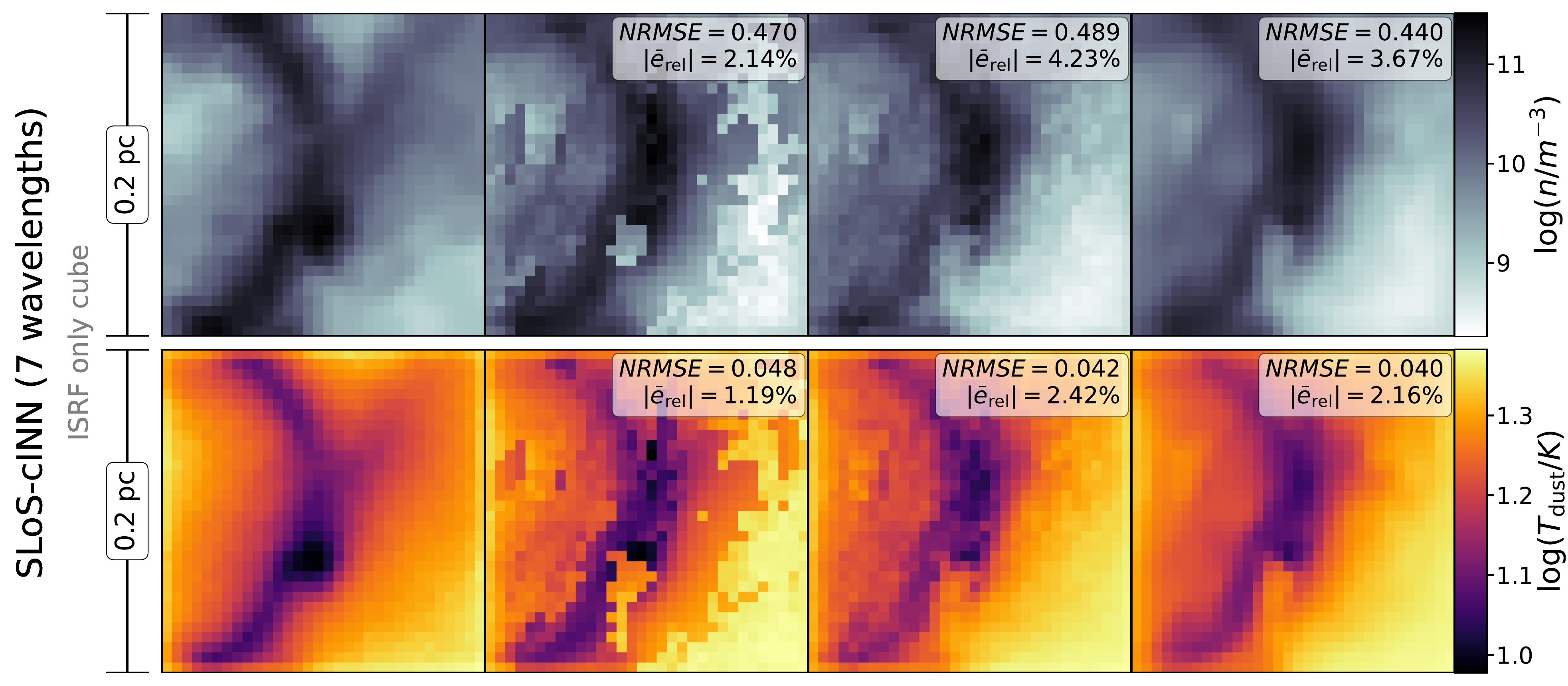}
    \caption{Predicted dust densities and temperatures for a single example cube slice perpendicular to the LoS. A comparison between the three point estimation methods to the ground truth is shown in the left column. The top two rows give the 23-wavelength SLoS-cINN result for the ISRF-only scenario, while rows 3\&4\ display the counterpart for the ISRF + star case. Rows 5\&6 present the NLoS-based outcome for the ISRF-only case, and the last two rows show the corresponding seven wavelengths SLoS-cINN prediction. The listed NRMSE and median absolute relative errors are averages over this slice only and not the entire cube.}
    \label{fig:SingleSliceComparison}
\end{figure*}

\begin{figure*}[ht!]
    \centering
    \includegraphics[width=0.8\textwidth]{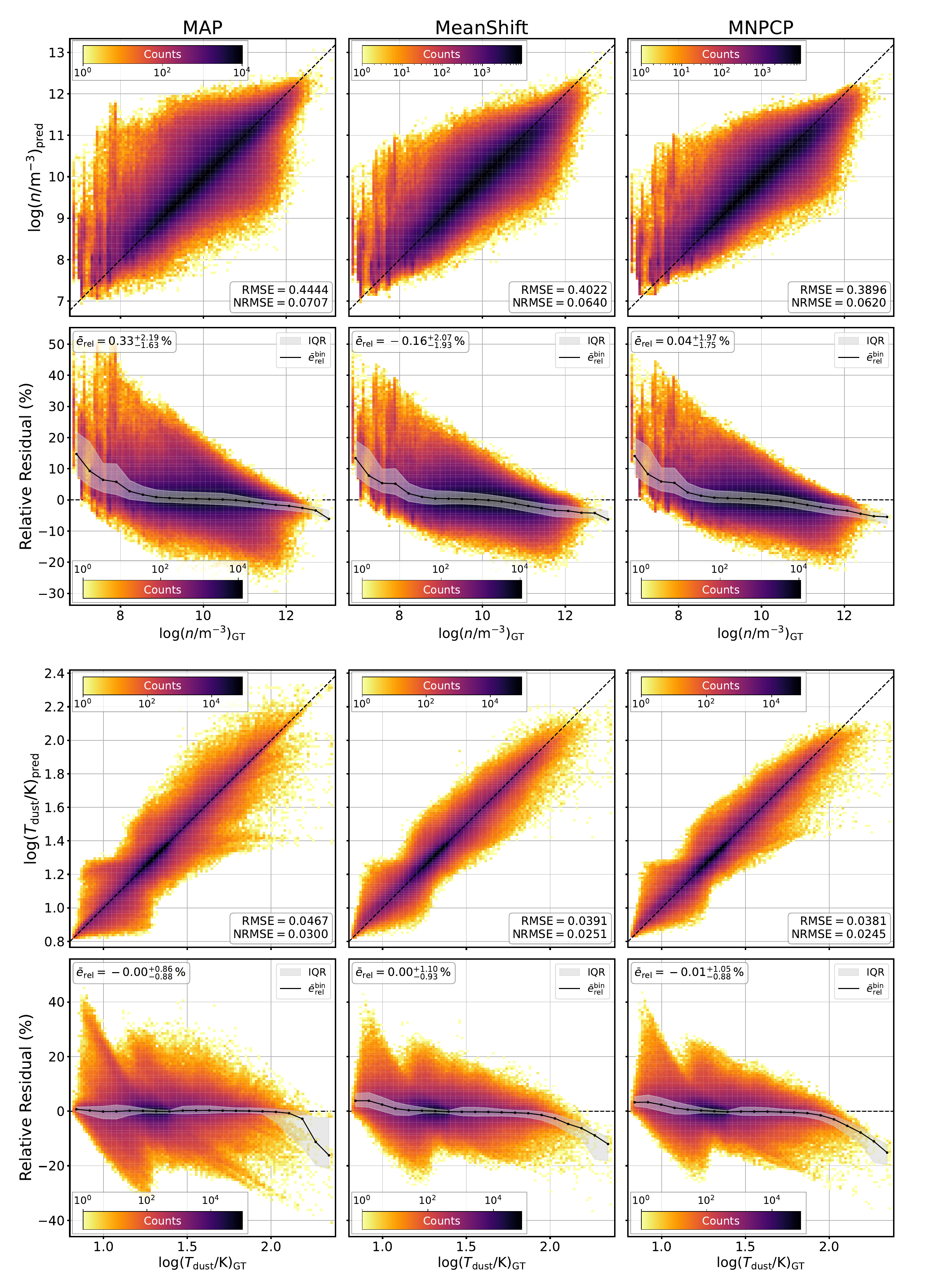}
    \caption{Performance breakdown of the SLoS-cINN using 23 wavelengths. 2D histograms comparing the cINN predictions for dust density (top two rows) and temperature (bottom two rows) to the ground truth across all pixels of the test set data are given, distinguishing the results of the three point estimation procedures: MAP, MeanShift, and MNPCP. Rows 1 and 3 present the direct one-to-one correlation of the predicted parameters to the ground truth, whereas rows 2 and 4 provide the corresponding relative residuals. In the latter panels, the black curve and grey shaded area indicates a binned median relative residual along with the interquantile range between the 25\% and 75\% quantile of these bins.}
    \label{fig:2D_histogram_STAR_ISRF_all}
\end{figure*}

\begin{figure*}
    \centering
    \includegraphics[width=0.49\textwidth]{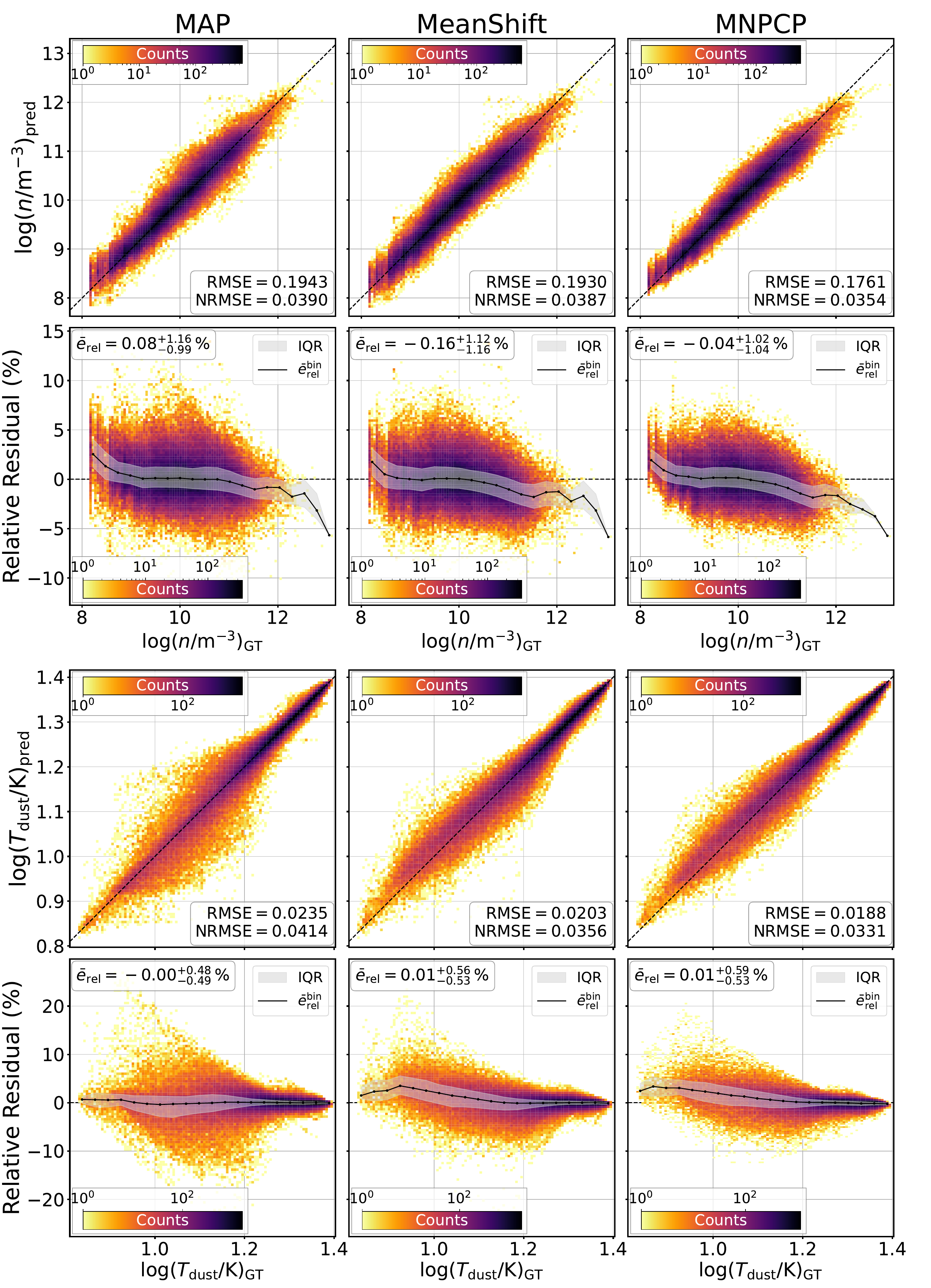}
    \includegraphics[width=0.49\textwidth]{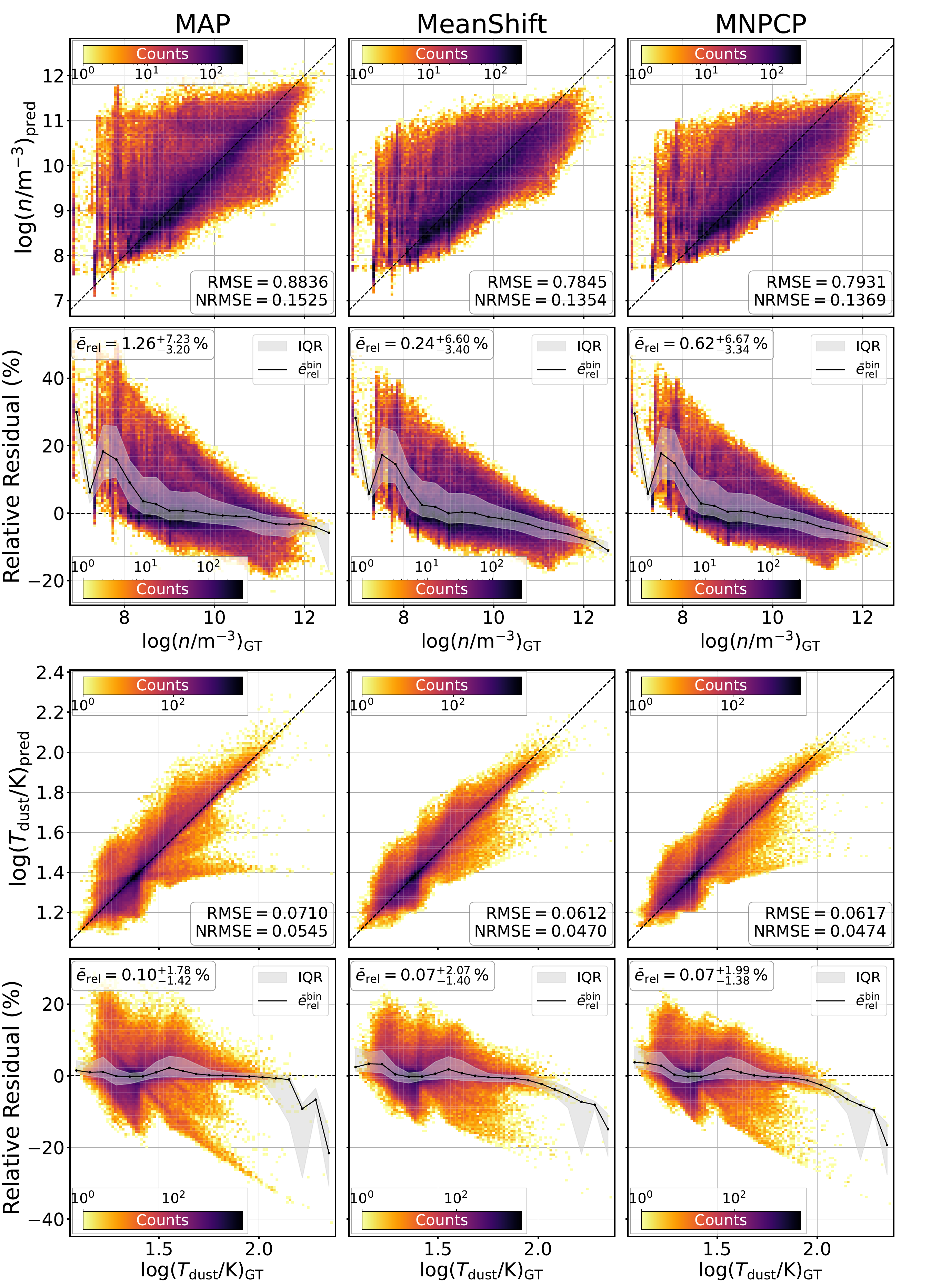}
    \caption{Breakdown of the predictive performance of the SLoS-cINN for the best and worst cases. Analogously to  Figure~\ref{fig:2D_histogram_STAR_ISRF_all}, we show the 2D histograms for the one-to-one comparison of the prediction results (rows one and three) and their respective residuals (rows 2\&4). The left three columns present the five best reconstructed cubes, whereas the five worst reconstructed ones are shown in the three right columns, respectively.}
    \label{fig:2D_histogram_STAR_ISRF_all_best_vs_worst}
\end{figure*}

In this section, we outline the evaluation results regarding the predictive performance of our trained cINN models on the held-out 100 test cubes. Table~\ref{tab:PerformanceSummary} provides a summary of the NRMSE and absolute relative errors achieved by our three different cINN setups with the three different point estimation approaches. In addition, it also shows a breakdown of the results between the two radiative transfer configurations, that is, ISRF-only and ISRF + star. In the following, we first discuss the influence of the point estimator choice on the prediction results on the example of the SLoS-cINN that accounts for all 23 wavelengths (Section~\ref{sec:Results_SLoS_cINN}). 
We then compare the outcomes of the NLoS approach to the SLoS setup (Section~\ref{sec:Results_NLoS_cINN}) and present an analysis of the SLoS performance for the more realistically limited wavelength coverage experiment (Section~\ref{sec:Results_SLoS_7Filter}). We conclude with a comparison of our approach with a classical SED fit to determine column densities (Section~\ref{sec:Results_SED_fit}), followed by discussions on the physical feasibility of the approach (Section~\ref{sec:Results_PhysFeasibility}) and on the application of our setup to real observational data (Section~\ref{sec:Results_ApplRealData}).

\subsection{Choice of the point estimator and influence of the radiation configuration}
\label{sec:Results_SLoS_cINN}

Figure~\ref{fig:SingleSliceComparison} shows a qualitative comparison of the point estimates for dust density and temperature to the ground truth for the MAP, MeanShift, and MNPCP estimators. In particular, we show the outcome for a single slice (perpendicular to the LoS) of one example cube of the test set. Here, the first four rows show the prediction results based on the SLoS-cINN, distinguishing the ISRF-only scenario (rows 1 and 2) and the ISRF + star radiation configuration (rows 3 and 4). 
In the ISRF scenario, we can see that all three point estimators provide a very decent reconstruction result for both dust density and temperature. 
However, the discontinuities in the MAP prediction (as previously discussed in Section~\ref{sec:Method}) are quite notable and give the prediction outcome a noisy character.
The MeanShift result in the third column also suffers from this effect, albeit to a slightly lesser degree. Given that both of these estimators have no spatial consistency guarantee perpendicular to the LoS, this is of course an expected result. 
Contrary to that, we can see that our MNPCP approach (fourth column) provides a much more consistent and smoothed prediction result than the other two estimators. 
Nevertheless, this can come at the expense of losing some of the finer, high-density features of the dust distribution. 
A full comparison of the prediction with the SLoS-cINN for all slices of this example cube in the ISRF-only configuration is given in Figure~\ref{fig:AllSliceComparison_ISRF_all_076412} in the appendix. 
We also refer to Figure~\ref{fig:IsodensitySurface_ISRF_all_076412} for a 3D visualisation of the prediction results in terms of isodensity surfaces.

The predictions for the same cube with the alternative radiation setup (rows 3\&4, Figure~\ref{fig:SingleSliceComparison}) indicate that the inclusion of a star affects not only the prediction of the dust temperature (as expected given the additional heating from the star), but the dust density estimates as well. Although the overall reconstruction of both dust density and temperature remain quite good in this example, it is obvious that the reconstructed dust density has lost accuracy in comparison to the cINN prediction for the same cube subject to only the ISRF. While the overall larger scale structures are still recovered well, a lot of the finer details of the dust distribution are lost compared to the prediction in the ISRF-only scenario. The inclusion of a star inside of the cube thus appears to add not only complexity to the prediction of the dust temperature, but also renders the recovery of the density more difficult. For a complete comparison of all slices of the example cube in the ISRF + star radiation configuration, we refer to Figure~\ref{fig:AllSliceComparison_STAR_all_076412} in the appendix. The corresponding 3D isodensity surface visualisation is provided in Figure~\ref{fig:IsodensitySurface_STAR_all_076412}. To provide additional insights into the difference between the predictions on the ISRF-only and ISRF+star cube, Figure~\ref{fig:PosteriorSliceComparison_CUBE_076412} in the appendix presents an extension to Figure~\ref{fig:SingleSliceComparison}, where some examples of the predicted posterior distributions are shown. Here, we find that the density posterior distributions become notably wider in the RT scenario that includes the star, which results in flat, plateau-like or even multi-peaked distributions for some pixels. In the latter cases, the peaks of the distributions may then no longer coincide with the ground truth (although the ground truth is always part of the distribution); for instance, the MAP estimator returns a suboptimal result in comparison to the prediction for the same cube in the ISRF-only scenario. There are two possible explanations for the broadening of the density posterior distributions in the second RT scenario. The first is that this an intrinsic degeneracy of the problem. While the cubes share the same density distribution between the two RT setups, the resulting dust temperatures naturally differ. It is not unreasonable to believe that recovering the density can become more or less ambiguous depending on the temperature given that opacity depends on temperature. It is also worth noting that the temperature posterior distributions do not suffer the same broadening effect and appear similarly well constrained in both RT configurations. The second possible explanation is a suboptimal convergence of the network related to the sampling of the training data, which is discussed in more detail further below.

To quantify the overall performance of the SLoS-cINN in combination with the three different point estimators, we present a direct one-to-one comparison of the predicted dust densities and temperatures to the respective ground truth values, and the corresponding relative residuals for all $100 \times 32^3$ pixels in the test set in Figure~\ref{fig:2D_histogram_STAR_ISRF_all}. As indicated by Table~\ref{tab:PerformanceSummary} and Figure~\ref{fig:2D_histogram_STAR_ISRF_all}, the overall predictive performance of the SLoS-cINN in the 23-wavelength configuration is quite excellent across all three point estimation approaches. Here, we achieved NRMSEs between $0.06$ to $0.07$ in $\log(n/\mathrm{m}^{-3})$ and between $0.025$ to $0.03$ in $\log(T_{\mathrm{dust}}/\mathrm{K})$, corresponding to median absolute relative residuals, $|\bar{e}_{\mathrm{rel}}|,$ in the ranges of $1.8$ to $2.0$\% and $0.9$ to $1.0$\%, respectively. Although there is a notable dispersion around a perfect one-to-one correlation between the estimated densities and temperatures and the corresponding ground truth, the binned median curve (and binned $25$ and $75$\% quantile curves) in the relative residual diagrams indicates that a majority of pixels is indeed close to a perfect recovery for most of the range covered in density and temperature. Nevertheless, the relative residuals can reach up to $40$ to $50$\% for a small number of individual pixels in terms of both density and temperature. 

The binned median relative residual curves also highlight some systematic trends in the point estimation outcome. For the dust density, we find a notable trend towards overestimation of the density for pixels with a true density below $\log(n/\mathrm{m}^{-3}) = 8$. There is also a tendency (although to a lesser extent) for underestimations at the high density end, starting at $\log(n/\mathrm{m}^{-3}) = 11$. For the dust temperature, we also observe a systematic underestimation at temperatures higher than $100$\,K, and for the MeanShift and MNPCP point estimates a slight tendency for overestimation below $10$\,K. It appears that the recovery of dust density and temperature in terms of the point estimation tends to struggle more overall towards the extreme ends of the respective parameter range. Part of this difficulty can likely be attributed to the relative complexity of these more extreme environments, but there might be a more direct issue with our training data that could explain this decrease in the predictive performance at the edges of the parameter space. Figure~\ref{fig:TS_priors_observables} in the appendix shows the prior distributions of dust density and temperature across all pixels in our training set. Comparing the thresholds at which the binned median relative error starts to show systematic offsets in Figure~\ref{fig:2D_histogram_STAR_ISRF_all}, that is, $\log(n/\mathrm{m}^{-3}) < 8$, $\log(n/\mathrm{m}^{-3}) > 11$ and $\log(T_{\mathrm{dust}}/K) > 2$, to the training set priors, we can see that there are comparatively a lot fewer pixels in these parameter ranges in the training data. This (relative) lack of training examples within these parameter ranges could lead to a suboptimal convergence of the cINN, so that it does not achieve the same robustness at the edges of the parameter space as it does within the intervals where a lot of training data is available. Given that the prior distributions of the dust densities and temperatures in our training data are (in part) dictated by the underlying dust cloud simulations, achieving a more even sampling across the parameter spaces is not a trivial matter and at this stage, this is beyond the scope of this proof of concept. It is also worth noting that very high temperature regions ($T_{\mathrm{dust}} > 100$\,K) are both rare in reality and likely affected by strong feedback, being either part of an HII region or related to strong outflow activity. This adds further complexity to these extreme environments, which is also currently not accounted for in the Cloud Factory as noted in Section~\ref{sec:TrainingData}. We plan to investigate these effects and the sampling strategy further in future optimisation of our training set generation.

In comparing the three point estimators more in detail, we find that both MeanShift and MNPCP are less prone to large outlier values, as evident by the smaller dispersion around the one-to-one correlation in Figure~\ref{fig:2D_histogram_STAR_ISRF_all} and the lower NRMSE in Table~\ref{tab:PerformanceSummary}. At the same time, it is the MAP estimator that returns the overall best $|\bar{e}_{\mathrm{rel}}|$ with $1.85$\% for $\log(n/\mathrm{m}^{-3})$ and $0.87$\% for $\log(T_{\mathrm{dust}}/\mathrm{K})$. Although the MNPCP estimator has a nominally better result for $|\bar{e}_{\mathrm{rel}}|$ with $1.84$\% for $\log(n/\mathrm{m}^{-3})$, it incurs a notably larger error in $\log(T_{\mathrm{dust}}/\mathrm{K})$ with $0.96$\%. The latter small performance decrease of the MNPCP approach in terms of $|\bar{e}_{\mathrm{rel}}|$ is likely a result of the effective smoothing that this estimator performs. As Figure~\ref{fig:2D_histogram_STAR_ISRF_all} shows, this enhances the underestimation tendencies at the high temperature end (also for high density but to a lesser degree) and, thus, the average error. For the MeanShift, which performs the worst in terms of $|\bar{e}_{\mathrm{rel}}|$, this might be a consequence of a suboptimally chosen bandwidth from our simplified bandwidth selection procedure (as described in Section~\ref{sec:Method}). If, for instance, the selected bandwidth is too large, the MeanShift kernel will overly smooth the density distribution and likely miss narrow peaks. This results in a comparable (over-) smoothing effect to the MNPCP, as evidenced by the similar behaviour of the two methods in Figure~\ref{fig:2D_histogram_STAR_ISRF_all}. 

Figure~\ref{fig:2D_histogram_STAR_ISRF_all_best_vs_worst} provides a breakdown of Figure~\ref{fig:2D_histogram_STAR_ISRF_all} for the best and worst case prediction outcomes, that is the five cubes with the best and five cubes with the worst NRMSE in the MAP point estimate. Averaged over the five best cubes $|\bar{e}_{\mathrm{rel}}|$ goes down to about $1$\% and $0.5$\% in $\log(n/\mathrm{m}^{-3})$ and $\log(T_{\mathrm{dust}}/\mathrm{K})$, respectively. In the worst cases on the other hand, $|\bar{e}_{\mathrm{rel}}|$ reaches up to $4.4$\% and $1.7$\% in $\log(n/\mathrm{m}^{-3})$ and $\log(T_{\mathrm{dust}}/\mathrm{K})$, respectively. What is interesting to note here is that the five best cubes are all only subject to the ISRF, whereas the five worst ones are all in the ISRF + star radiation configuration. This reaffirms our earlier assessment that the presence of a star in the cube notably complicates the problem. We further quantified this by breaking down the overall performance on the test set between the two radiation setups in Table~\ref{tab:PerformanceSummary}. As we can see, $|\bar{e}_{\mathrm{rel}}|$ increases by almost a factor of two for the cubes with ISRF + star setup in comparison to the ISRF-only configuration cubes regardless of the choice of the point estimator. We also want to emphasise that the observed performance is not dependent on the selected viewing angle of the cubes. We have confirmed in a test limited to the 50 ISRF-only cubes that the cINN returns a similarly excellent reconstructive performance, when the cubes are observed from different directions. Thus, our choice to only generate synthetic observations from one direction in the training data has not introduced a bias in the form of a preferred viewing angle (for more details, see Appendix~\ref{app:RotationTest}).

In Section~\ref{sec:Method}, we specifically introduce the MNPCP approach, because the MAP and MeanShift estimators per construction of our inverse problem do not have a spatial consistency guarantee perpendicular to the LoS (as demonstrated in Figure~\ref{fig:SingleSliceComparison}). To quantify whether the MNPCP approach improves upon this situation (beyond the qualitative comparison in Figure~\ref{fig:SingleSliceComparison}), we computed the median difference in density and temperature for neighbouring pixels following:
\begin{equation}
        \Delta_{\parallel \mathrm{LoS}} = \mathrm{median} \left\{x_{i+1, j, k} - x_{i, j, k} \right\}
        \begin{cases}
            \forall i \in \{1, \ldots, N-1\}\\
            \forall j, k \in \{1, \ldots, N\}
        \end{cases}
\end{equation}
and perpendicular to the LoS:
\begin{equation}
    \Delta_{\perp \mathrm{LoS}} = \mathrm{median} \left\{\Delta_j, \Delta_k\right\},
\end{equation}
where
\begin{equation*}
    \begin{split}
        \Delta_j & = \left\{x_{i, j+1, k} - x_{i, j, k} \right\}
            \begin{cases}
                \forall j \in \{1, \ldots, N-1\}\\
                \forall i, k \in \{1, \ldots, N\}
            \end{cases}\\ 
        \Delta_k & = \left\{x_{i, j, k+1} - x_{i, j, k} \right\} 
            \begin{cases}
                \forall k \in \{1, \ldots, N-1\}\\
                \forall i, j \in \{1, \ldots, N\},
            \end{cases}
    \end{split}
\end{equation*}
for our three different point estimators. We then compared these results to the respective values obtained from the ground truth. The results of this analysis on the test set are summarised in Table~\ref{tab:delta_pix}. As expected, there is no preferred direction in the ground truth, with values of about $0.075$ in $\log(n/\mathrm{m}^{-3})$ and $0.01$ in $\log(T_{\mathrm{dust}}/\mathrm{K})$ for both $\Delta_{\parallel \mathrm{LoS}}$ and $\Delta_{\perp \mathrm{LoS}}$. In the MAP and MeanShift prediction results, on the other hand, we find $\Delta_{\perp \mathrm{LoS}}$ to be about twice as large as $\Delta_{\parallel \mathrm{LoS}}$ on average, confirming again the spatial consistency issue. In contrast, the MNPCP estimator offers a much more balanced result, achieving about even $\Delta_{\perp \mathrm{LoS}}$ and $\Delta_{\parallel \mathrm{LoS}}$ for $\log(T_{\mathrm{dust}}/\mathrm{K})$, and at least reducing $\Delta_{\perp \mathrm{LoS}}$ to about $1.5 \Delta_{\parallel \mathrm{LoS}}$ for $\log(n/\mathrm{m}^{-3})$. Yet even with that outcome, the MNPCP estimate does not quite achieve the balance of the ground truth results. It is also interesting to note that all three point estimators return solutions where $\Delta_{\parallel \mathrm{LoS}}$ is notably smaller than in the ground truth. This indicates that the cINN prediction tends to return a smoother transition along the LoS than the ground truth. This is likely a result of the fact that the cINN returns a smooth continuous output, whereas the ground truth is limited by the coarseness of the simulation resolution. 

In summary, we find an overall very satisfactory performance of the SLoS-cINN in the 23-wavelength configuration, providing a fairly robust recovery of dust density and temperature for most of the tested parameter range. We do note, however, a systematic decrease in performance towards the lower and upper limits of the trained range in terms of density and temperature, which can potentially be traced back to a relative lack of examples in these regimes in the training data. We also identified a dependence of the performance on the radiation setup of the test cubes, where the ones also hosting a star in addition to the ISRF appear more difficult to reconstruct. Regarding the choice of the point estimator, there is no clear winner in terms of the NRMSE and $|\bar{e}_{\mathrm{rel}}|$ performance indicators. Nevertheless, we believe that the MNPCP approach appears as the most reasonable solution because it can provide smooth reconstruction solutions both along and perpendicular to the LoS. One caveat to keep in mind is the fact that the MNPCP point estimator does amplify the systematic error tendencies at the lower and upper limits of the density and temperature ranges due to its inherent smoothing effect. 

\begin{table*}
    \centering
    \caption{Overview of the median difference, $\Delta,$ in dust density and temperature between neighbouring pixels along and perpendicular to the LoS for the three different point estimators and two inverse problem setups. The ground truth values differ slightly between the two setups, as the NLoS approach does not contain the edge LoSs of each cube.}
    \begin{tabular}{llcccc}
        \toprule
         & & \multicolumn{2}{c}{$\log(n/\mathrm{m}^{-3})$} & \multicolumn{2}{c}{$\log(T_{\mathrm{dust}}/\mathrm{K})$}\\
        \cmidrule(rl){3-4}\cmidrule(rl){5-6}
        Inverse problem setup & Measure & $\Delta_{\parallel \mathrm{LoS}}$ & $\Delta_{\perp \mathrm{LoS}}$ & $\Delta_{\parallel \mathrm{LoS}}$ & $\Delta_{\perp\mathrm{LoS}}$ \\
        \midrule
        & Ground Truth & $0.075_{-0.050}^{+0.086}$ & $0.073_{-0.048}^{+0.082}$ & $0.010_{-0.006}^{+0.010}$ & $0.010_{-0.006}^{+0.009}$ \\
        \cmidrule(rl){2-6}
        \multirow{3}{*}{Single LoS (23 wavelengths)} & MAP & $0.040_{-0.024}^{+0.049}$ & $0.106_{-0.059}^{+0.099}$ & $0.005_{-0.003}^{+0.008}$ & $0.011_{-0.006}^{+0.012}$ \\
        & MeanShift & $0.050_{-0.029}^{+0.043}$ & $0.098_{-0.054}^{+0.084}$ & $0.007_{-0.004}^{+0.007}$ & $0.011_{-0.006}^{+0.010}$ \\
        & MNPCP & $0.045_{-0.027}^{+0.045}$ & $0.064_{-0.035}^{+0.055}$ & $0.007_{-0.004}^{+0.007}$ & $0.008_{-0.005}^{+0.008}$ \\
        \cmidrule(rl){2-6}
        \multirow{3}{*}{Single LoS (7 wavelengths)} & MAP & $0.032_{-0.018}^{+0.037}$ & $0.107_{-0.060}^{+0.106}$ & $0.003_{-0.002}^{+0.006}$ & $0.012_{-0.007}^{+0.014}$ \\
        & MeanShift & $0.041_{-0.025}^{+0.036}$ & $0.087_{-0.048}^{+0.078}$ & $0.006_{-0.004}^{+0.007}$ & $0.011_{-0.006}^{+0.011}$ \\
        & MNPCP & $0.038_{-0.024}^{+0.041}$ & $0.063_{-0.035}^{+0.056}$ & $0.006_{-0.004}^{+0.007}$ & $0.008_{-0.005}^{+0.008}$ \\
        \midrule
        & Ground Truth & $0.077_{-0.051}^{+0.085}$ & $0.077_{-0.050}^{+0.083}$ & $0.011_{-0.006}^{+0.010}$ & $0.010_{-0.005}^{+0.009}$ \\
        \cmidrule(rl){2-6}
        \multirow{3}{*}{Neighbour LoS} & MAP & $0.045_{-0.027}^{+0.055}$ & $0.103_{-0.057}^{+0.095}$ & $0.006_{-0.004}^{+0.008}$ & $0.010_{-0.006}^{+0.010}$ \\
        & MeanShift & $0.056_{-0.033}^{+0.048}$ & $0.095_{-0.052}^{+0.082}$ & $0.008_{-0.005}^{+0.008}$ & $0.010_{-0.005}^{+0.009}$ \\
        & MNPCP & $0.051_{-0.030}^{+0.049}$ & $0.066_{-0.037}^{+0.058}$ & $0.008_{-0.008}^{+0.008}$ & $0.007_{-0.004}^{+0.007}$\\
        \bottomrule
    \end{tabular}
    \label{tab:delta_pix}
\end{table*}

\subsection{Taking the neighbouring LoSs into account}
\label{sec:Results_NLoS_cINN}

\begin{figure*}
    \centering
    \includegraphics[width=0.8\textwidth]{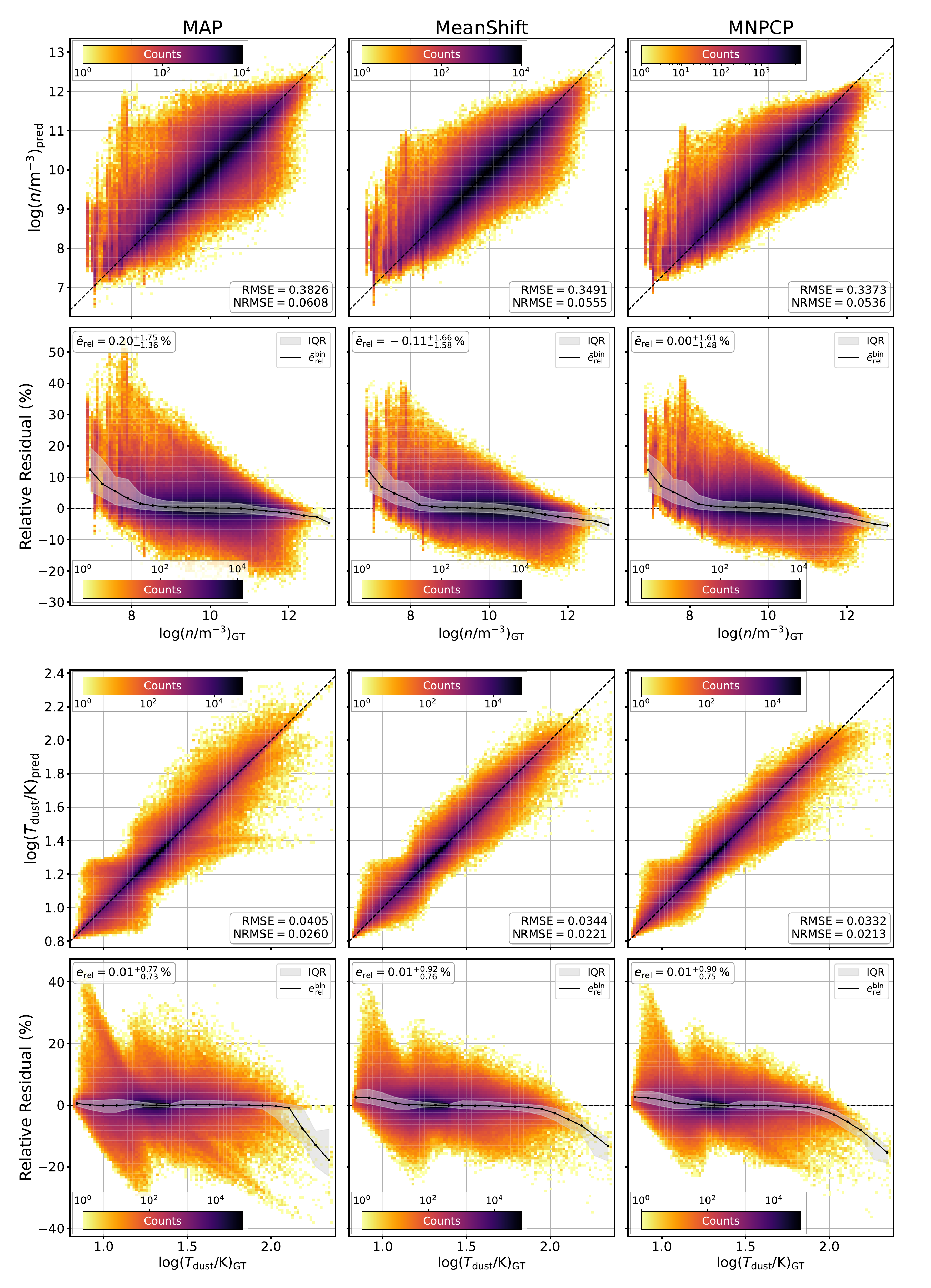}
    \caption{Prediction performance summary for the NLoS-cINN on the held-out test data. 2D histograms of the direct one-to-one comparison of prediction results to the ground truth (rows one and three) and the respective residuals (rows 2\&4) are shown for the three different MAP estimators (analogously to Figure~\ref{fig:2D_histogram_STAR_ISRF_all}).}
    \label{fig:2D_histogram_STAR_ISRF_NN_LoS_all}
\end{figure*}

Rows 5 and 6 of Figure~\ref{fig:SingleSliceComparison} provide the qualitative comparison between the prediction outcomes of the three point estimation approaches for the NLoS-cINN, which is the model trained for the alternative formulation of the inverse problem described in Section~\ref{sec:Method_SpatialConsistency}. For direct compatibility the shown slice and example cube are the same as for the SLoS-cINN outcomes in Figure~\ref{fig:SingleSliceComparison}, except that the $124$ border pixels, which lack the required number of neighbouring LoSs for the prediction with the NLoS-cINN, are missing. At first glance the MAP and MeanShift results appear less noisy, which is subject to fewer strong discontinuities, with the NLoS-cINN in comparison to the SLoS-cINN outcome (first two rows of Figure~\ref{fig:SingleSliceComparison}). In addition, the NLoS-cINN based reconstructions seem overall to be slightly more faithful to the ground truth in terms of the recovered details. 

Looking at the spatial discontinuities perpendicular to the LoSs in the MAP and MeanShift estimates (see Table~\ref{tab:delta_pix}), it appears that the NLoS-cINN suffers on average from the same issue as the SLoS-based prediction results. $\Delta_{\perp \mathrm{LoS}}$ is still twice as large as $\Delta_{\parallel \mathrm{LoS}}$ for both density and temperature. Again, only the MNPCP approach achieves a more balanced result, but not quite at the level of the ground truth. Thus, accounting for the fluxes in the neighbouring LoSs does not appear to lead to a significant improvement of the spatial consistency perpendicular to the LoS in the prediction of dust density and temperature. It is worth noting, however, that this observation may only hold in the fully resolved dust emission map scenario that we have posed in this study, which renders neighbouring LoSs effectively independent. In real observations, however, where the PSF of a given instrument is larger than a single pixel, neighbouring pixels in the dust emission maps may become correlated. In the latter case, it is possible that the NLoS-cINN may perform better with regards to the spatial consistency. We will conduct a corresponding test in our subsequent work, once we have established a proper treament of the instrument-related resolution effects during training.

Looking at the overall performance of the NLoS-cINN in comparison to the SLoS-cINN (see Table~\ref{tab:PerformanceSummary} and Figure~\ref{fig:2D_histogram_STAR_ISRF_NN_LoS_all}), we do find a general improvement. In particular, the NRMSE goes down to values as low as $0.054$ and $0.0213$ for $\log(n/\mathrm{m}^{-3})$ and $\log(T_{\mathrm{dust}}/\mathrm{K})$ with the MNPCP estimator, compared to the SLoS-cINN results of $0.062$ and $0.0245$. $|\bar{e}_{\mathrm{rel}}|$ with, for instance, the MNPCP estimator improves to $1.54$\% and $0.82$\% opposed to the SLoS-based performance of $1.84$\% and $0.96$\%, respectively. It is possible that this performance improvement is only a data selection effect, since the NLoS and SLoS test sets are not fully identical, with the former missing the edge LoSs of every cube. To test this hypothesis, we recomputed the SLoS-cINN performance, limited to the LoSs of the NLoS test set. This experiment reveals that the observed average performance improvement of the NLoS-cINN is real, as the SLoS-cINN performs even slightly worse on this limited test set, returning for instance $|\bar{e}_{\mathrm{rel}}|$ values of $1.86$\% and $0.98$\% for $\log(n/\mathrm{m}^{-3})$ and $\log(T_{\mathrm{dust}}/\mathrm{K})$ with the MNPCP estimator, respectively.

In summary, accounting for the fluxes of the neighbouring LoSs in the input has not achieved its initial goal of improving the spatial consistency of the predicted dust densities and temperatures perpendicular to the LoS. It has, however, demonstrated a slight improvement of the predictive performance of the model, indicating that knowledge of the fluxes in the neighbouring LoSs provides additional constraints for the prediction of the dust properties. However, this comes at a cost of flexibility, as all query LoSs  now also require observations of the eight adjacent LoSs in this approach. All in all, the NLoS-cINN does not increase the spatial consistency and despite a slight performance improvement does not appear to be a markedly superior approach.   

\subsection{Realistic wavelength coverage test}
\label{sec:Results_SLoS_7Filter}

\begin{figure*}
    \centering
    \includegraphics[width=0.8\textwidth]{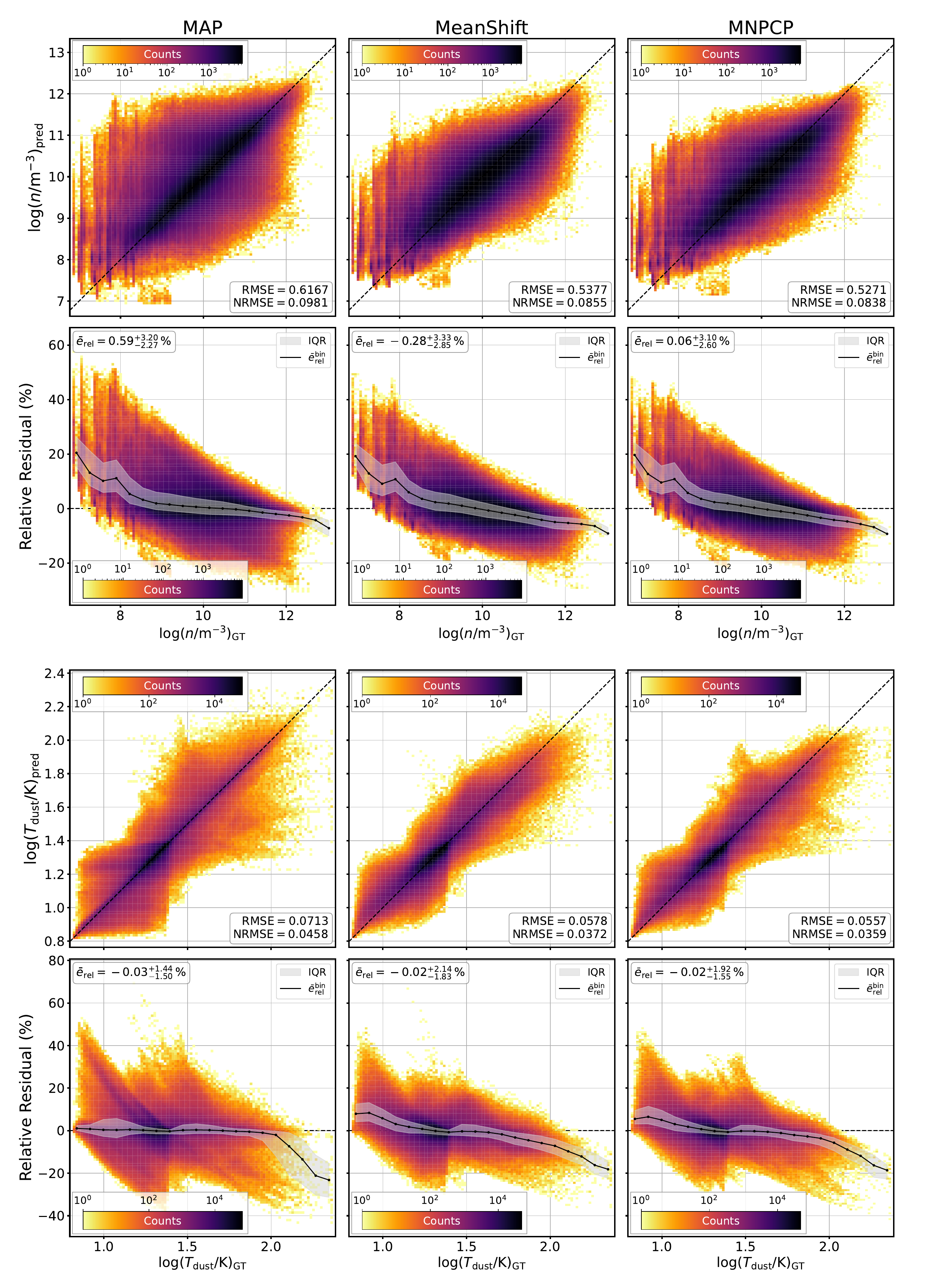}
    \caption{Summary of the predictive performance on the held-out test data for the SLoS-cINN using only seven wavelenghts as input. Shown are 2D histograms of the direct one-to-one comparison of prediction results to the ground truth (rows one and three) and the respective residuals (rows 2\&4) for the three different MAP estimators (analogously to Figure~\ref{fig:2D_histogram_STAR_ISRF_all}).}
    \label{fig:2D_histogram_STAR_ISRF_rho_oph}
\end{figure*}

As our final experiment, we trained and tested an SLoS-cINN for a more realistically limited wavelength coverage, corresponding to the central wavelengths of the following seven bands: WISE $22\ \mu\mathrm{m}$, SOFIA $89\ \mu\mathrm{m}$ and $154\ \mu\mathrm{m}$, Herschel PACS $100\ \mu\mathrm{m}$ and $160\ \mu\mathrm{m}$, Herschel SPIRE $350\ \mu\mathrm{m,}$ and LABOCA $870\ \mu\mathrm{m}$. The last two rows in Figure~\ref{fig:SingleSliceComparison} provide the qualitative example of the dust density and temperature prediction results for a single cube slice in comparison to the ground truth (see also Figure~\ref{fig:EmissionMapsExample} in the appendix for the corresponding input emission maps at the seven selected wavelengths, as indicated by the highlighted panels). It is immediately evident that the large reduction in wavelength coverage leads to a notably decreased quality in the reconstruction. While  larger scale and more diffuse features of the density and temperature distributions are still being recovered very well, this is no longer true for narrow details at high density and low temperature, in particular, which tend to be only partially reconstructed. This is even more apparent in the full prediction summary of this cube in Figure~\ref{fig:AllSliceComparison_ISRF_7filter_076412} (and the corresponding isodensity surface diagram in Figure~\ref{fig:IsodensitySurface_ISRF_7filter_076412}) in the appendix. In addition, the MAP point estimator appears to produce a lot more and larger spatial discrepancies in the predicted dust densities and temperatures perpendicular to the LoS. Interestingly, where the MeanShift algorithm appeared to show similar behaviour to the MAP estimator in the previous SLoS-cINN setup in terms of spatial inconsistencies, it seems to provide more consistent prediction results here. Looking at $\Delta_{\perp \mathrm{LoS}}$ and $\Delta_{\parallel \mathrm{LoS}}$ in Table~\ref{tab:delta_pix}, this can be quantitatively confirmed at least for the dust density as well, with $\Delta_{\perp \mathrm{LoS}}$ being only about two times greater than $\Delta_{\parallel \mathrm{LoS}}$ in the MeanShift result, compared to $\Delta_{\perp \mathrm{LoS}} \approx 3 \Delta_{\parallel \mathrm{LoS}}$ in the MAP outcome. With this outcome, it seems that  a determination of the most likely solution in the joint target parameter space in this setup is more robust in terms of the spatial consistency of the dust density than the marginalisation approach in the MAP estimator. 

Looking at the overall performance on the test set in Table~\ref{tab:PerformanceSummary}, the limitation to only seven wavelengths in this setup increases the NRMSE to $0.084$ -- $0.098$ and $0.036$ -- $0.046$ for $\log(n/\mathrm{m}^{-3})$ and $\log(T_{\mathrm{dust}}/\mathrm{K})$, respectively. The corresponding $|\bar{e}_{\mathrm{rel}}|$ values rise to  $2.6$ -- $3.1$\% and $1.5$ -- $2.0$\%. Although this is a notable performance decrease with $|\bar{e}_{\mathrm{rel}}|$ increasing by a factor of $1.5$ -- $2$ in comparison to the 23-wavelength setup, considering that this cINN has 16 wavelengths fewer to work with, this is still a quite satisfactory performance. This indicates that the 3D reconstruction approach can be feasible even for more limited observational coverage.

Investigating the prediction performance in more detail in Figure~\ref{fig:2D_histogram_STAR_ISRF_rho_oph} (which provides a breakdown, analogously to Figures~\ref{fig:2D_histogram_STAR_ISRF_all} and \ref{fig:2D_histogram_STAR_ISRF_NN_LoS_all}), we find that the change in wavelength coverage also influences the systematic tendencies of the prediction results. Where for instance the binned median relative residual curve for the dust density in the 23-wavelength SLoS-cINN outcome exhibits a plateau between $10^8$ and $10^{11}\,\mathrm{m}^{-3}$ and only really leans into systematic behaviour below and above this range, the systematic offsets are amplified in the seven wavelength prediction outcomes. In particular for the MeanShift and MNPCP point estimates we now find a pivot point at around $10^{10}\,\mathrm{m}^{-3}$, below which the density tends to be overestimated and above which it is systematically underestimated. We find a similar amplification of the earlier observed systematic behaviour in the MeanShift and MNPCP estimates of the dust temperature, where there is now a pivot point between over- and underestimation at about $19$ K. Nevertheless, for most pixels, the error remains comparatively small, as indicated by $|\bar{e}_{\mathrm{rel}}|$, and only few individual pixels will manage to reach the maximum relative residual of about $60$\,\%. Part of this systematic behaviour can likely be attributed to the relative lack of examples in the training data towards the lower and upper limits of density and temperature, as we discuss in Section~\ref{sec:Results_SLoS_cINN}. However, given that this cINN also exhibits systematic offsets within the parameter ranges, where a lot of training data is available, this also points towards the increased complexity of the recovery task with much less observational information; in particular in the intermediate density and temperature ranges. Evidence for the intrinsic increase in complexity can be found in the shape of the predicted posterior distributions, a few examples of which are shown in Figure~\ref{fig:PosteriorSliceComparison_CUBE_076412} in the appendix. Compared to the full 23-wavelength SLoS-cINN, the predicted posteriors of the seven wavelength limited model appear notably broader and often exhibit double or multi-peaked distributions for both density and dust temperature even in the ISRF-only RT configuration. This shows that the cINN has both a harder time to constrain the prediction and finds more degeneracy in the more limited problem.

In summary, although the 7-wavelength SLoS-cINN exhibits an (expected) reduction in predictive performance compared to the 23-wavelength counterpart, this experiment proves that the 3D reconstruction of the dust distributions is quite feasible even when subject to more realistic observational constraints. We also want to emphasise again that our wavelength selection does not necessarily preserve the most information for the given inverse problem. A different, more optimal selection of wavelengths might retain more of the predictive performance of the full 23-wavelength cINN. We plan to look further into both the optimal choice of wavelength combination and relative importance of each wavelength (for the reconstruction) for realistic applications in our continued development of this 3D reconstruction approach.

\subsection{Comparison with classical SED fit}
\label{sec:Results_SED_fit}
The dust emission is often modelled by a modified black body (MBB) in order to construct a column density map and the temperature distribution from multi-wavelength observations. As a final test, we compared the predictive performance of our cINN method to the results of a classical MBB fit. For the purposes of this test, we employed the full 23 wavelengths of coverage and focussed on the simpler RT scenario, comparing SED fit and cINN approach on the 50 cubes that are only subject to the ISRF. The dust emission, $I_\nu$, may be approximated as an MBB by:
\begin{equation}
    I_\nu = B_\nu\left(T_{\mathrm{dust}}\right) \left( 1-e^{-\tau_\nu} \right) \approx B_\nu\left(T_{\mathrm{dust}}\right) \tau_\nu = \mu_{\mathrm{g}}m_{\mathrm{H}} N_{\mathrm{H}}\delta_{\mathrm{gd}}\kappa_\nu B_\nu\left(T_{\mathrm{dust}}\right)
,\end{equation}
where $\tau_\nu$ is the optical depth, $m_{\mathrm{H}}$ is the hydrogen mass, and the black body $B_\nu\left(T_{\mathrm{dust}}\right)$ spectrum is modified by the dust opacity:
\begin{equation}
    \kappa_\nu =  2 \left( \frac{\nu}{\nu_0} \right)^\beta\ \mathrm{cm}^2\ \mathrm{g}^{-1}\, .
\end{equation}
For the opacity, we take a characteristic frequency of $\nu_0=600\ \mathrm{GHz}$ and a spectral index of $\beta=2$. Here, the dust temperature $T_{\mathrm{dust}}$ and the gas column density $N_{\mathrm{H}}$ are the fit parameters to be determined by a least-squares fit.

\begin{figure*}
    \centering
    \includegraphics[width=0.75\textwidth]{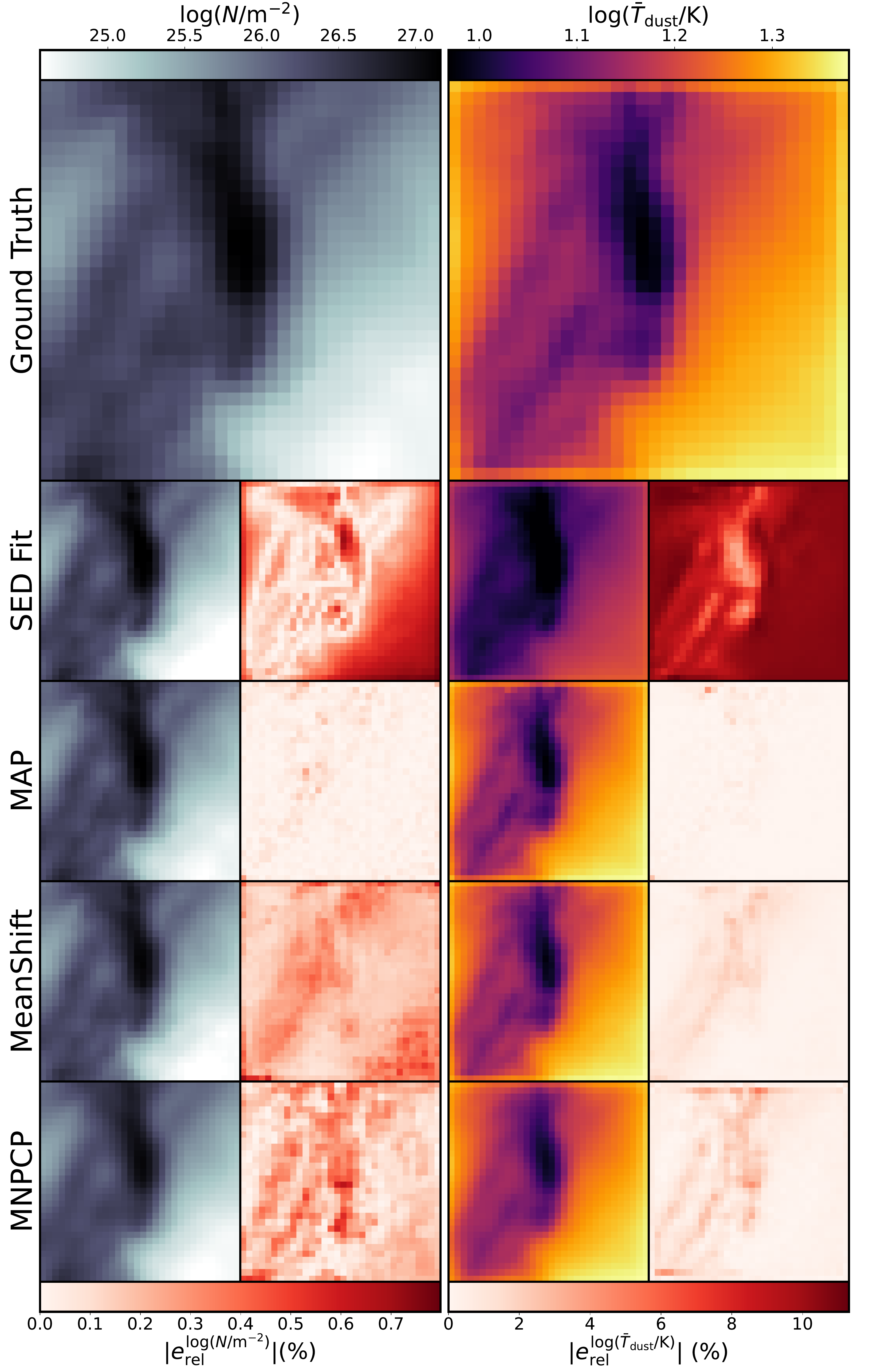}
    \caption{Comparison of estimated column density and average temperature maps for one example cube in the ISRF-only RT configuration between a classical SED fit and our cINN approach. The large panels on the top show the ground truth for the column density (left) and density weighted average temperature (right), respectively. The smaller panels below each ground truth panel present a map of the estimates from the respective method on the left and a map of the absolute relative error $|e_{\mathrm{rel}}|$ on the right. The example cube as in Figures~\ref{fig:SingleSliceComparison}, \ref{fig:AllSliceComparison_ISRF_all_076412}, and \ref{fig:IsodensitySurface_ISRF_all_076412} is shown.}
    \label{fig:ColumnDensityComparison_CubeID_076412}
\end{figure*}

\begin{figure*}
    \centering
    \includegraphics[width=\textwidth]{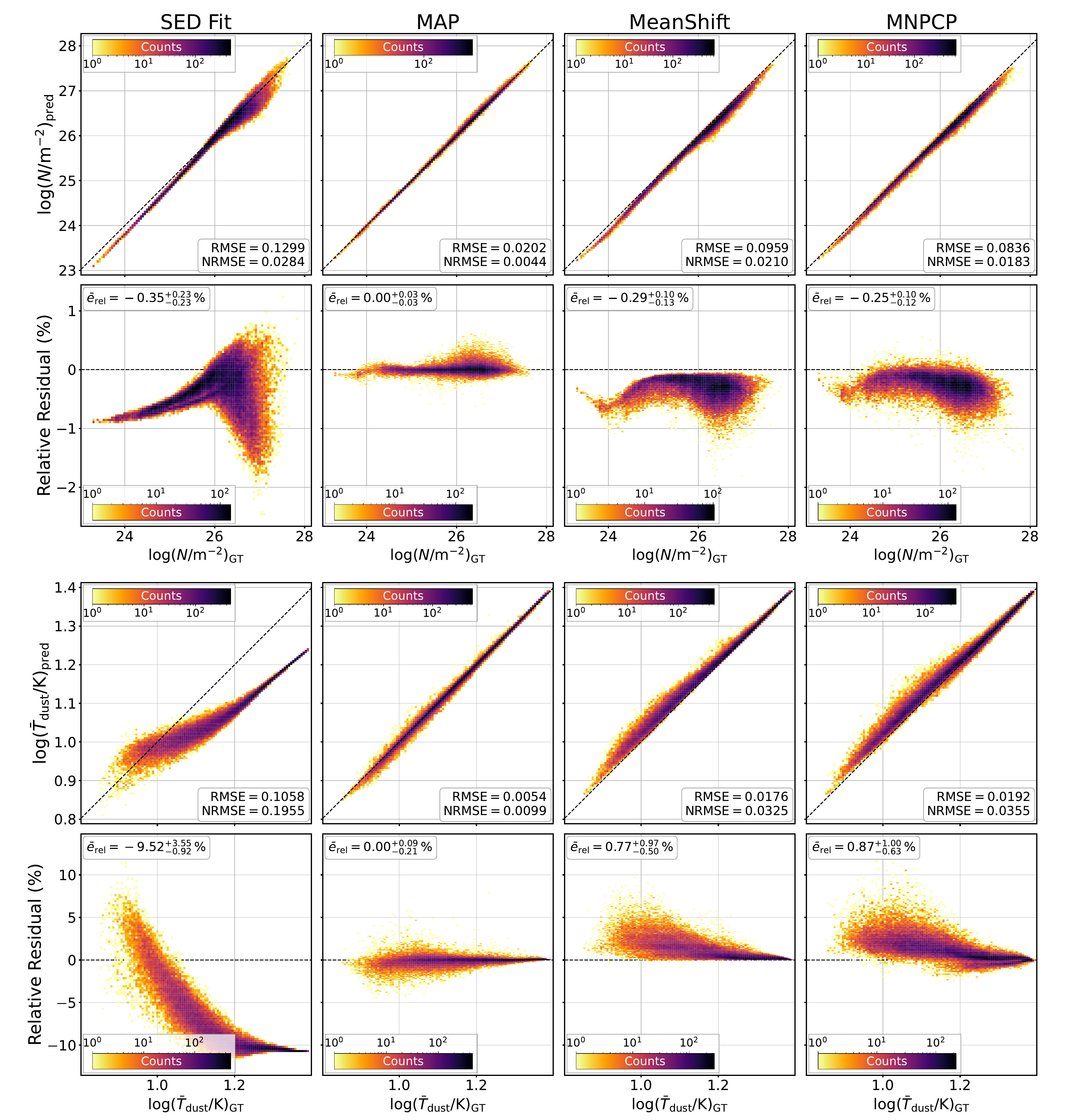}
    \caption{Comparison of the predictive performance for column density and density averaged temperature between a classical SED fitting technique and our cINN approach across all pixels of the 50 test cubes that are only subject to the ISRF. The top two rows show the results for the estimated column density, whereas the bottom two present the density averaged dust temperature, $\bar{T}_{\mathrm{dust}}$. In each group, the upper panels provide a direct one-to-one comparison of the predicted values to the respective ground truth, whereas the lower panels show the corresponding relative residuals.}
    \label{fig:ColumnDensity_1to1_2D_hist_ISRF_eval}
\end{figure*}

To compare the cINN outcome with the SED fit results we compute the column number density, $N,$ by integrating the density along each LoS for both the ground truth and SLoS-cINN prediction outcome, that is:
\begin{equation}
    N = \int_0^{L} n\ \mathrm{d}l = \sum_{i=1}^{32} n_i\ \Delta l, 
\end{equation}
where $L$ indicates the depth of our test cubes of $0.2\,\mathrm{pc}$ and $\Delta l = L / 32$ is our cube spatial resolution. As a proxy for the temperature determined by the SED fit, we can compute a density-weighted average temperature $\bar{T}_{\mathrm{dust}}$ following:
\begin{equation}
    \bar{T}_{\mathrm{dust}} = \frac{\sum_{i=1}^{32} n_i T_{\mathrm{dust}, i}}{\sum_{i=1}^{32} n_i}.
\end{equation}
Figure~\ref{fig:ColumnDensityComparison_CubeID_076412} provides an example comparison of the column number densities and density weighted average temperatures between the outcome of the SED fit, the cINN prediction and the ground truth for one example cube that is only subject to the ISRF. As this example shows, both the SED fit and cINN approach manage to reproduce the column number density map for this cube quite accurately with overall absolute relative errors below 1\%. However, the result based on the MAP cINN estimates returns notably smaller residual errors than the SED fit. The maps based on the MeanShift and MNPCP point estimators on the other hand are about on par with the SED fit, if not slightly better. For MeanShift this is likely explained by the on average higher absolute relative error of the method itself in comparison to the MAP estimator (as we describe in Section~\ref{sec:Results_SLoS_cINN}). The MNPCP estimator on the other hand performs worse here because of its inherent smoothing action perpendicular to the LoS. As such small amounts of material may get mixed between adjacent LoSs, increasing the error on the column density. Looking now at the density averaged temperature $\bar{T}_{\mathrm{dust}}$, we find excellent results for the cINN based estimates, whereas the SED fit outcome notably underestimates the temperature by up to 10\% or more. However, it is important to note that the density-weighted average temperature is only an approximation of the temperature that an SED fit would return, so a discrepancy is to be expected. Lastly, Figure~\ref{fig:ColumnDensity_1to1_2D_hist_ISRF_eval} provides the performance comparison between the SED fit and cINN approach across all pixels of the 50 considered test cubes. This figure confirms that the cINN based estimates of both column number density and density weighted average temperature are overall quite excellent. In addition, the cINN approach outperforms the SED fit in all cases, although the margin is comparatively small in the column number density for the MeanShift-based outcome. We also note a slight tendency of the MeanShift and MNPCP derived results towards overestimation of the density averaged temperature. We emphasise that the MBB SED fit may only provide a 2D estimate of the underlying density and temperature distributions, whereas the advantage of the cINN approach is the reconstruction of the full 3D information along the LoS. We also note that while this test indicates the MAP point estimate as the best suited choice to recover the column density, for the full 3D reconstruction, we still recommend the MNPCP estimate because of its higher degree of spatial consistency in the 3D structure.

\subsection{On the physical feasibility of the 3D reconstruction}
\label{sec:Results_PhysFeasibility}
The physical reason why our approach of reconstructing the underlying 3D density and temperature structure works so well is that the dust opacity strongly depends on the energy of the incident electromagnetic wave. In addition, there is a dependence on the chemical composition and the grain size distribution, but we keep those fixed here to the standard Milky Way values (Section \ref{sec:TrainingData_SyntheticImages}). For further details we refer to \citet{Tielens2010} and \citet{Draine2011}. Overall, this means that different wavelengths trace different optical depths within the cloud. As a consequence, the observed radiation is accumulated in different ways along the LoS and the resulting SED encodes critical information about the density and temperature structure of the cloud, as also discussed in the  Section \ref{sec:Results_SED_fit}. This process is highly degenerate, as varying cloud configurations can lead to very similar SEDs; thus, solving the inverse problem of reconstructing the cloud properties from the SED is a challenging task. This is where the INN architecture can play out its full potential since it belongs to the group of normalising flow methods. It offers direct access to the full posterior distribution function and enables us to approach this challenge in a statistically reliable way. 

\subsection{Application to real data}
\label{sec:Results_ApplRealData}
An application to real observational data of cloud cores is at this stage not currently feasible with our setup. The main reason for this is our treatment of the synthetic dust emission observations. The choice to only use monochromatic dust emission and to not include instrument-related effects, such as a consideration of the PSF or noise renders our synthetic dust emission observations notably different to what the actual observatories would measure. This results in a gap between the synthetic observable space learned by our cINN models and the actual observable space spanned by real measurements, so that prediction attempts are bound to fail (for an example see Appendix~\ref{app:BeamConvTest}). For an application to real data, the generation of the synthetic dust observations thus needs to be further refined by accounting for instance for the actual band width of real instruments and ensuring a proper sampling and modelling of their respective PSFs. We plan to address these challenges in our subsequent development of our method.

Apart from the current limitations of our synthetic dust emission observations, we also identified another area that may require improvements towards an application to real data.
As we describe in Section~\ref{sec:TrainingData}, our training data generation is aimed at creating simplified, synthetic analogues to objects such as $\rho$ Oph A.
However, we suspect that our simplified prescription for adding a star to the cubes, that is, randomly placing it in a region of low dust density to emulate stellar feedback, is not likely to produce truly realistic results. In this approach, the dust clouds are merely illuminated by the star, rather than being mechanically affected by the stellar feedback. Consequently, we do not actually model clouds that are actively shaped by their nearby stars, as is (in particular) the case for cores such as that of $\rho$ Oph A. For that reason, we plan to move to more sophisticated MHD simulations, where stellar feedback is properly accounted for before we apply our method to real observations.

\section{Summary and conclusions}
\label{sec:Summary}
In this paper, we present a proof-of-concept deep-learning approach to reconstructing interstellar dust distributions in 3D from observed dust emission maps, focussing on the sub-parsec length scales of individual star-forming clumps. To train this approach, we modelled simplified analogues of the $\rho$ Oph A star-forming clump, employing the Cloud Factory dust cloud simulations by \cite{Smith2020_CF1, Smith2021_CF2}. For the basis of our training database, we selected $11.065$ cubes from the Cloud Factory, centered on clump-like dust aggregations, with a side length of $0.2$~pc and $32 \times 32 \times 32$ pixel resolution. We simulated the corresponding dust emission observations with the {\sc POLARIS} \citep{Reissl2016, Reissl2019} radiative transfer code at 23 different wavelengths between $12\ \mu\mathrm{m}$ and $1300\ \mu\mathrm{m}$, matching the central wavelengths of the observational instruments of WISE, MSX, Spitzer, SOFIA, Herschel, ALMA, APEX, and CSO. For the radiative transfer, we considered two irradiation scenarios. In the first, the dust is only subject to the interstellar radiation field (ISRF) with an amplitude that matches conditions in nearby star-forming cores. In the second, we randomly inserted one B4-type star in a low density area inside the cube in addition to the ISRF in order to emulate cloud cores that are subject to the radiation of nearby stars. This procedure yields a final training dataset of $22.130$ mock dust clumps, including their dust density and temperature distributions on a 3D grid and the corresponding simulated dust emission fluxes in 23 wavelengths.

To reduce the complexity and dimensionality of the inverse problem posed by the 3D reconstruction task, we broke the problem down to individual LoS under the assumption that (to first order) the emission measured in a given pixel only depends on the material along the corresponding LoS. Specifically, we aimed to recover the 32 dust densities and temperatures along a given LoS from the measured fluxes at different wavelengths in the corresponding pixel of the dust emission map. For this purpose, we trained a conditional invertible neural network (cINN), which is a deep-learning approach that can efficiently estimate full posterior distributions for the target parameters conditioned on the observations. To recover point estimates from the posterior distributions predicted by the cINN (and compare these to the ground truth), we tested three different methods. 
The first uses a 1D kernel density estimate to determine the maximum a posteriori (MAP) prediction (that is the most likely value) for the dust density and temperature from the marginalised 1D posterior distributions of the individual pixels.  
The second method employs the MeanShift algorithm to determine the most likely solutions as the point with the highest (probability) density in the full 64-dimensional space of the predicted posterior distributions.
As both the MAP and MeanShift estimators have no spatial consistency guarantee perpendicular to the LoS, as per our reduction of the reconstruction task, we introduced a third estimator to rectify this circumstance. This method, dubbed median neighbour pixel combined posterior (MNPCP), determines the point estimates for a given pixel by accumulating the posterior samples of all neighbouring pixels and computing the median of dust density and temperature on this collection.

In total, we trained and tested cINNs for three different formulations of the reconstruction task. The first consists of a perfect information scenario, where we have access to the dust emission maps at all 23 wavelengths. In the second, we extended the network input to also account for the observed fluxes in the neighbouring pixels of the query LoS to evaluate whether this improves the spatial consistency of the prediction outcome. Lastly, we considered a more realistically limited observational scenario, where observations are only available in a combination of seven wavelengths (matching the coverage of real observational data that is for instance available for the star-forming core $\rho$~Oph~A). We then evaluated the performance of our trained cINN models on a synthetic test set that consists of 50 mock dust clouds in both irradiation scenarios. Our main findings are summarised in the following. 

The model trained for the 23-wavelength scenario achieves an excellent overall predictive performance, returning median absolute relative errors $|\bar{e}_{\mathrm{rel}}|$ of $1.84$ to $2.01$\% in $\log(n/m^{-3})$ and $0.87$ to $1.01$\% in $\log(T_{\mathrm{dust}}/\mathrm{K}),$ depending on the choice of the point estimator. Averaged over the five best reconstructed cloud cores, $|\bar{e}_{\mathrm{rel}}|$ even goes as low as $1$\% and $0.5$\% in $\log(n/m^{-3})$ and $\log(T_{\mathrm{dust}}/\mathrm{K})$, respectively. In general, we find that the reconstructive performance is better for clouds that are only subject to the ISRF, indicating that the presence of a star in the vicinity of a mock dust cloud adds a notable level of complexity to the reconstruction task. Breaking the predictive performance down to the individual pixels, we also identified some systematic behaviours. Although the dust densities and temperatures were  recovered well overall, we found trends for overestimating the density in the low density regime ($n < 10^8\,\mathrm{m}^{-3}$) and of slight underestimations for very-high-density regions ($n > 10^{11}\,\mathrm{m}^{-3}$). A similar tendency for underestimation also appears in the dust temperature point estimates for $T_{\mathrm{dust}} > 100$ K. We identified a bias in the training data as a potential explanation for this behaviour, as the training database contains significantly fewer pixels in these parameter ranges. It should be noted, however, that (in particular) the $T_{\mathrm{dust}} > 100$ K regime represents more extreme conditions that are more typically found in HII regions or in regions of strong outflow activity; however, these are neither accounted for in the Cloud Factory simulations nor the focus of our analysis. This incomplete modelling of the high temperature regime may also tie into the comparatively worse reconstructive performance.

The comparison of the three point estimators reveals no clear favourite in terms of the quantitative performance measures. Although the MeanShift and MNPCP estimators achieve a slightly better NRMSE than the MAP and the MAP estimator provides the smallest $|\bar{e}_{\mathrm{rel}}|$, the difference between the three methods is only on the order of $10^{-3}$ in NRMSE and $0.1$\% in $|\bar{e}_{\mathrm{rel}}|$. The MAP and MeanShift estimates do suffer from the aforementioned spatial inconsistencies perpendicular to the LoS (as per construction). Although the MeanShift results appear to be slightly less affected by this, we find that determining an optimal bandwidth for MeanShift in the 64-dimensional target parameter space can be tricky and may lead to oversmoothing. The MNPCP approach provides much better results in term of spatial consistency, but also exhibits oversmoothing behaviour, which amplifies the systematic trends of the method at the edges of the parameter ranges. Ultimately, we would generally recommend the MNPCP solution (despite its flaws), as it quite effectively avoids the more severe and unphysical spatial inconsistencies perpendicular to the LoS that the MAP and MeanShift approaches can hardly avoid (by construction).

For the cINN model that also accounts for the emission in the neighbouring LoSs, we found an overall slight increase in the predictive performance, but no significant improvement in the spatial consistency of the prediction outcome. This experiment indicates that accounting for the measured fluxes in the neighbouring pixels may help with constraining the dust density and temperatures overall, but cannot counteract the inherent spatial consistency bias of the LoS approach. In addition, this setup is less flexible as it requires a query LoS to have measurements for all eight neighbouring pixels, which excludes (for instance) edge pixels. Although the slight performance improvement of this approach is certainly desirable, it is overall not large enough to clearly distinguish this setup as a superior method in comparison to the single LoS approach. 

Limiting the input to a more realistic coverage of seven wavelengths -- in our case corresponding to the central wavelengths of the WISE $22\ \mu\mathrm{m}$, SOFIA $89\ \mu\mathrm{m}$ and $154\ \mu\mathrm{m}$, Herschel PACS $100\ \mu\mathrm{m}$ and $160\ \mu\mathrm{m}$, Herschel SPIRE $350\ \mu\mathrm{m,}$ and LABOCA $870\ \mu\mathrm{m}$ bands -- naturally leads to a decrease in predictive performance. Nevertheless, the cINN overall still achieves a satisfactory performance with $|\bar{e}_{\mathrm{rel}}|$ on the order of $2.6$ to $3.1$\% for $\log(n/\mathrm{m}^{-3})$ and $1.5$ to $2.0$\% in $\log(T_{\mathrm{dust}}/\mathrm{K})$. The most notable change in the behaviour of the seven wavelength cINN is an amplification of the average systematic errors in the predicted point estimates, leading to  more difficulties overall in reconstructing the dust distributions in very dense structures or extremely diffuse regions. It is worth emphasising, however, that the tested selection of wavelengths was inspired by available data for $\rho$ Oph A and not curated to maximise the network performance. A more optimal wavelength configuration is certainly likely to achieve even better reconstructive power.
Lastly, we find that an application of our approach to real observational data is not yet feasible at this stage, as our simplified simulation setup is not able to correctly model all the nuances of real dust emission observations.
We conclude that a more in-depth treatment of the synthetic dust emission observations and improvements to our simulation basis (to e.g.~properly model the interaction between stars and the dust through stellar feedback) may be required.

In summary, we have shown that a cINN-based approach for the 3D reconstruction of dust distributions produces quite excellent results in a perfect information scenario on synthetic data and still retains a satisfactory performance, even for more realistly limited observational coverage. For future applications to real data, however, the method still requires a further refinement of the simulation setup for the training data to resolve the mismatch between real and synthetic observations, with some adjustment to the training set sampling to reduce the potential bias, and an analysis of the most informative band combinations to maximise performance in realistic coverage scenarios. Once these points have been resolved over the course of our follow-up studies and we can demonstrate a successful application to real observational data, we plan to make this approach available to the community in the form of an open-source tool. Access to a work-in-progress code may be provided upon reasonable request beforehand.

\begin{acknowledgements}
The team in Heidelberg acknowledges funding from the European Research Council via the ERC Synergy Grant ``ECOGAL'' (project ID 855130), from the German Excellence Strategy via the Heidelberg Cluster of Excellence (EXC 2181 - 390900948) ``STRUCTURES'', and from the German Ministry for Economic Affairs and Climate Action in project ``MAINN'' (funding ID 50OO2206). They also thank for computing resources provided by {\em The L\"{a}nd} and DFG through grant INST 35/1134-1 FUGG and for data storage at SDS@hd through grant INST 35/1314-1 FUGG. RJS gratefully acknowledges an STFC Ernest Rutherford fellowship (grant ST/N00485X/1).
\end{acknowledgements}

\bibliographystyle{aa}
\bibliography{./bibtex}

\begin{appendix}
\section{Training set generation}
\label{app:TS_generation}

\begin{figure*}
    \centering
    \includegraphics[width=0.8\textwidth]{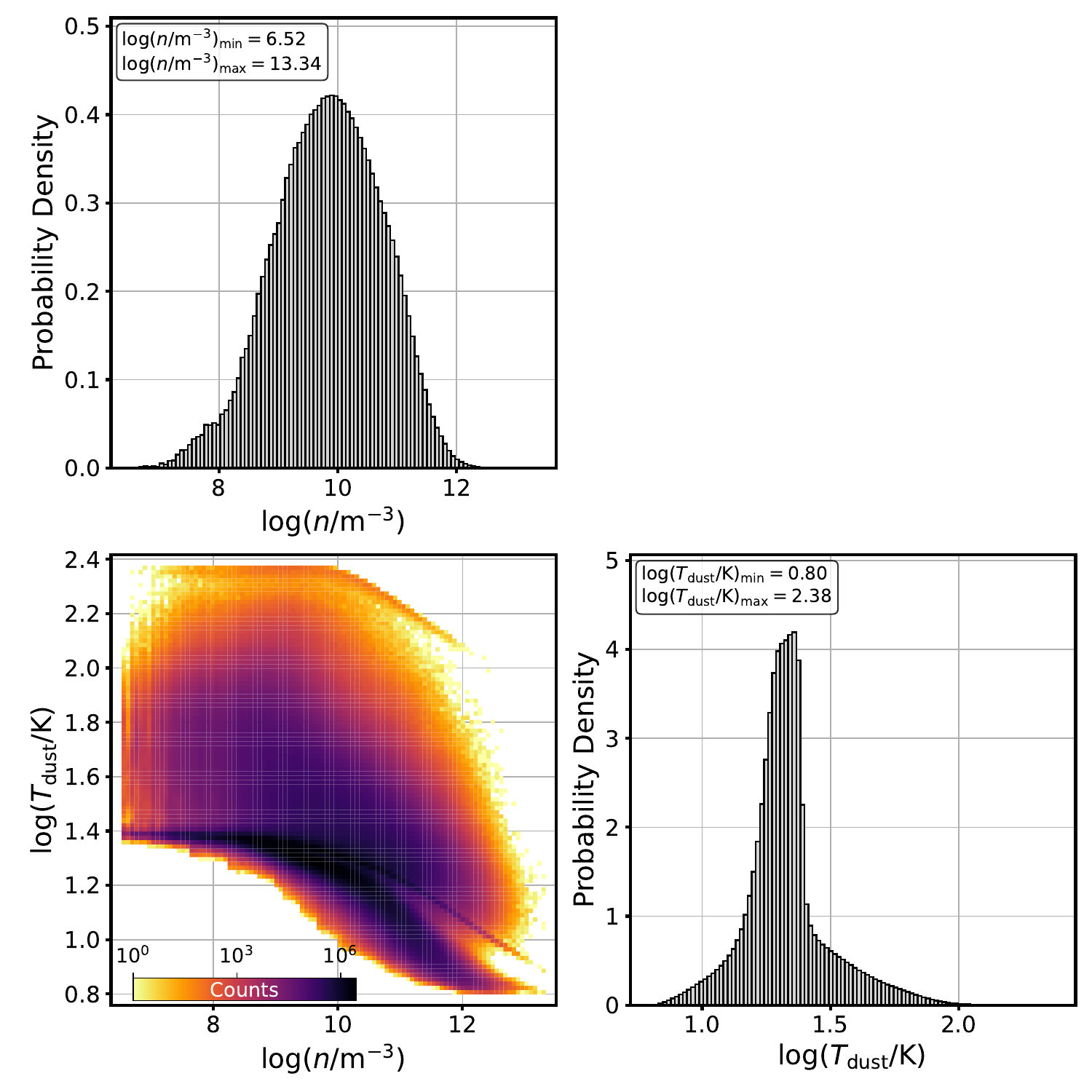}
    \caption{Histograms of the prior distributions and correlation of the target parameters in the training data over all pixels. The top-left and bottom-right panels show the 1D distributions of the number density and dust temperature, respectively. In both panels, the boxes at the top left provide the minimum and maximum of the respective parameter. The bottom left panel presents a 2D histogram of the effective prior distribution in the combined density-temperature space.}
    \label{fig:TS_priors_targets}
\end{figure*}

\begin{figure*}
    \centering
    \includegraphics[width=0.85\textwidth]{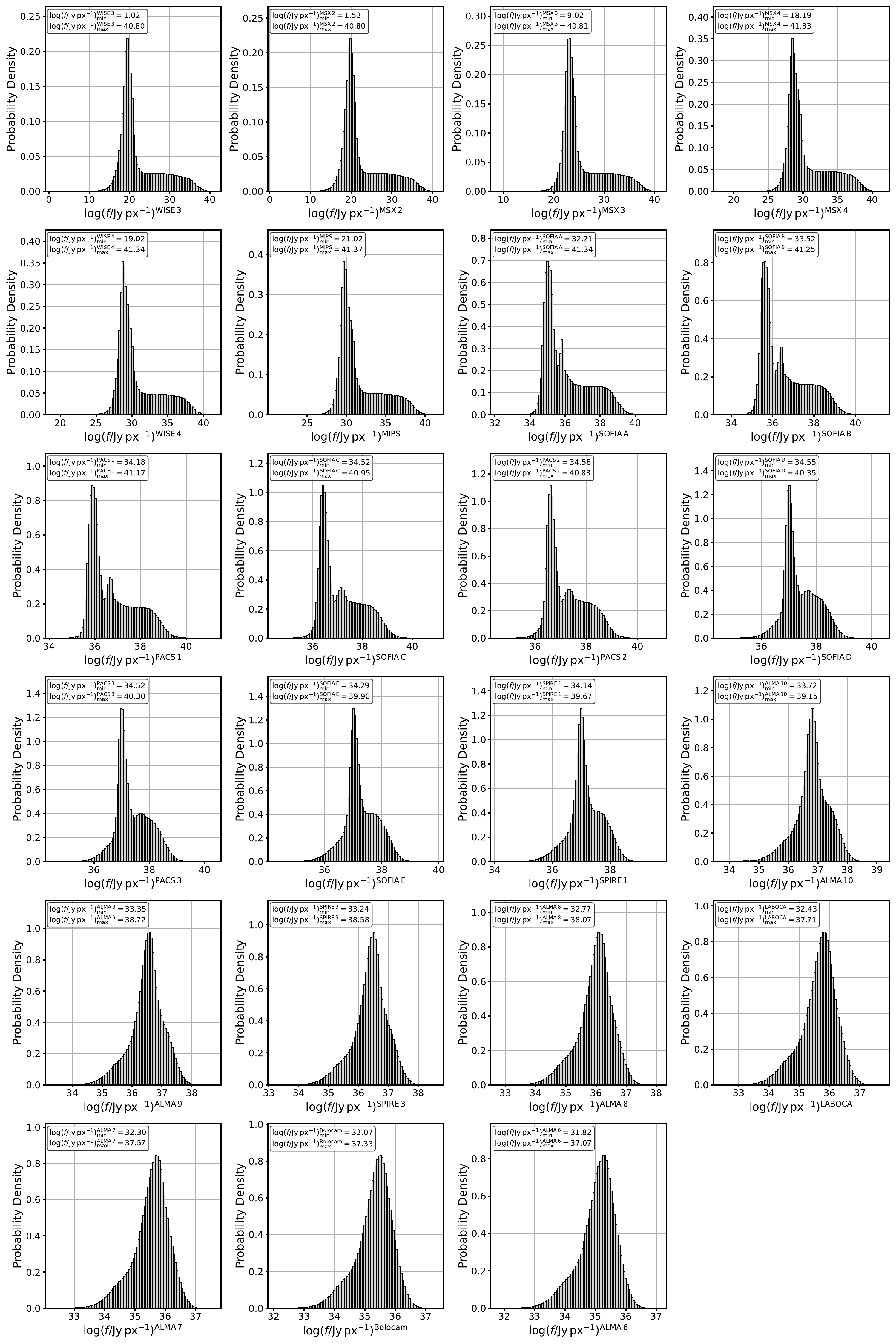}
    \caption{Histograms of the prior distributions for the observables in the training data over all cubes and pixels}
    \label{fig:TS_priors_observables}
\end{figure*}

\begin{figure*}
    \centering
    \includegraphics[width=\textwidth]{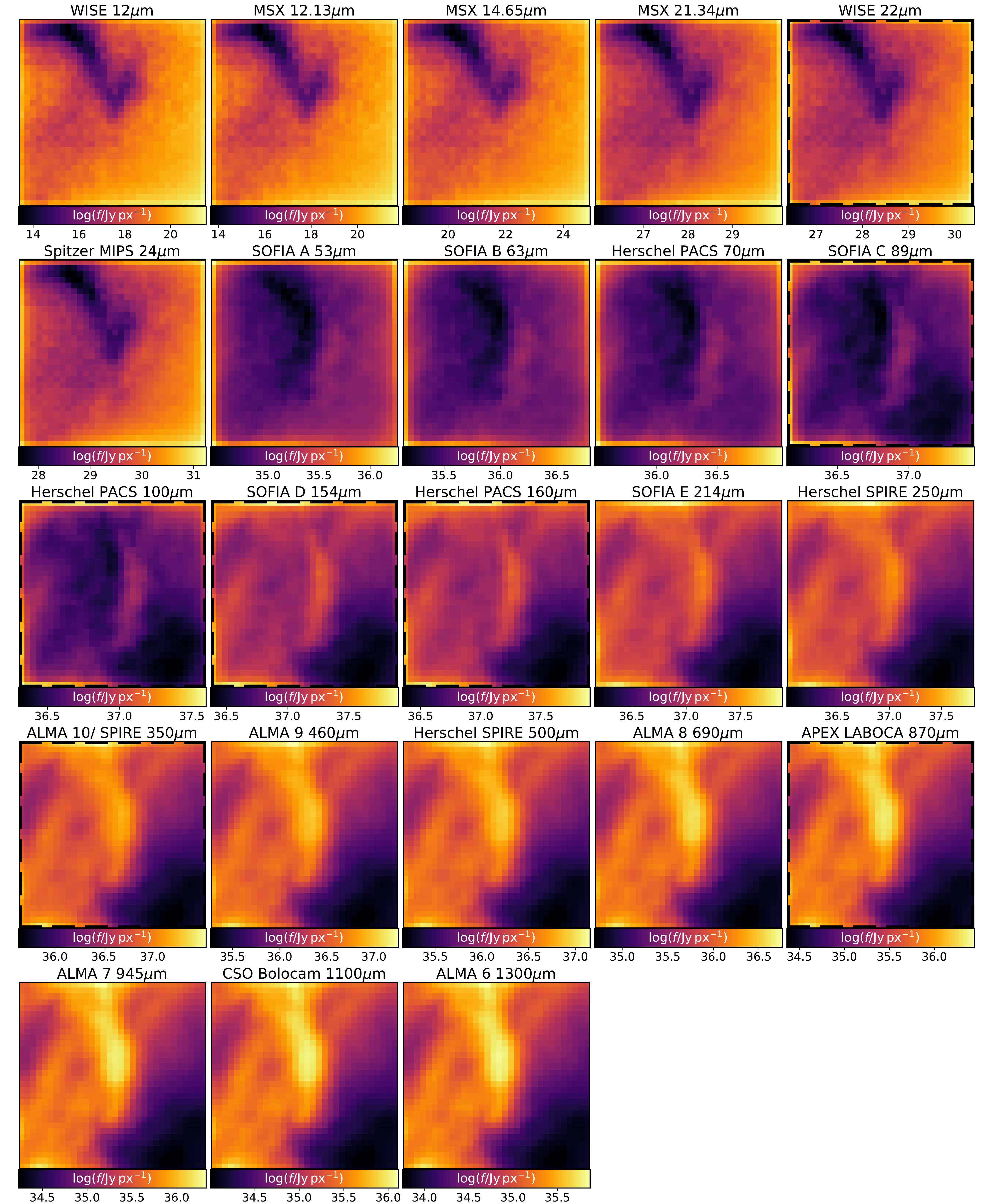}
    \caption{Example of the synthetic dust emission maps at all 23 considered wavelengths that serve as the input to our cINN approach. These correspond to bands of specific instruments (as labelled on the top of each panel), but do not account for PSF-related resolution effects of the respective telescopes. The shown example corresponds to the example cube in the ISRF-only configuration that is also the subject of Figures~\ref{fig:SingleSliceComparison}, \ref{fig:AllSliceComparison_ISRF_all_076412}, and \ref{fig:IsodensitySurface_ISRF_all_076412}. The panels outlined with the black dashed lines correspond to the seven wavelengths considered in our more limited experiment in Section~\ref{sec:Results_SLoS_7Filter}. We emphasise that the presented flux maps are corrected for distance.}
    \label{fig:EmissionMapsExample}
\end{figure*}

This appendix provides additional material with regard to the construction of our training data. Table~\ref{tab:FilterBeamSizes} provides a summary of the different bands we consider in our generation of the synthetic dust emission maps in Section~\ref{sec:Method}. Listed are the instrument name, band ID, the central wavelength we assume for the band, and the reference indicating from where we have extracted the respective information. Figures~\ref{fig:TS_priors_targets} and \ref{fig:TS_priors_observables} provide histograms for the effective prior distributions for the dust density and temperature, and the simulated fluxes in all wavelengths that the cINN sees during training, summarised over all the pixels in our training data set. We emphasise that these effective priors are an outcome of our training data selection, depicting the actual distributions of the respective parameters in the final training data set, and not a prescription after which the training data is selected. Lastly, Figure~\ref{fig:EmissionMapsExample} shows an example of the emission maps generated by POLARIS at the 23 different wavelengths that we consider for the example cube used in Figures~\ref{fig:SingleSliceComparison} (top row), \ref{fig:AllSliceComparison_ISRF_all_076412}, \ref{fig:IsodensitySurface_ISRF_all_076412}, and~\ref{fig:PosteriorSliceComparison_CUBE_076412} (top row).

\begin{table}
    \centering
    \caption{Overview of filter bands and instruments considered in this study. For each instrument and band, we list the central wavelength $\lambda$  and the literature reference for the quoted values. We note that we treat ALMA Band 10 and Herschel SPIRE 2 as one band in our analysis, because they share the same central wavelength and we are not considering any instrument related effects. Consequently, there is no difference in the corresponding synthetic dust emission maps between SPIRE 2 and ALMA Band 10 in our analysis.} 
    \label{tab:FilterBeamSizes}
    \resizebox{0.5\textwidth}{!}{
        \begin{tabular}{lccc}
        \toprule
        Instrument & Band & $\lambda (\mu \mathrm{m})$ & Reference \\ 
        \midrule
        WISE &  3 &  $12$ & \cite{Wright2010} \\
        WISE &  4 &  $22$ & \cite{Wright2010} \\
        MSX/SPIRIT-III &  2 &  $12.13$ & \cite{Egan1999} \\
        MSX/SPIRIT-III &  3 &  $14.65$ & \cite{Egan1999} \\
        MSX/SPIRIT-III &  4 &  $21.34$ & \cite{Egan1999} \\
        Spitzer &  MIPS  &  $24$ & Table 1, \cite{Dole2003} \\
        SOFIA / HAWC+ & A & $53$ & \cite{Harper2018} \\ 
        SOFIA / HAWC+ & B & $63$ & \cite{Harper2018} \\ 
        SOFIA / HAWC+ & C & $89$ & \cite{Harper2018} \\
        SOFIA / HAWC+ & D & $154$ & \cite{Harper2018}\\
        SOFIA / HAWC+ & E & $214$ & \cite{Harper2018} \\
        Herschel &  PACS 1 & $70$ & PACS Photometer Quickstart Guide, \cite{PACS_Quickstart2017}  \\
        Herschel &  PACS 2 & $100$ & PACS Photometer Quickstart Guide, \cite{PACS_Quickstart2017}  \\
        Herschel &  PACS 3 & $160$ & PACS Photometer Quickstart Guide, \cite{PACS_Quickstart2017}  \\
        Herschel &  SPIRE 1 & $250$ & Table 5.2, SPIRE Handbook, \cite{SPIRE_Handbook2018} \\
        Herschel &  SPIRE 2 & $350$ & Table 5.2, SPIRE Handbook, \cite{SPIRE_Handbook2018} \\
        Herschel &  SPIRE 3 & $500$ & Table 5.2, SPIRE Handbook, \cite{SPIRE_Handbook2018} \\
        ALMA &  Band 10 & $350$ & Table 7.1, \cite{ALMA_Handbook2022} \\
             &         &       & Table 7.1, \cite{ALMA_Handbook2022} \\
        ALMA &  Band 9 & $460$ & Table 7.1, \cite{ALMA_Handbook2022} \\
             &         &       & Table 7.1, \cite{ALMA_Handbook2022} \\ 
        ALMA &  Band 8 & $690$ & Table 7.1, \cite{ALMA_Handbook2022} \\
             &         &       & Table 7.1, \cite{ALMA_Handbook2022} \\
        ALMA &  Band 7 & $945$ & Table 7.1, \cite{ALMA_Handbook2022} \\
             &         &       & Table 7.1, \cite{ALMA_Handbook2022} \\
        ALMA &  Band 6 & $1300$ & Table 7.1, \cite{ALMA_Handbook2022} \\
             &         &        & Table 7.1, \cite{ALMA_Handbook2022} \\
        APEX &  LABOCA &  $870$ & APEX LABOCA website, \cite{LABOCA_Website} \\
        CSO &  Bolocam &  $1100$ & Bolocam website, \cite{Bolocam2004} \\
        \bottomrule
        \end{tabular}
    }
\end{table}

\section{Additional material for the performance analysis}

This appendix provides complementary diagrams for the performance analysis of our cINN approach outlined in Section~\ref{sec:Results}. Figure~\ref{fig:AllSliceComparison_ISRF_all_076412} provides an extended comparison of the point estimation results to the ground truth for the example cube shown in the first to rows of Figure~\ref{fig:SingleSliceComparison}, depicting all 32 slices of the cube. As an alternative visualisation of this comparison, Figure~\ref{fig:IsodensitySurface_ISRF_all_076412} shows 3D isodensity surface diagrams for densities of $10^{10}$ (grey) and $10^{11}\,\mathrm{m}^{-3}$ (red) for two different rotation angles of the example cube, comparing again the results of our three point estimators to the ground truth. As discussed in Section~\ref{sec:Results_SLoS_cINN}, this diagram illustrates the very good recovery of the 3D dust structure at intermediate densities, whereas some finer, high-density structures may be subject to underestimation and are, thus, lost in this visualisation.
As an extension to Figure~\ref{fig:SingleSliceComparison}, Figure~\ref{fig:PosteriorSliceComparison_CUBE_076412} provides examples for the predicted posterior distributions in the different network and cube configurations analysed in Figure~\ref{fig:SingleSliceComparison}.

Analogously to Figures~\ref{fig:AllSliceComparison_ISRF_all_076412} and \ref{fig:IsodensitySurface_ISRF_all_076412}, Figures~\ref{fig:AllSliceComparison_STAR_all_076412} and \ref{fig:IsodensitySurface_STAR_all_076412} show the respective results for the cube in the ISRF + star radiation configuration (rows 2\&3 of Figure~\ref{fig:SingleSliceComparison}), Figures~\ref{fig:AllSliceComparison_ISRF_NLoS_076412} and \ref{fig:IsodensitySurface_ISRF_NLoS_076412} present the outcome as derived from the NLoS-cINN (Section~\ref{sec:Results_NLoS_cINN}), and Figures~\ref{fig:AllSliceComparison_ISRF_7filter_076412} and \ref{fig:IsodensitySurface_ISRF_7filter_076412} illustrate the predictions from the SLoS-cINN that is limited to the seven wavelengths configuration (Section~\ref{sec:Results_SLoS_7Filter}). 

\begin{figure*}
    \centering
    \includegraphics[width=\textwidth]{figures/AllSlicesComparison_ISRF_076412_V2.pdf}
    \caption{Comparison of the point estimate prediction results to the ground truth for all slices of one example cube that is only subject to the ISRF. In each panel, the subpanels show from top left to bottom right the cube slices going along the LoS from front to back. The left and right columns show the dust density and dust temperature respectively. From top to bottom, the rows indicate the ground truth and the MAP, MeanShift, and MNPCP estimates based on the outcome of the single LoS cINN using all 23 wavelengths. This diagram shows the full cube of the single slices shown in the first two rows of Figure~\ref{fig:SingleSliceComparison}. }
    \label{fig:AllSliceComparison_ISRF_all_076412}
\end{figure*}

\begin{figure*}
    \centering
    \includegraphics[width=\textwidth]{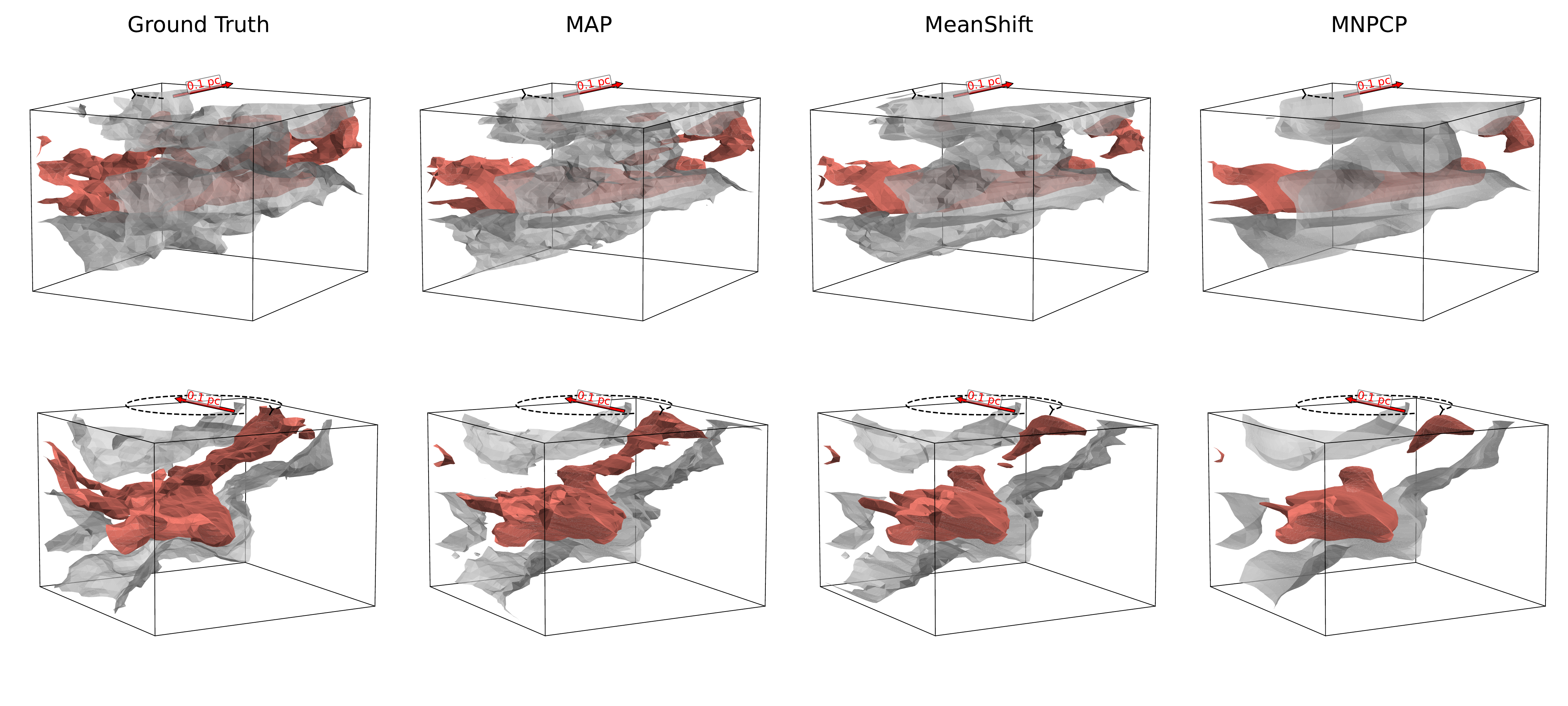}
    \caption{3D isodensity surface diagrams for one example test cube in the ISRF-only radiation configuration, comparing the SLoS-cINN prediction results to the ground truth. Here the two rows show different rotation angles of the given cube. The grey surfaces indicate a density of $10^{10}\,\mathrm{m}^{-3}$, whereas red surfaces mark the $10^{11}\,\mathrm{m}^{-3}$ density level.}
    \label{fig:IsodensitySurface_ISRF_all_076412}
\end{figure*}

\begin{figure*}
    \centering
    \includegraphics[width=\textwidth]{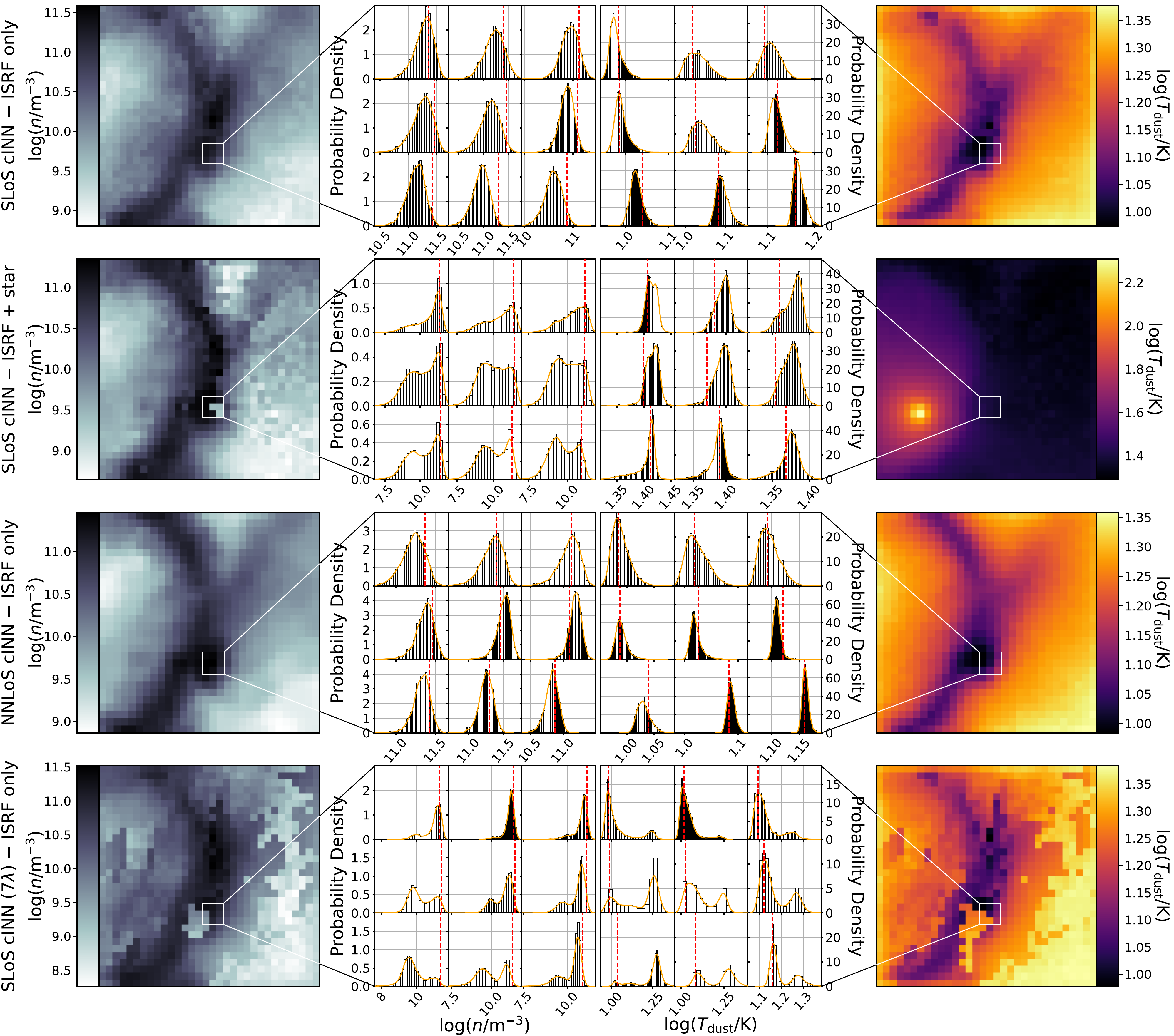}
    \caption{Comparison of the predicted posterior distributions for a few example pixels of the cubes shown in Figure~\ref{fig:SingleSliceComparison}. The panels in the leftmost and right-most column show the MAP prediction results for density and dust temperature for the same slice as in Figure~\ref{fig:SingleSliceComparison}, respectively. The middle two columns provide histograms of the predicted posterior distributions for the nine pixels indicated by the white square in the left-most and right-most columns. The orange curve represents the kernel density estimate used to determine the MAP values, whereas the red dashed line mark the respective ground truth value of each example pixel. We note that within each of the posterior histogram panels, the subpanels share the same x axis column-wise and the same y axis row-wise.}
    \label{fig:PosteriorSliceComparison_CUBE_076412}
\end{figure*}

\begin{figure*}
    \centering
    \includegraphics[width=\textwidth]{figures/AllSlicesComparison_STAR_076412_V2.pdf}
    \caption{Comparison of the point estimate prediction results to the ground truth for all slices of one example cube that is subject to the ISRF and one B4 star. In each panel, the subpanels show (from top-left to bottom-right) the cube slices going along the LoS from front to back. The left and right column show the dust density and dust temperature respectively. From top to bottom, the rows indicate the ground truth and the MAP, MeanShift and MNPCP estimates,  based on the outcome of the SLoS-cINN using all 23 wavelengths. This diagram shows the full cube of the single slice shown in rows 3 and 4 of Figure~\ref{fig:SingleSliceComparison}. The white star symbol in the right column indicates the approximate position of the star, defined as the hottest pixel in the cube for both the ground truth and the prediction results.}
    \label{fig:AllSliceComparison_STAR_all_076412}
\end{figure*}

\begin{figure*}
    \centering
    \includegraphics[width=\textwidth]{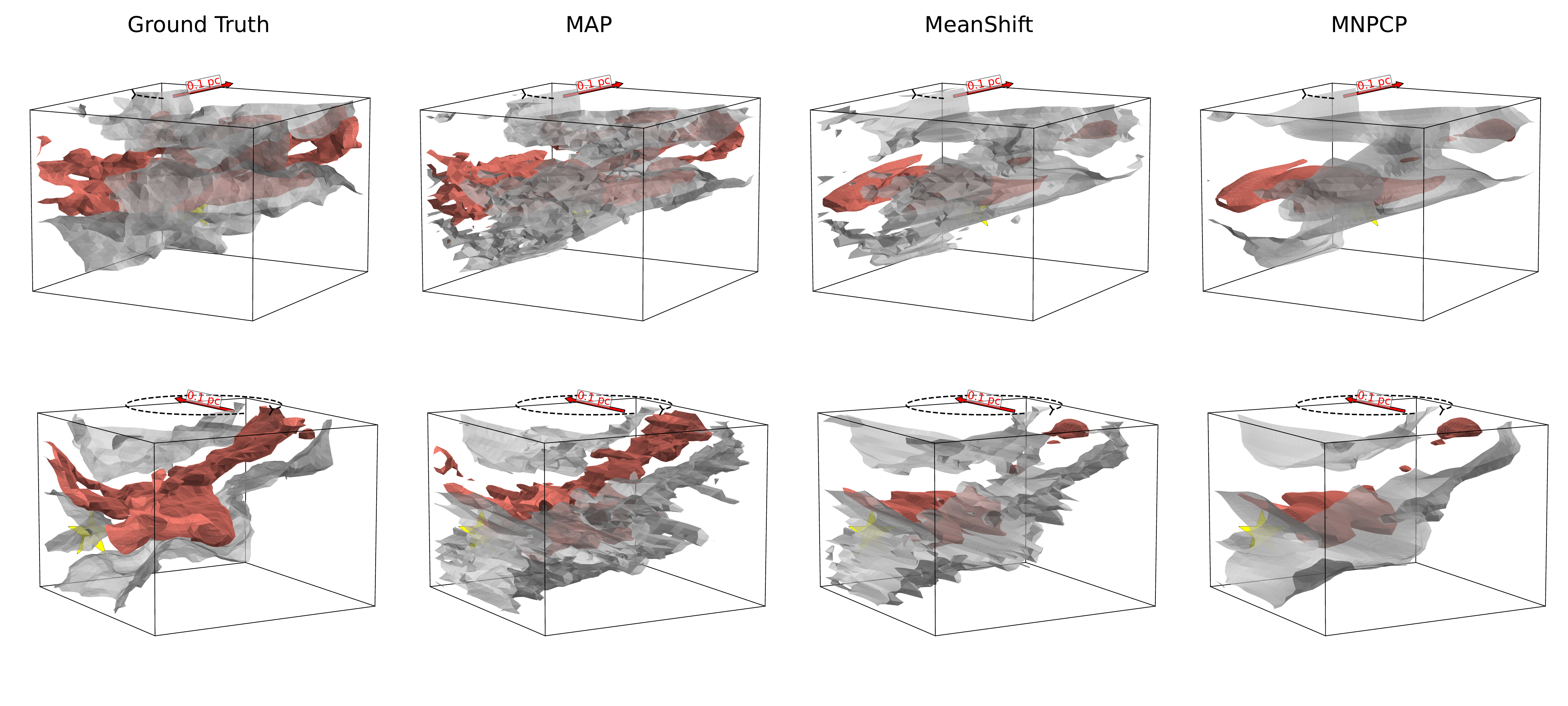}
    \caption{3D isodensity surface diagrams for one example test cube in the ISRF + star radiation configuration, comparing the SLoS-cINN prediction results to the ground truth. Here the two rows show different rotation angles of the given cube. The grey surfaces indicate a density of $10^{10}\,\mathrm{m}^{-3}$, whereas red surfaces mark the $10^{11}\,\mathrm{m}^{-3}$ density level. The yellow star indicates the position of the B4 anaologue placed inside the cube.}
    \label{fig:IsodensitySurface_STAR_all_076412}
\end{figure*}

\begin{figure*}
    \centering
    \includegraphics[width=\textwidth]{figures/AllSlicesComparison_ISRF_NLoS_076412_V2.pdf}
    \caption{Comparison of the point estimate prediction results to the ground truth for all slices of one example cube that is only subject to the ISRF. In each panel, the subpanels show from top left to bottom right the cube slices going along the LoS from front to back. The left and right columns show the dust density and dust temperature, respectively. From top to bottom, the rows indicate the ground truth and the MAP, MeanShift, and MNPCP estimates based on the outcome of the NLoS-cINN using all 23 wavelengths (as opposed to the SLoS-cINN outcome shown in Fig.~\ref{fig:AllSliceComparison_ISRF_all_076412}). This diagram shows the full cube of the single slice shown in rows 5\&6\ of Figure~\ref{fig:SingleSliceComparison}.}
    \label{fig:AllSliceComparison_ISRF_NLoS_076412}
\end{figure*}

\begin{figure*}
    \centering
    \includegraphics[width=\textwidth]{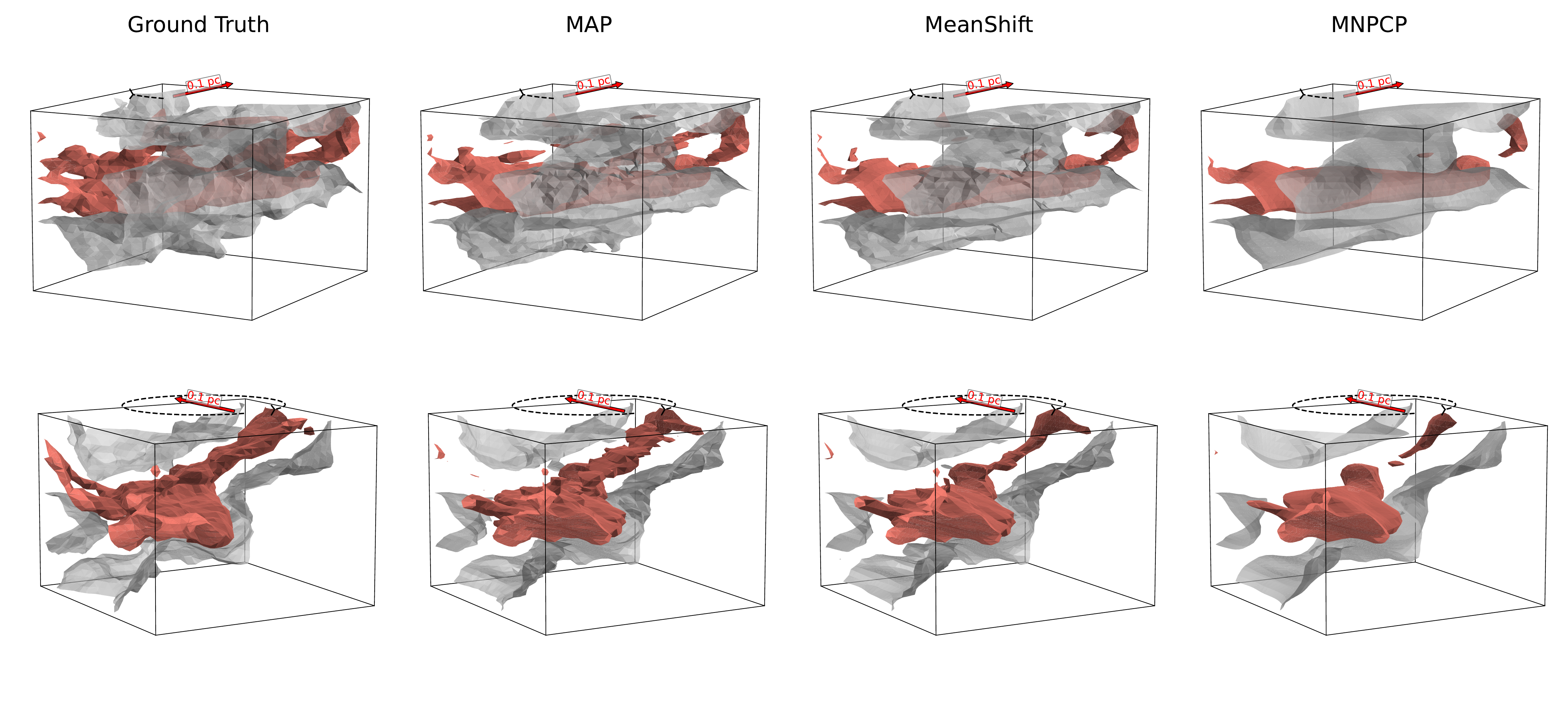}
    \caption{3D isodensity surface diagrams for one example test cube in the ISRF-only radiation configuration, comparing the NLoS-cINN prediction results to the ground truth in the left column. Here, the two rows show different rotation angles of the given cube. The grey surfaces indicate a density of $10^{10}\,\mathrm{m}^{-3}$, whereas red marks the $10^{11}\,\mathrm{m}^{-3}$ density level. These data are analogous to the SLoS-cINN results in Figure~\ref{fig:IsodensitySurface_ISRF_all_076412}.}
    \label{fig:IsodensitySurface_ISRF_NLoS_076412}
\end{figure*}

\begin{figure*}
    \centering
    \includegraphics[width=\textwidth]{figures/AllSlicesComparison_ISRF_7filter_076412_V2.pdf}
    \caption{Comparison of the point estimate prediction results to the ground truth for all slices of one example cube that is only subject to the ISRF. In each panel, the subpanels show (from top-left to bottom-right) the cube slices going along the LoS from front to back. The left and right columns show the dust density and dust temperature respectively.  From top to bottom the rows indicate the ground truth and the MAP, MeanShift, and MNPCP estimates based on the SLoS-cINN that uses only seven wavelengths. This diagram shows the full cube of the single slice shown in the last two rows of Figure~\ref{fig:SingleSliceComparison}.}
    \label{fig:AllSliceComparison_ISRF_7filter_076412}
\end{figure*}

\begin{figure*}
    \centering
    \includegraphics[width=\textwidth]{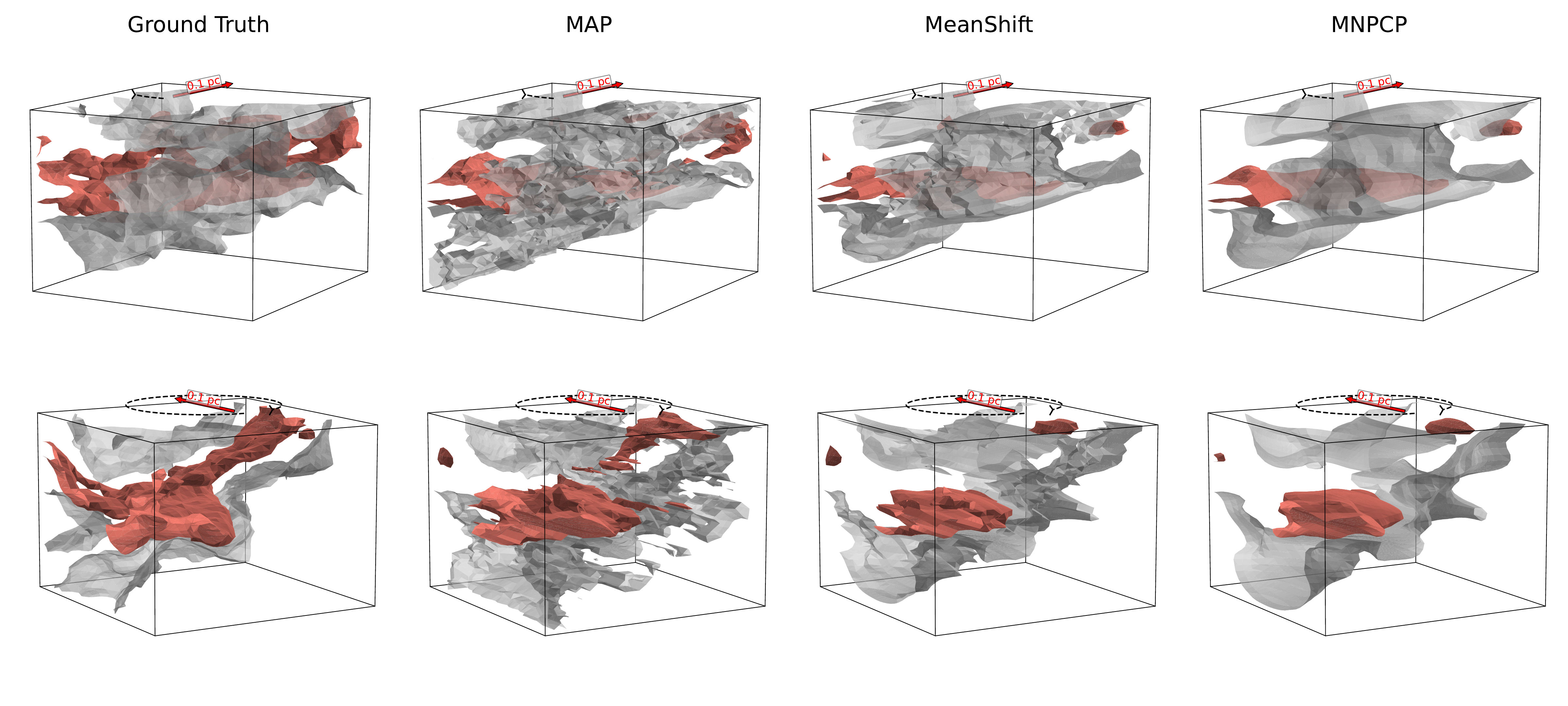}
    \caption{3D isodensity surface diagrams for one example test cube in the ISRF-only radiation configuration, comparing the SLoS-cINN prediction results, based on only seven wavelengths, to the ground truth. Here, the two rows show different rotation angles of the given cube. The grey surfaces indicate a density of $10^{10}\,\mathrm{m}^{-3}$, whereas red surfaces mark the $10^{11}\,\mathrm{m}^{-3}$ density level. These data are analogous to the 23-wavelength SLoS results shown in Figure~\ref{fig:IsodensitySurface_ISRF_all_076412}.}
    \label{fig:IsodensitySurface_ISRF_7filter_076412}
\end{figure*}

\subsection{Influence of the observation angle}
\label{app:RotationTest}
In Section~\ref{sec:TrainingData_SyntheticImages}, we outline that during our training set generation we only simulate synthetic observations for the simulation cubes from one viewing angle. In particular, it our simulated observations are performed from the same direction for all training cubes. Although somewhat unlikely, this circumstance begs the question, whether the chosen viewing angle may have introduced a bias in our trained method, so the reconstruction from other viewing angles may differ. Naturally, the simulated synthetic observations depend on the viewing angle of the cloud as the contribution of different parcels of gas in the cloud to the observed dust emission depends on the 3D position after all. Therefore, it is to be expected that a reconstruction of the same cloud from two different viewing angles is not 100\% identical. To test how the viewing angle may affect the reconstructive performance and rule out a direction bias in our method, we conducted an experiment where we rotated the 50 cubes that are only subject to the ISRF, resimulating the corresponding synthetic dust emission observations and then reconstructing again the 3D distribution of the dust with the 23-wavelength SLoS-cINN. Compared to the main viewing angle, for which we present the performance in Section~\ref{sec:Results_SLoS_cINN}, we tested two additional observation directions. In the first, the cubes are rotated by 90° clockwise around the vertical axis in the plane of the primary viewing angle; namely, we are now observing the right face of the cube instead of the front face. For the second test, we rotated the cube around the horizontal axis by 90° clockwise, so that we are now observing the bottom face of the cube. 

\begin{figure*}
    \includegraphics[width=\textwidth]{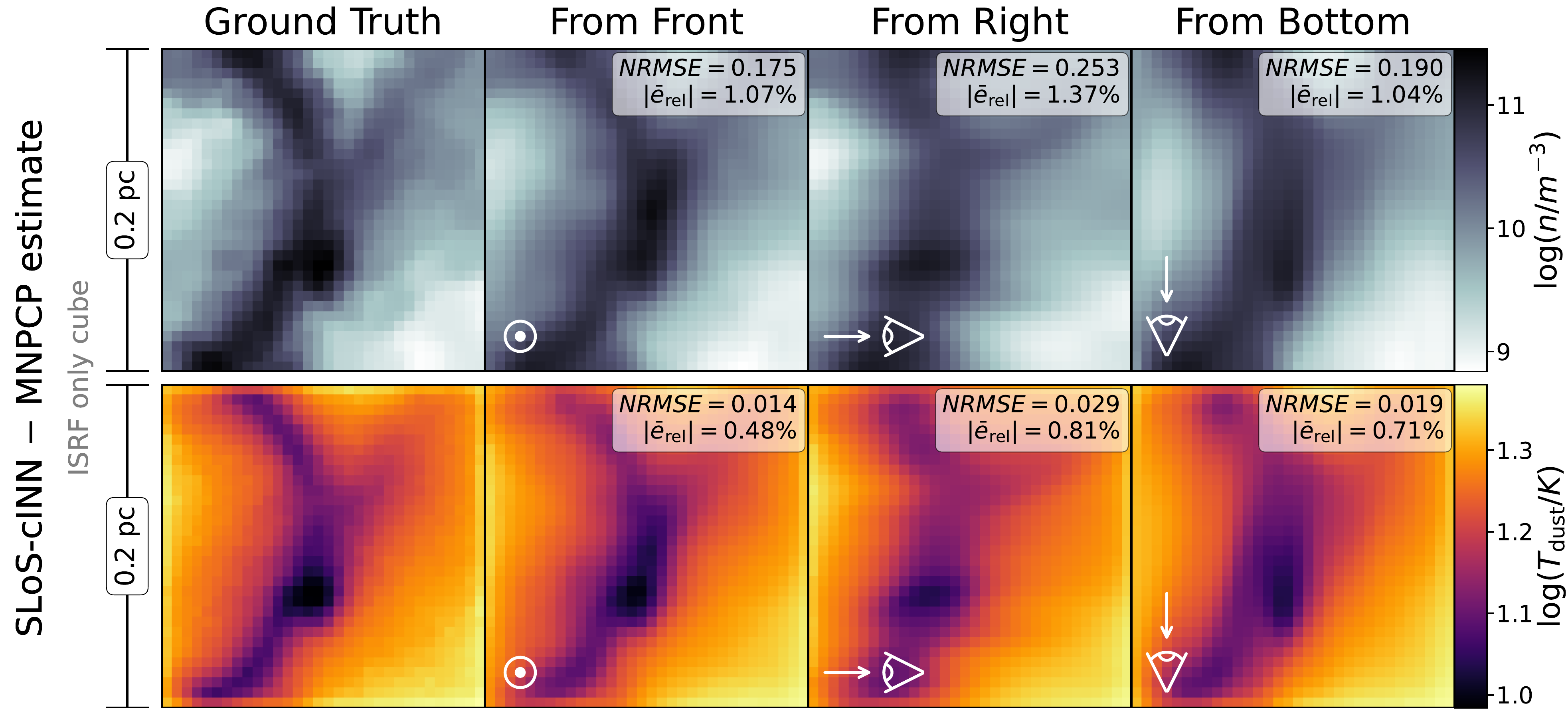}
    \caption{Comparison of MNPCP prediction results from different viewing angles to the ground truth for a single slice of an example cube subject only to the ISRF. Shown is the same example cube as in Figure~\ref{fig:SingleSliceComparison} and all predictions are determined with the 23-wavelength SLoS-cINN. The presented slice is taken along the main viewing angle analysed in this work (Section~\ref{sec:Results}). The results in the second column correspond to the nominal prediction outcome derived from the main viewing angle (i.e. the cube is observed from the front). The arrow and eye symbol in the third and fourth columns indicate the direction from which the cube is observed instead to generate the corresponding prediction result. The NRMSEs and median absolute relative errors listed in the three right-most panels are determined over the depicted slice only.}
    \label{fig:RotationTest_SingleSlice}
\end{figure*}

Figure~\ref{fig:RotationTest_SingleSlice} provides an example comparison of the reconstruction result for one slice (along the original LoS) of one example cube (the same as in Figure~\ref{fig:SingleSliceComparison}) from the three different viewing angles that we have tested. As we can see, the cINN recovers the bulk of the structure in this slice quite well, independently of the observation direction. Naturally, smaller details do differ a bit, but as mentioned before this is to be expected, both because the information encoded in the synthetic dust emission maps may not be identical and because the spatial consistency of the predicted posterior distributions depends on the viewing angles. Where the original viewing angles produces posteriors that are consistent along the LoS going into the plane of the shown slice, the other observation directions provide consistency along the perpendicular LoSs now. As a consequence, features in the reconstruction may appear a little more elongated along the given LoS compared to the other viewing angles, as becomes apparent when comparing the two right columns in Figure~\ref{fig:RotationTest_SingleSlice}. Table~\ref{tab:RotationTest_Performance} provides a summary of the predictive performance of the 23-wavelength SLoS-cINN for the three different viewing angles across all 50 test cubes. As we can see,  the performance is very similar, independently of the direction from where the cubes were observed. As such, we can conclude that the cINN has not been biased towards a prefered observational direction and can reconstruct the synthetic dust clouds quite reliably even from different viewing angles.

\begin{table*}[]
    \centering
    \caption{Performance summary for the viewing angle comparison experiment.  NRMSE and the median absolute relative error $|\bar{e}_{\mathrm{rel}}|$ (along with the 25\% and 75\% quantiles) are given, respectively, for the dust density and temperature prediction evaluated across all pixels of the 50 test cubes subject to the ISRF-only case. All predictions are made with the 23-wavelength SLoS-cINN here (Section~\ref{sec:Results_SLoS_cINN}).}
    \begin{tabular}{cccccc}
         & & & \multicolumn{3}{c}{Viewing angle}  \\
         \cmidrule(rl){4-6}
         Point estimator & Measure & Parameter & Front face & Right face & Bottom face \\
         \midrule
         \multirow{4}{*}{MAP} & \multirow{2}{*}{NRMSE} & $\log(n/\mathrm{m}^{-3})$ & $0.0510$ & $0.0531$ & $0.0517$ \\
         && $\log(T_{\mathrm{dust}}/\mathrm{K})$ & $0.0633$  & $0.0682$ & $0.0648$\\
         \cmidrule(rl){2-6}
         & \multirow{2}{*}{$|\bar{e}_{\mathrm{rel}}|\,(\%)$} & $\log(n/\mathrm{m}^{-3})$ & $1.41_{-0.83}^{+1.45}$ & $1.53_{-0.90}^{+1.58}$ & $1.38_{-0.81}^{+1.39}$ \\
         && $\log(T_{\mathrm{dust}}/\mathrm{K})$ & $0.61_{-0.39}^{+1.03}$ & $0.66_{-0.43}^{+1.16}$ & $0.58_{-0.37}^{+0.96}$ \\
         \midrule
         \multirow{4}{*}{MeanShift} & \multirow{2}{*}{NRMSE} & $\log(n/\mathrm{m}^{-3})$ & $0.0480$ & $0.0499$ & $0.0490$ \\
         && $\log(T_{\mathrm{dust}}/\mathrm{K})$ & $0.0522$ & $0.0561$ & $0.0455$ \\
         \cmidrule(rl){2-6}
         & \multirow{2}{*}{$|\bar{e}_{\mathrm{rel}}|\,(\%)$} & $\log(n/\mathrm{m}^{-3})$ & $1.53_{-0.88}^{+1.42}$ & $1.62_{-0.94}^{+1.52}$ & $1.48_{-0.85}^{+1.38}$ \\
         && $\log(T_{\mathrm{dust}}/\mathrm{K})$ & $0.70_{-0.45}^{+1.12}$ & $0.73_{-0.47}^{+1.23}$ & $0.65_{-0.41}^{+1.03}$\\
         \midrule
        \multirow{4}{*}{MNPCP} & \multirow{2}{*}{NRMSE} & $\log(n/\mathrm{m}^{-3})$ & $0.0456$ & $0.0476$ & $0.0455$ \\
         && $\log(T_{\mathrm{dust}}/\mathrm{K})$ & $0.0506$ & $0.0545$ & $0.0501$ \\
         \cmidrule(rl){2-6}
         & \multirow{2}{*}{$|\bar{e}_{\mathrm{rel}}|\,(\%)$} & $\log(n/\mathrm{m}^{-3})$ & $1.40_{-0.80}^{+1.33}$ & $1.50_{-0.87}^{+1.44}$ & $1.37_{-0.79}^{+1.30}$ \\
         && $\log(T_{\mathrm{dust}}/\mathrm{K})$ & $0.70_{-0.42}^{+1.02}$ & $0.74_{-0.45}^{+1.15}$ & $0.67_{-0.40}^{+0.96}$\\
         \bottomrule
    \end{tabular}
    \label{tab:RotationTest_Performance}
\end{table*}

\subsection{Influence of the instrument PSF}
\label{app:BeamConvTest}
In Section~\ref{sec:TrainingData_SyntheticImages}, we are ignoring any instrument related effects in our generation of the synthetic dust emission observations used throughout the rest of this study. Our reasoning for this is that proper modelling of these effects is not trivial. Nevertheless, the latter will be a necessary step in the ongoing development of our approach in order to ultimately apply it to real observational data. In this appendix, we perform a simplified test to demonstrate how a cINN that is trained on fully resolved data behaves on data that is subject to resolution effects. For this purpose, we took the 50 test cubes subject only to the ISRF and modified their corresponding synthetic dust emission maps to simulate the influence of the PSFs of the respective telescope instruments. For simplicity, we approximated this effect by convolving the fully resolved synthetic dust emission maps with a Gaussian, where the standard deviations are matched to the full width half maxima (FWHM) of the PSFs. Notably, this includes the implicit assumption that all considered telescopes would have the same pixel size; in this case, matching that of our input simulated data. 

\begin{figure*}
    \centering
    \includegraphics[width=\textwidth]{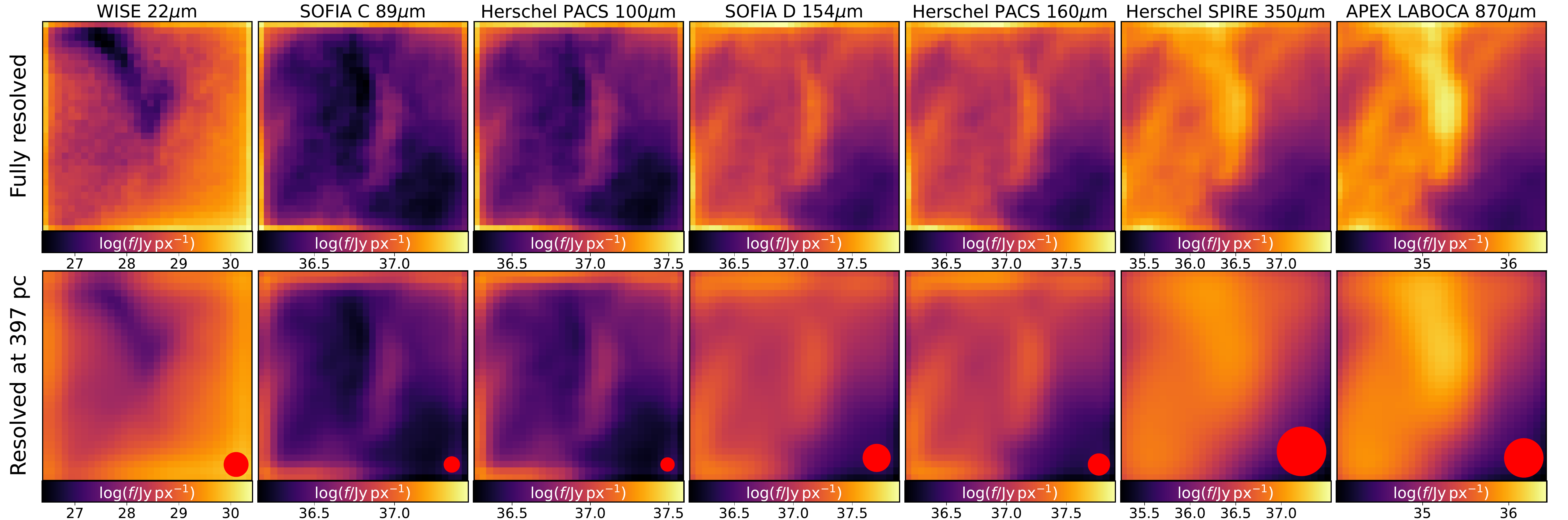}
    \caption{Comparison of dust emission maps between the perfect resolution scenario and a case that accounts for the PSFs of the respective instruments for the limited filter configuration used in Section~\ref{sec:Results_SLoS_7Filter}. The red circle in the bottom row panels indicates the FWHM of the respective PSFs. We note that this example is based on the assumption that all considered telescopes share the same pixel size, matching our simulation resolution at a query distance of $d=397$ pc.}
    \label{fig:BeamConvTest_EmissionMaps}
\end{figure*}

For the convolution, we assumed a distance to the observed cubes, at which the PSFs of all considered instruments are sampled by at least two pixels in the dust emission maps (i.e.  to fulfil the Nyquist sampling criterion). Given the excellent spatial resolution of ALMA, this distance would be quite large if we were to consider all 23 simulated wavelengths (namely larger than $8500$\,pc). Therefore, we conduct this test only on the seven wavelength subset used in Section~\ref{sec:Results_SLoS_7Filter}, for which the PSF sampling criterion is fulfilled at a distance of $d = 397$\,pc. Figure~\ref{fig:BeamConvTest_EmissionMaps} shows an example for how this PSF consideration affects the input dust emission maps (analogously to Figure~\ref{fig:EmissionMapsExample}, we select the same example cube here). As we can see, if we require that all seven considered filters are Nyquist sampled, meaning that the selected distance is naturally dictated by the instrument with the best resolution out of the seven (in this case Herschel PACS at $100\,\mu\mathrm{m}$), then there is already a notable loss in detail in the input dust emission maps for instruments with a worse resolution. 

\begin{figure*}
    \centering
    \includegraphics[width=\textwidth]{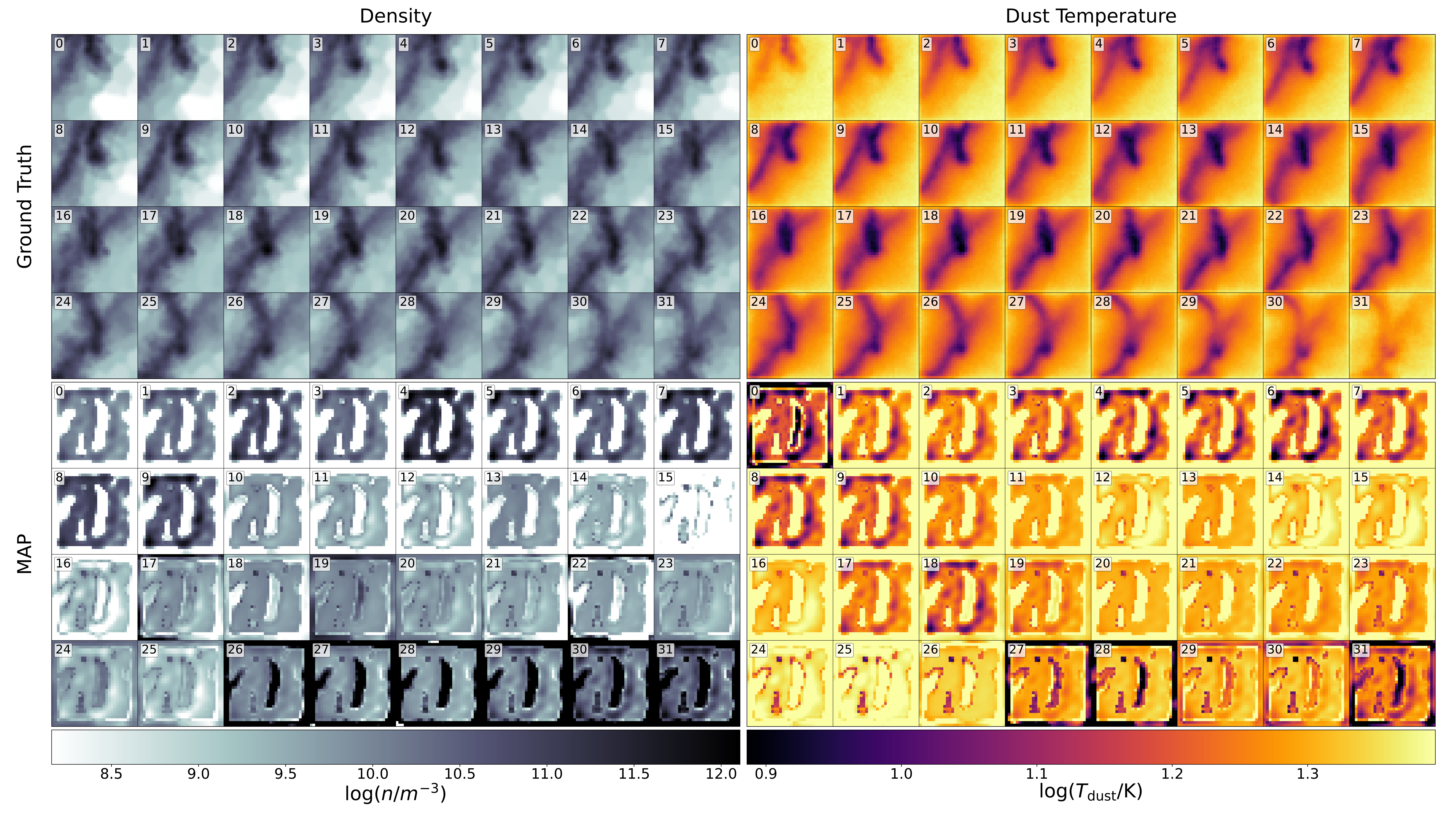}
    \caption{Comparison of the seven wavelength SLoS-cINN MAP prediction result to the ground truth for one example cube that is only subject to the ISRF. Same cube as in Figure~\ref{fig:AllSliceComparison_ISRF_7filter_076412} is shown, with the difference being that here the input dust emission maps were first convolved with the PSFs of the respective instruments, corresponding to the input shown in the bottom row in Figure~\ref{fig:BeamConvTest_EmissionMaps}.}
    \label{fig:BeamConvTest_PredResult}
\end{figure*}

We went on to test how a cINN (namely the seven wavelength SLoS-cINN discussed in Section~\ref{sec:Results_SLoS_7Filter}) that is trained on perfectly resolved data performs on input observations that are subject to these varying resolution effects at the $397$\,pc distance on the 50 ISRF-only test cubes. Figure~\ref{fig:BeamConvTest_PredResult} presents an example MAP prediction result in comparison to the ground truth for the same cube shown in Figure~\ref{fig:AllSliceComparison_ISRF_all_076412}. As we can see, the cINN can no longer recover the dust densities and temperatures at all, a result that we find for all 50 tested cubes. This outcome is not surprising, however, as this resolution effect is not small, rendering data affected by it likely quite far outside of the observable domain that this cINN has learned, where it is bound to fail. While this particularly strong failure here might be exacerbated by the limited wavelength coverage or filter choice, this still confirms the necessity to account for instrumental effects already during training. 

\begin{figure*}
    \centering
    \includegraphics[width=\textwidth]{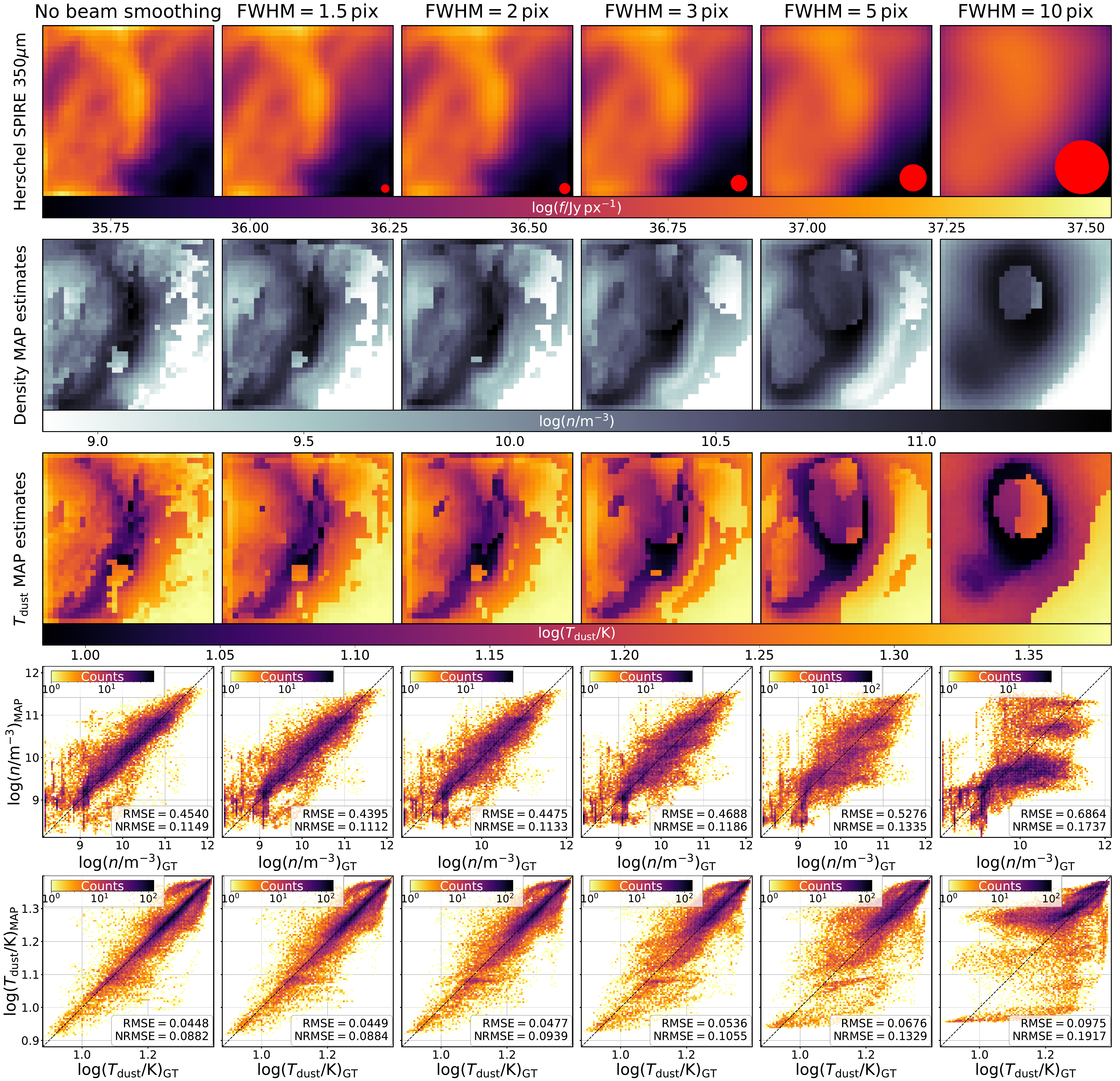}
    \caption{Comparison of prediction results with the seven wavelength SLoS-cINN for different amounts of smoothing applied to the input observations at all wavelengths equally. The top row shows the emission maps at the central wavelength of the Herschel SPIRE 350 $\mu$m filter as a an example to illustrate the effect of the convolution with a Gaussian beam. The red circle in each of these panels indicates the FWHM of the respective Gaussian kernel. The second and third row show the corresponding MAP estimates for dust density and temperature, respectively, for one example slice of the same cube analysed throughout the paper (i.e.~the same slice as in Figures~\ref{fig:SingleSliceComparison}, \ref{fig:PosteriorSliceComparison_CUBE_076412}, \ref{fig:RotationTest_SingleSlice}). The fourth and the fifth rows present a 2D histogram that directly compares the ground truth to the predicted density and temperature MAP estimates, respectively, for all pixels of this test cube, summarising the predictive performance.}
    \label{fig:BeamConvTest_SameResResults}
\end{figure*}

To identify at which point the prediction breaks down, we followed up with an additional test that further simplifies the consideration of the PSF effects. Instead of convolving the emission maps in each wavelength with the (approximate) PSF of the corresponding instrument filter, we applied the same PSF resolution to all wavelengths and investigated the beam size at which the prediction starts to deteriorate. For simplicity, we directly specified the FWHM of the Gaussian convolution kernel in pixels for this experiment. For reference, the tested FWHM values of 1.5, 2, 3, 5, and 10 pixels correspond to distances of $79.9$, $106.5$, $159.8$, $266.4,$ and $532.7$\,pc, respectively, when using (for instance) the Herschel SPIRE $350\,\mu$m filter. Figure~\ref{fig:BeamConvTest_SameResResults} provides a summary of this test for the same example cube (in the ISRF-only configuration) used in Figure~\ref{fig:BeamConvTest_PredResult}, comparing the prediction results on the smoothed emission maps for varying FWHMs to the reference outcome on the original synthetic observations. As we can see, the cINN predictions are somewhat robust with respect to the smoothing of the input emission maps up to a FWHM of the Gaussian beam of 3 pixels, exhibiting only small changes in the NRMSEs for this example. However, the prediction outcome rapidly deteriorates for the larger tested FWHMs of 5 and 10 pixels. Given this outcome, the failure in the previous test might be explained by a combination of the complexity of varying resolution at each wavelength and some rather large PSFs at the tested distance for the central wavelengths of Herschel SPIRE 350$\mu$m and APEX LABOCA. In conclusion, this test shows that our approach can compensate a minor degree of unaccounted for resolution effects if they are the same at all wavelengths. Nevertheless, for realistic applications with variations between observational instruments, proper modelling of the instrumental effects within the training data will be necessary to build a truly robust reconstruction model.

\end{appendix}
\end{document}